\documentclass[final,5p,times,twocolumn,longtitle]{elsarticle-la}

\usepackage[dvips]{color}
\usepackage{NIMpaper}
\usepackage{float}
\usepackage{subfig}
\usepackage[abs]{overpic}
\usepackage{lineno}
\usepackage{graphicx}
\DeclareGraphicsExtensions{{eps}{epsi}}

\usepackage{amssymb}
\usepackage{url}

\journal{Nuclear Instruments and Methods in Physics}

\begin{document}

%\begin{frontmatter}



%% Add symbol footnote def
\long\def\symbolfootnote[#1]#2{\begingroup%
\def\thefootnote{\fnsymbol{footnote}}\footnote[#1]{#2}\endgroup}

%\title{The T2K Long Baseline Neutrino Oscillation Experiment}
\title{The T2K Experiment}


%\author{}
%\address{}
%%%%%%%%%%%%%%%%%%%%%%%%%%%%%%%%%%%%%%%%%%%%%%%%%%%%%%%%%%%%%%
% T2K author list generated on Fri, 24 Dec 2010 05:19:35 +0900
% setting: extra = 0
% Initial Number of authors = 318
% continuously being modified by C.K. Jung
%%%%%%%%%%%%%%%%%%%%%%%%%%%%%%%%%%%%%%%%%%%%%%%%%%%%%%%%%%%%%%

\author[19]{K.\,Abe}
\author[46]{N.\,Abgrall}
\author[27]{H.\,Aihara}
\author[21]{Y.\,Ajima}
\author[57]{J.B.\,Albert}
\author[47]{D.\,Allan}
\author[1]{P.-A.\,Amaudruz}
\author[47]{C.\,Andreopoulos}
\author[11]{B.\,Andrieu}
\author[55]{M.D.\,Anerella}
% ...

% Footnotes
\cortext[deceased]{Deceased}
\fntext[fn1]{Also at JINR, Dubna, Russia}
\fntext[fn3]{Also at Institute of Particle Physics, Canada} 
\fntext[fn2]{Also at BMCC/CUNY, New York, New York, U.S.A.}

\begin{collab}
(The T2K Collaboration)\\
\end{collab}

% ...
\address[55]{Brookhaven National Laboratory, Physics Department, Upton, New York, U.S.A.}
% ...
\address[57]{Duke University, Department of Physics, Durham, North Carolina, U.S.A.}
% ...
\address[46]{University of Geneva, Section de Physique, DPNC, Geneva, Switzerland}
% ...
\address[21]{High Energy Accelerator Research Organization (KEK), Tsukuba, Ibaraki, Japan}
% ...
\address[11]{UPMC, Universit\'e Paris Diderot, CNRS/IN2P3, Laboratoire de Physique Nucl\'eaire et de Hautes Energies (LPNHE), Paris, France}
% ...
\address[47]{STFC, Rutherford Appleton Laboratory, Harwell Oxford, United Kingdom}
\address[27]{University of Tokyo, Department of Physics, Tokyo, Japan}
\address[19]{University of Tokyo, Institute for Cosmic Ray Research, Kamioka Observatory, Kamioka, Japan}
% ...
\address[1]{TRIUMF, Vancouver, British Columbia, Canada}
% ...

\begin{abstract}
The T2K experiment is a long-baseline 
neutrino oscillation experiment. Its main goal is to measure 
the last unknown lepton sector mixing angle $\theta_{13}$ by observing $\nu_e$
appearance in a $\nu_\mu$ beam. It also aims to make a precision 
measurement of the known
%\numu{}\goesto{}\nutau{} 
oscillation parameters,
$\dms{}_{23}$ and $\sstt{}_{23}$, via 
\numu{} disappearance studies. 
Other goals of the experiment 
include various neutrino cross section measurements and sterile 
neutrino searches. 
The experiment uses an intense proton beam generated by the J-PARC 
accelerator in Tokai, Japan, and is composed of a neutrino beamline, 
a near detector complex (ND280), and a far detector (Super-Kamiokande)  
located 295~km away from 
%the proton beam target in 
J-PARC.
This paper provides a comprehensive review of the 
instrumentation aspect of the 
T2K experiment and a summary of the vital information for each subsystem.

\end{abstract}


%\end{frontmatter}

\maketitle

%\linenumbers

%\twocolumn[]

\section{Introduction}
\label{introduction} 
% \section{Introduction}
% Lead Author C.K. Jung

The T2K (Tokai-to-Kamioka) experiment~\cite{:2003zr}
is a 
long baseline neutrino oscillation experiment
designed to probe the mixing of the muon neutrino with other species and
shed light on the neutrino mass scale. It is the first long baseline
neutrino oscillation experiment proposed and approved to look explicitly
for the electron neutrino appearance from the muon neutrino, thereby
measuring $\theta_{13}$, the last unknown mixing angle in the lepton
sector.

T2K's physics goals include the measurement of the
%\numu{}\goesto{}\nutau{} 
neutrino oscillation parameters with precision of
$\delta{}(\dms{}_{23}) \sim 10^{-4}\eVs{}$ and 
$\delta{}(\sstt{}_{23}) \sim 0.01$ via \numu{} disappearance studies, 
and achieving a factor of about 20 better sensitivity compared to
the current best limit on $\theta_{13}$
from the CHOOZ experiment~\cite{Apollonio:2002gd} through
the search for \numu{}\goesto{}\nue{} appearance 
($\sstt{}_{\mu{}e} \simeq \frac{1}{2}\sstt{}_{13} > 0.004$ 
at 90\% CL for CP violating phase $\delta=0$).  In
addition to neutrino oscillation studies, the T2K neutrino beam (with
$E_\nu \sim 1~\GeV{}$) will enable a rich fixed-target physics program of
neutrino interaction studies at energies covering the transition between
the resonance production and deep inelastic scattering regimes.

\begin{figure}[htb]
\centering\includegraphics[width=\linewidth]{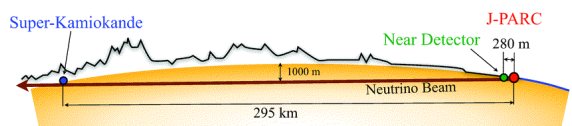}
\caption{A schematic of a neutrino's journey from the neutrino beamline 
at J-PARC, through the near detectors (green dot) which are used to 
determine the properties of the neutrino beam, and then 295 km underneath 
the main island of Japan to Super-Kamiokande.}
\label{fig:T2Klayout}
\end{figure}

T2K uses Super-Kamiokande~\cite{Fukuda:2002uc} as the far
detector to measure neutrino rates at a distance of 295~km from the
accelerator, and near detectors to sample the beam just after production. 
The experiment includes a neutrino beamline and a near
detector complex at 280~m (ND280), both of which were newly constructed. 
Super-Kamiokande was upgraded
and restored to 40\% photocathode coverage (the same as the original
Super-Kamiokande detector) with new photomultiplier tubes in 2005--06,
following the accident of 2001.
Fig.~\ref{fig:T2Klayout} shows a schematic layout of the T2K experiment 
as a whole.

T2K adopts the off-axis method \cite{Beavis:1995up} to generate the
narrow-band neutrino beam using the new MW-class proton
synchrotron at
J-PARC\footnote{Japan Proton Accelerator Research Complex 
jointly constructed and operated by KEK and JAEA.}. 
In this method the neutrino beam is purposely 
directed at an angle with respect to the baseline connecting the 
proton target and the far detector, Super-Kamiokande.
The off-axis angle is set at 2.5\degree{} so that the narrow-band 
muon-neutrino beam generated toward the far detector has a peak energy at 
$\sim$0.6~GeV, which maximizes the effect of the neutrino oscillation 
at 295~km and minimizes the background to electron-neutrino
appearance detection.
The angle can be reduced to $2.0$\degree{}, allowing variation of
the peak neutrino energy, if necessary.

The near detector site
at $\sim$280~m from the production target
houses on-axis and off-axis detectors.
The on-axis detector (INGRID), composed of an array of 
iron/scintillator sandwiches, measures the neutrino beam
direction and profile.
The off-axis detector, immersed in a magnetic
field, measures the muon neutrino flux and 
energy spectrum, and intrinsic electron neutrino
contamination in the beam in the direction of the far detector, along with measuring
rates for exclusive neutrino reactions. These measurements are essential
in order to characterize signals and backgrounds that are observed in the
Super-Kamiokande far detector.

The off-axis detector is composed of: a water-scintillator 
detector optimized to identify
$\pi^0$'s (the \pod{}); the tracker
consisting of time projection chambers (TPCs) and 
fine grained detectors (FGDs) 
optimized to study charged current interactions; 
and an electromagnetic calorimeter (ECal) 
that surrounds the \pod{} and the tracker. 
The whole off-axis detector is placed in 
a 0.2~T magnetic field provided by the
recycled UA1 magnet, which also serves as part of a side muon range
detector (SMRD).  

The far detector, Super-Kamiokande, is located in the Mozumi
mine of the Kamioka Mining and Smelting
Company, near the village of Higashi-Mozumi,
Gifu, Japan. The detector cavity lies under the
peak of Mt. Ikenoyama, with 1000~m of rock, or
2700 meters-water-equivalent (m.w.e.) mean overburden.
It is a water Cherenkov detector
consisting of a welded stainless-steel tank,
39~m in diameter and 42~m tall, with a total nominal
water capacity of 50,000~tons. 
The detector contains approximately 13,000 photomultiplier
tubes (PMTs) that image neutrino interactions in pure water.
Super-Kamiokande has been
running since 1996 and has had four distinctive running periods. 
The latest period, SK-IV, is running stably and features upgraded PMT
readout electronics. A detailed description of the detector 
can be found elsewhere~\cite{Fukuda:2002uc}.

%The Japanese government approved the T2K experiment in 2003.  
%The total
%approved budget for the five-year construction plan was \textyen{}16B
%($\sim$\$160M), which included funds for the neutrino beamline and the
%ND280 detector. 
Construction of the neutrino beamline started 
in April 2004. 
The complete chain of accelerator and neutrino beamline was
successfully commissioned during 2009, and 
T2K began 
accumulating neutrino beam data for physics analysis in January 2010.
%neutrino data taking in January 2010.  
%By the end of 2010  greater than 100\unit{\kilo\watt{}} 
%beam power operation was achieved. 
%In this
%period, we achieved a continuous beam power of 50\unit{\kilo\watt{}},
%demonstrated, albeit briefly, a maximum power of 100\unit{\kilo\watt{}},
%and collected a total of $3.36\times{}10^{19}$ protons on target for use
%in physics analysis.  

%The SK-IV detector has collected a total of 33 fully
%contained T2K events (with 23 inside the fiducial volume and with 
%$E_{vis} > 30\unit{\MeV{}}$.

Construction of the majority of the ND280 detectors was completed in 2009 with the full
installation of INGRID,
the central ND280 off-axis sub-detectors 
(\pod{}, FGD, TPC and downstream ECal) and the SMRD. 
The ND280 detectors began stable operation in February 2010.
The rest of the ND280 detector (the ECals) was completed in the fall of 2010.
%and have recorded
%more than 95\% of the delivered beam.

The T2K collaboration consists of over 500 physicists and technical staff
members from 59 institutions in 12 countries 
(Canada, France, Germany, Italy, Japan, Poland, Russia, 
South Korea, Spain, Switzerland, 
the United Kingdom and the United States).  

This paper provides a comprehensive 
review of the instrumentation aspect of the
T2K experiment and a summary of the vital information for each subsystem.
Detailed descriptions of some of the major subsystems, and their performance,
will be presented in separate technical papers.
%More detailed descriptions of some of the major 
%subsystems and their performance 
%will be presented in separate technical papers, and 
%are not included in this paper. 
 %Jung
\section{J-PARC Accelerator}
\label{accelerator}
% \section{J-PARC Accelerator}
% Lead Author Unknown, J.-M. POUTISSOU and C. K. Jung

J-PARC, which
was newly constructed at Tokai, Ibaraki,
consists of three
accelerators~\cite{J-PARC}: a linear accelerator (LINAC), a
rapid-cycling synchrotron (RCS) and the main ring (MR) synchrotron. 
An H$^{-}$ beam is accelerated up to 400~MeV\footnote{
Note that from here on all accelerator beam energies 
given are kinetic energies.} 
(181~MeV at present) by the LINAC, and is converted to an H$^{+}$ beam by 
charge-stripping foils at the RCS injection. The beam is accelerated up to
3~GeV by the RCS with a 25~Hz cycle. The harmonic number of the RCS is two, and
there are two bunches in a cycle.  
About 5\% of these bunches are supplied to the MR. 
The rest of the bunches are supplied to 
the muon and neutron beamline in the Material and Life Science Facility. 
The proton beam injected into the MR is accelerated up to 30~GeV. 
The harmonic number of the MR is nine, and the number of
bunches in the MR is eight (six before June 2010). There are two extraction
points in the MR: slow extraction for the hadron beamline and
fast extraction for the neutrino beamline.

In the fast extraction mode, the eight circulating proton
bunches are extracted 
within a single turn by a set of five kicker magnets. 
The time structure of the 
extracted proton beam is key to 
discriminating various backgrounds, including cosmic rays,
in the various neutrino detectors.
The parameters of the J-PARC MR for the fast extraction are 
listed in Tab.~\ref{tab:nubeam parameter}.
%Neutrino-induced events are searched for in 
%a 10 $\mu$sec 
%window centered upon the arrival time of the protons on 
%target plus the time of flight 
%of the relativistic neutrinos to the detector. Beam-induced neutrino 
%events are expected
% in narrow time bins corresponding to each proton bunch ($\sim$58~ns wide). 
%An effective duty 
%factor of $1.3 \times 10^{-7}$ is realized.
%
%Continuing efforts to upgrade the MR beam power are planned, and the T2K
%neutrino beamline, which will be described in the following sections, is
%designed accordingly. The proton beam transport, production target 
%and neutrino facilities have been designed to
%accommodate the full design power of the MR accelerator 
%(750 kW) while the shielding of the
% target area, decay tunnel and beam dump itself 
%will be sufficient to accommodate future
% power upgrades  up to 2 MW.
 %???
\section{T2K Neutrino Beamline}
\label{beamline}
% \section{T2K Neutrino Beam Line}
% Lead Author: Nakadaira/Ichikawa/Marchionni

Each proton beam spill consists of eight proton bunches
extracted from the MR to the T2K neutrino beamline,
which produces the neutrino beam.

\begin{figure}[t]
  \begin{center}
    \includegraphics[keepaspectratio=true,width=80mm]{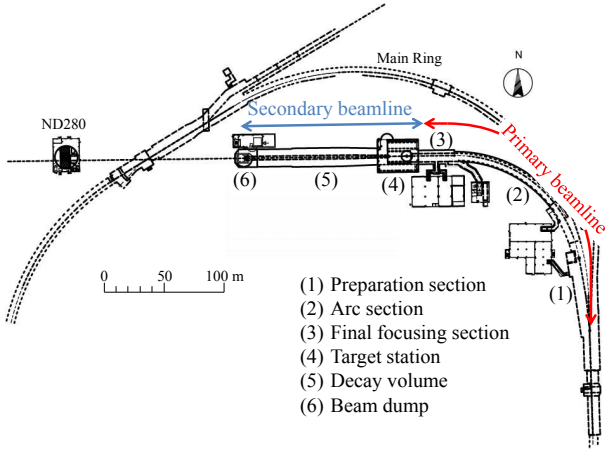}
    \caption[]
    {\label{fig:nu beamline}
    Overview of the T2K neutrino beamline.}
  \end{center}
\end{figure}

The neutrino beamline is composed of two sequential sections: the primary and
secondary beamlines.  In the primary beamline, the extracted proton
beam is transported to point toward Kamioka.  
In the secondary beamline,
the proton beam impinges on a target to produce secondary pions, which
are focused by magnetic horns and decay into neutrinos.  An overview
of the neutrino beamline is shown in Fig.~\ref{fig:nu beamline}.
Each component of the beamline is described in this section.

\begin{figure}[t]
  \begin{center}
    \includegraphics[keepaspectratio=true,width=80mm]{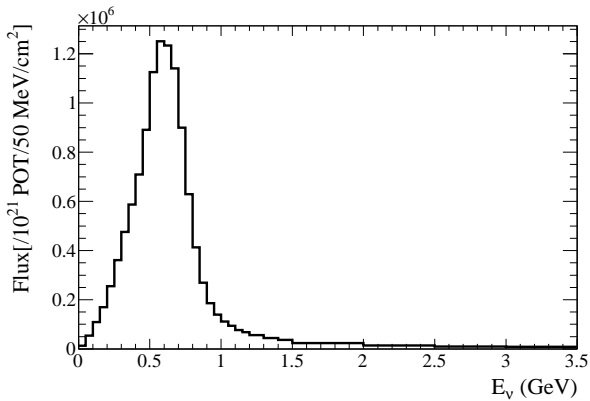}
    \caption[]
    {\label{fig:nu_flux_at_sk} The unoscillated $\nu_\mu$ flux at Super-Kamiokande
      with an off-axis angle of 2.5$^\circ$ when the electromagnetic horns are operated at 250~kA.}
  \end{center}
\end{figure}

The neutrino beamline is designed so that
the neutrino energy spectrum at Super-Kamiokande can be tuned
by changing the off-axis angle down to
a minimum of $\sim$2.0$^\circ$,
from the current (maximum) angle of $\sim$2.5$^\circ$.
The unoscillated $\nu_{\mu}$ flux at Super-Kamiokande with this off-axis angle is shown
in Fig.~\ref{fig:nu_flux_at_sk}.
Precise measurements of the baseline distance and off-axis angle were
determined by a GPS survey, described in Section~\ref{globalsurvey}.

\begin{table}[t]
  \caption[]
  {\label{tab:nubeam parameter}
  Machine design parameters of the J-PARC MR for the fast extraction.}
  \begin{center}
    \begin{tabular}{lc}
      \hline
        Circumference           & 1567~m \\
        Beam power              & $\sim$750~kW  \\
        Beam kinetic energy     & 30~GeV \\
        Beam intensity          & $\sim$$3\times 10^{14}$~p/spill \\
        Spill cycle             & $\sim$0.5~Hz \\
        Number of bunches       & 8/spill \\
        RF frequency            & 1.67 -- 1.72~MHz \\
%        Bunch interval          & 581~ns \\
%        Bunch width             & 58~ns \\
        Spill width             & $\sim$5~$\mu{}$sec \\
      \hline
    \end{tabular}
  \end{center}
\end{table}

 %Nakadaira/Ichikawa/Marchionni
\subsection{Primary Beamline}
\label{primary}
% \subsection{Primary Beam Line}
% Lead Author: Nakadaira/Ichikawa/Marchionni

The primary beamline consists of the preparation section (54~m long),
arc section (147~m) and final focusing section (37~m).  In the
preparation section, the extracted proton beam is tuned with a series
of 11~normal conducting magnets (four steering, two dipole and five
quadrupole magnets) so that the beam can be accepted by the arc
section.  In the arc section, the beam is bent toward the direction of
Kamioka by 80.7\degree{}, with a 104~m radius of curvature, using
14~doublets of superconducting combined function magnets
(SCFMs) \cite{Nakamoto:2004nj, Nakamoto:2005dq, Ogitsu:2005}.
There are also three pairs of horizontal and vertical
superconducting steering magnets to correct the beam orbit.  
In the final focusing section, ten normal conducting magnets
(four steering, two dipole and four quadrupole magnets) guide and
focus the beam onto the target, while directing the beam downward by
3.637\degree{} with respect to the horizontal.  

\begin{figure}[]
  \begin{minipage}{0.49\hsize}
    \begin{center}
      \includegraphics[keepaspectratio=true,width=30mm]{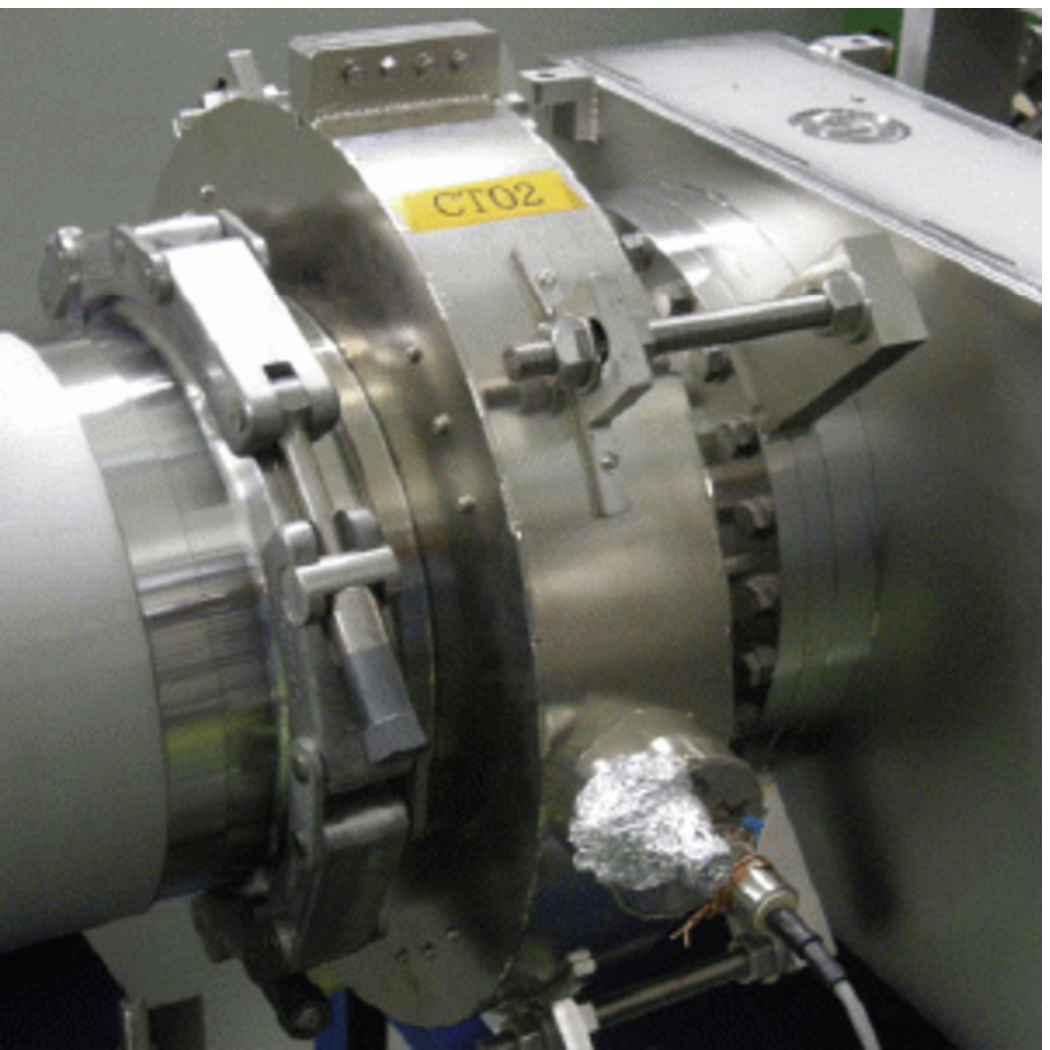}
    \end{center}
  \end{minipage}
  \begin{minipage}{0.49\hsize}
  \begin{center}
      \includegraphics[keepaspectratio=true,width=36mm]{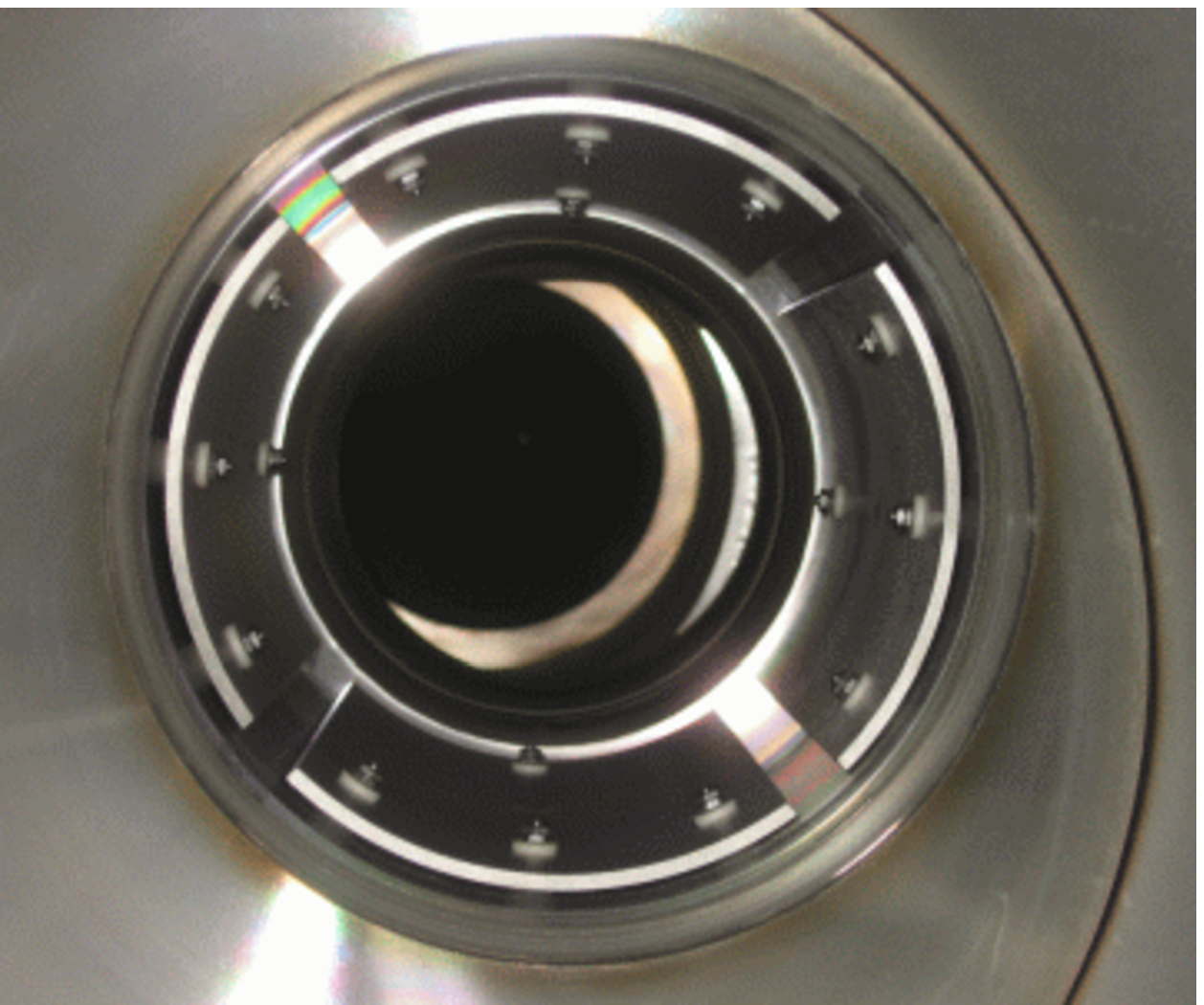}
    \end{center}
  \end{minipage}

  \begin{minipage}{0.49\hsize}
    \begin{center}
      \includegraphics[keepaspectratio=true,width=36mm]{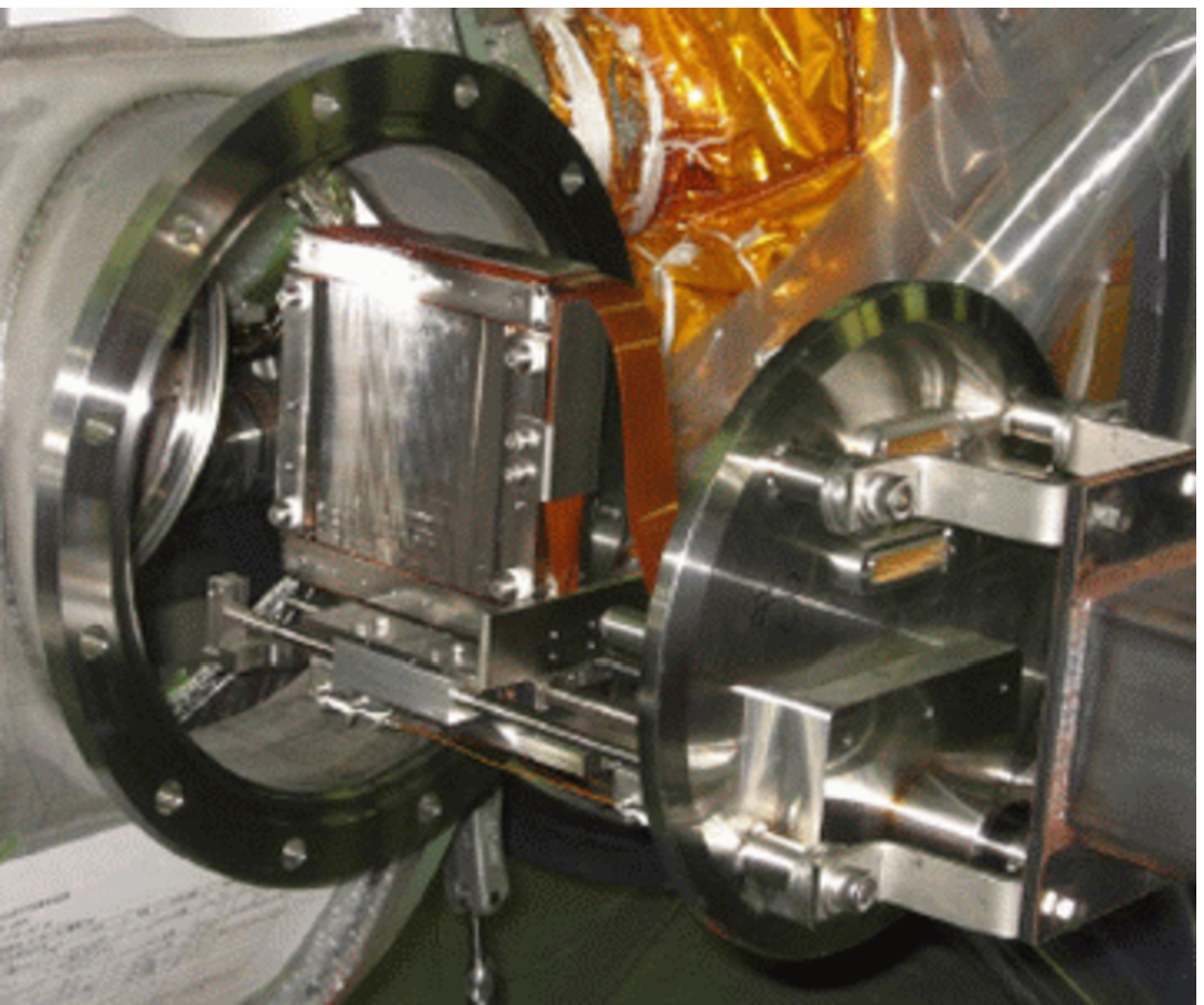}
    \end{center}
  \end{minipage}
  \begin{minipage}{0.49\hsize}
    \begin{center}
      \includegraphics[keepaspectratio=true,width=36mm]{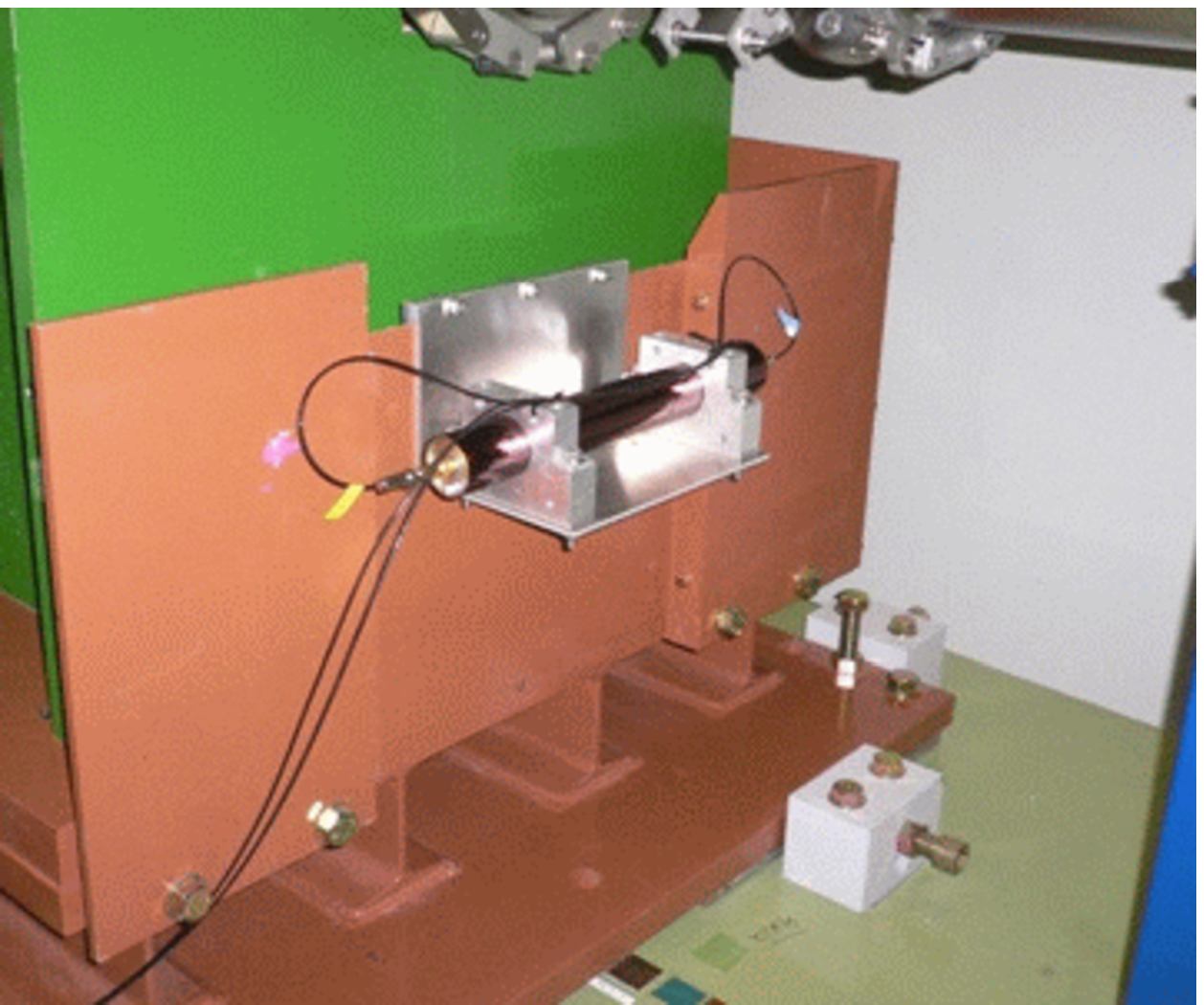}
    \end{center}
  \end{minipage}
  \caption[]
  {\label{fig:photo monitor}
  Photographs of the primary beamline monitors.
  Upper left: CT. Upper right: ESM. Lower left: SSEM. Lower right: BLM.}
\end{figure}

\begin{figure}[t]
  \begin{center}
    \includegraphics[width=\linewidth]{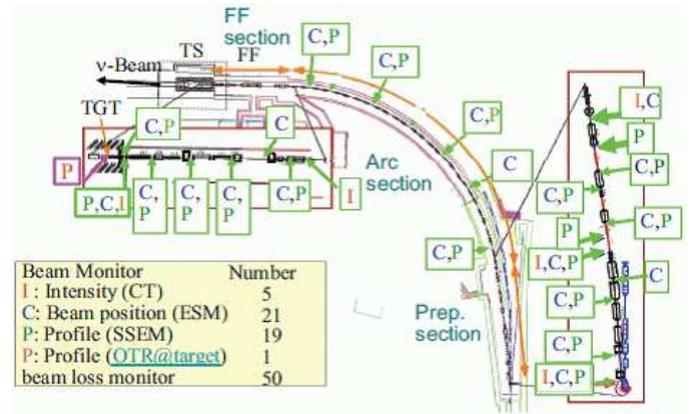}
    \caption[]
    {\label{fig:primary beamline}
    Location of the primary beamline monitors.}
  \end{center}
\end{figure}

A well-tuned proton beam is essential for stable neutrino beam
production, and to minimize beam loss in order to achieve high-power
beam operation.  Therefore, the intensity, position, profile and loss
of the proton beam in the primary sections are precisely monitored by
five current transformers (CTs), 21~electrostatic monitors (ESMs),
19~segmented secondary emission monitors (SSEMs) and 50~beam loss
monitors (BLMs), respectively.  Photographs of the monitors are shown
in Fig.~\ref{fig:photo monitor}, while the monitor locations are shown
in Fig.~\ref{fig:primary beamline}.  Polyimide cables and ceramic
feedthroughs are used for the beam monitors, because of their
radiation tolerance.

The beam pipe %, inside of the primary beamline, 
is kept at $\sim{}3\times{}10^{-6}$~Pa using ion pumps, 
in order to be connected
with the beam pipe of the MR and to reduce the heat load to the SCFMs.  The
downstream end of the beam pipe is connected to the ``monitor stack'':
the 5~m tall vacuum vessel embedded within the 70~cm thick wall
between the primary beamline and secondary beamline. The most
downstream ESM and SSEM are installed in the monitor stack. Because of
the high residual radiation levels, the monitor stack is equipped with
a remote-handling system for the monitors.

\subsubsection{Normal Conducting Magnet}

The normal conducting magnets are designed to be tolerant of radiation
and to be easy to maintain in the high-radiation environment.
For the four most upstream magnets in the preparation
section, a mineral insulation coil is used because of its radiation
tolerance.  To minimize workers' exposure to radiation, remote
maintenance systems are installed such as twistlock slings, alignment
dowel pins, and quick connectors for cooling pipes and power lines.

For the quadrupole magnets, ``flower-shaped'' beam pipes, whose surfaces
were made in the shape of the magnetic pole surface, are adopted to maximize
their apertures.

\subsubsection{Superconducting Combined Function Magnet (SCFM)}

In total, there are 28 SCFMs \cite{SCFM1,SCFM2,SCFM3,SCFM4},
each with a coil aperture of 173.4~mm. 
The operating current for a 30~GeV proton beam is 4360~A,
while the magnets themselves were tested up to 7500~A, which
corresponds to a 50~GeV proton beam.
%%%CKJ
%Add these citations
%[\cite{SCFM1}, \cite{SCFM2}, \cite{SCFM3}, \cite{SCFM4}]
%%%CKJ

The combined field is generated with a left-right asymmetric single
layer Rutherford-type coil, made of NbTi/Cu.  Two SCFMs are
enclosed in one cryostat in forward and backward directions to
constitute a defocus-focus doublet, while each dipole field is kept in
the same direction.  All the SCFMs are cooled in series with
supercritical helium at 4.5~K and are excited with a power supply
(8~kA, 10~V).

There are also three superconducting corrector dipole magnets,
which are cooled by conduction, in the SCFM section. 
Each magnet has two windings, one for vertical and one for
horizontal deflections. These magnets allow the beam to be precisely
positioned along the beamline (to minimize losses).

The magnet safety system (MSS) protects the magnets and the bus-bars
of the primary beamline in the case of an abnormal condition, and
supplements the passive safety protection provided by cold diodes
mounted in parallel with the superconducting magnets.  The MSS is
based on the detection of a resistive voltage difference across the
magnet that would appear in the case of a quench. It then secures the
system by shutting down the magnet power supply and issuing a beam
abort interlock signal. Most units of the MSS are dual redundant.
This redundancy increases the reliability of the system.  
The MSS is based on 33 MD200 boards \cite{MD200}.

% 1) Expand the current text roughly at least a factor of two by including,
% why SCFM were needed/what it achieves.
% 2) Include a short description of the corrector coils including why they 
% are needed. This could be provided by Peter Wanderer at BNL.
% 3) Add a figure (photo) SCFM that shows the uneven coil winding, which is 
% really a beautiful thing.

\subsubsection{Beam Intensity Monitor}

Beam intensity is measured with five current transformers (CTs).  Each
CT is a 50-turn toroidal coil around a cylindrical ferromagnetic core.
% The five current transformers (CTs) are toroidal coils; each
% consisting of 50~turns of a copper wire around a ferromagnetic core.
To achieve high-frequency response up to 50~MHz for the short-pulsed
bunches and to avoid saturation caused by a large peak current of 200~A,
CTs use a FINEMET\textsuperscript{\textregistered} (nanocrystalline
Fe-based soft magnetic material) core, which has a high saturation flux
density, high relative permeability and low core loss over a wide
frequency range.  The core's inner diameter is 260~mm, its outer diameter
is 340~mm and it has a mass of 7~kg.  It is impregnated with epoxy resin.  To
achieve high radiation hardness, polyimide tape and alumina fiber tape
are used to insulate the core and wire.
% Between the core and the copper wire, radiation tolerant insulators
% are used: polyimide and alumina fiber tapes.
Each CT is covered by an iron shield to block electromagnetic noise.

Each CT's signal is transferred through about 100~m of 20D colgate cable
and read by a 160~MHz Flash ADC (FADC).  The CT is calibrated using another coil
around the core, to which a pulse current,
shaped to emulate the passage of a beam bunch, is applied.
The CT measures the absolute proton beam
intensity with a 2\% uncertainty and the relative intensity with a
0.5\% fluctuation.  It also measures the beam timing with precision
better than 10~ns.

\subsubsection{Beam Position Monitor}

Each electrostatic monitor (ESM)
has four segmented cylindrical electrodes surrounding the
proton beam orbit (\mbox{80\degree{}} coverage per electrode).  By
measuring top-bottom and left-right asymmetry of the beam-induced
current on the electrodes, it monitors the proton beam center position
nondestructively (without direct interaction with the beam).

The longitudinal length of an ESM is 125~mm for the 15~ESMs in the
preparation and final focusing sections, 210~mm for the five~ESMs in
the arc section and 160~mm for the ESM in the monitor stack.
The signal from each ESM is read by a 160~MHz FADC.

The measurement precision of the beam position is less than 450~$\mu$m 
(20--40~$\mu$m for the measurement fluctuation, 
100--400~$\mu$m for the alignment precision 
and 200~$\mu$m for the systematic uncertainty other than the alignment), 
while the requirement is 500~$\mu$m.

\subsubsection{Beam Profile Monitor}

Each segmented secondary emission monitor (SSEM)
has two thin (5~$\mu$m, $10^{-5}$~interaction lengths)
titanium foils stripped horizontally and vertically, and an anode HV
foil between them. The strips are hit by the proton beam and emit
secondary electrons in proportion to the number of protons that go
through the strip. The electrons drift along the electric field and
induce currents on the strips. The induced signals are transmitted to
65~MHz FADCs through twisted-pair cables. The proton beam profile is
reconstructed from the corrected charge distribution on a bunch-by-bunch
basis. The strip width of each SSEM ranges from 2 to 5~mm,
optimized according to the expected beam size at the installed position.
The systematic uncertainty of the beam width measurement is
200~$\mu$m while the requirement is 700~$\mu$m.  Optics parameters of
the proton beam (Twiss parameters and emittance) are reconstructed
from the profiles measured by the SSEMs, and are used to estimate the
profile center, width and divergence at the target.

Since each SSEM causes beam loss (0.005\% loss), they are remotely
inserted into the beam orbit only during beam tuning, and
extracted from the beam orbit during continuous beam operation.

\subsubsection{Beam Loss Monitor}

To monitor the beam loss, 19~and~10~BLMs are installed near the beam
pipe in the preparation and final focusing sections respectively,
while 21~BLMs are positioned near the SCFMs in the arc section.  Each
BLM (Toshiba Electron Tubes \& Devices E6876-400) is a wire
proportional counter filled with an Ar-CO$_2$
mixture~\cite{TOSHIBATETD}.

The signal is integrated during the spill and if it exceeds a
threshold, a beam abort interlock signal is fired.  The raw signal before
integration is read by the FADCs with 30~MHz sampling for the software
monitoring.

By comparing the beam loss with and without the SSEMs in the beamline, 
it was shown that the BLM has a sensitivity down to a 16~mW beam loss.
In the commissioning run, it was confirmed 
that the residual dose and BLM data integrated during the period 
have good proportionality. 
This means that the residual dose can be monitored by watching the BLM data.
 
 %Nakadaira/Ichikawa/Marchionni
\subsection{Secondary Beamline}
\label{secondary}
% \subsection{Secondary Beamline}
% Lead Author: Nakadaira/Ichikawa/Marchionni

Produced pions decay in flight inside a single
volume of $\sim$1500~$\mathrm{m}^3$, filled with helium gas (1~atm) to
reduce pion absorption and to suppress tritium and NO$_{\textrm x}$
production by the beam. The helium vessel is connected to the
monitor stack via a titanium-alloy beam window which separates the
vacuum in the primary beamline and the helium gas volume in the secondary
beamline. Protons from the primary beamline are directed to the target via the
beam window.

\begin{figure}[t]
  \begin{center}
    \includegraphics[width=\linewidth]{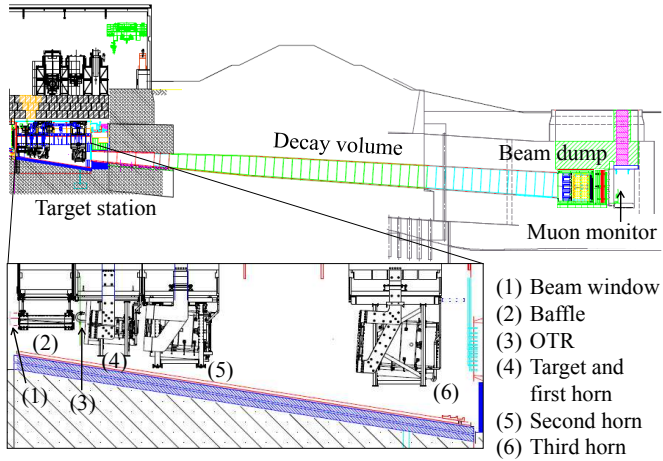}
    \caption[]
    {\label{fig:secondarybeamline}
    Side view of the secondary beamline. The length of the decay volume is $\sim$96~m.}
  \end{center}
\end{figure}

The secondary beamline consists of three sections: the target station,
decay volume and beam dump (Fig.~\ref{fig:secondarybeamline}). The
target station contains: a baffle which is a collimator to protect the
magnetic horns; an optical transition radiation monitor (OTR) to
monitor the proton beam profile just upstream of the target; the
target to generate secondary pions; and three magnetic horns excited
by a 250~kA (designed for up to 320~kA) current pulse to focus the pions. The produced pions
enter the decay volume and decay mainly into muons and muon
neutrinos. 
All the hadrons, as well as muons below $\sim$5~GeV/c,
are stopped by the beam dump.
The neutrinos pass through the beam dump and are used for physics experiments.
Any muons above $\sim$5~GeV/c that also pass
through the beam dump are monitored to characterize the neutrino beam.

\subsubsection{Target Station}

The target station consists of 
the baffle, OTR, target, and horns,
all located inside a helium vessel.
The target station is separated from the primary beamline by
a beam window at the upstream end,
and is connected to the decay volume at the downstream end.

The helium vessel, which is made of 10~cm thick steel,
is 15~m long, 4~m wide and 11~m high. 
It is evacuated down to 50~Pa before it is filled with helium gas. Water
cooling channels, called plate coils, are welded to the surface of the
vessel, and $\sim$30$^\circ$C water cools the vessel to prevent its
thermal deformation. An iron shield with a thickness of $\sim$2~m and
a concrete shield with a thickness of $\sim$1~m are installed
above the horns
inside the helium vessel. Additionally, $\sim$4.5~m thick concrete
shields are installed above the helium vessel. 

The equipment and shields inside the vessel are removable by remote
control in case of maintenance or replacement of the horns or target.
Beside the helium vessel, there is a maintenance area where
manipulators and a lead-glass window are installed, as well as a
depository for radio-activated equipment.

\subsubsection{Beam Window}

The beam window, comprising two helium-cooled 0.3~mm thick titanium-alloy
skins, separates the primary proton beamline vacuum from the target
station.  The beam window assembly is sealed both upstream and
downstream by inflatable bellows vacuum seals to enable it to be
removed and replaced if necessary.

\subsubsection{Baffle}
The baffle is located between the beam window and OTR.  It is a 1.7~m
long, 0.3~m wide and 0.4~m high graphite block, with a beam hole of
30~mm in diameter.  The primary proton beam goes through this hole.
It is cooled by water cooling pipes.

\subsubsection{Optical Transition Radiation Monitor}

The OTR has a thin titanium-alloy foil, which is placed at 45\degree{}
to the incident proton beam.  As the beam enters and exits the foil,
visible light (transition radiation) is produced in a narrow cone
around the beam. The light produced at the entrance transition is
reflected at 90\degree{} to the beam and directed away from the target
area. It is transported in a dogleg path through the iron and concrete
shielding by four aluminum \mbox{90\degree{}} off-axis parabolic mirrors
to an area with lower radiation levels.
It is then collected by a charge injection device
camera to produce an image of the proton beam profile.

\begin{figure}[htb]
  \begin{center}
     \includegraphics[width=0.6\linewidth]{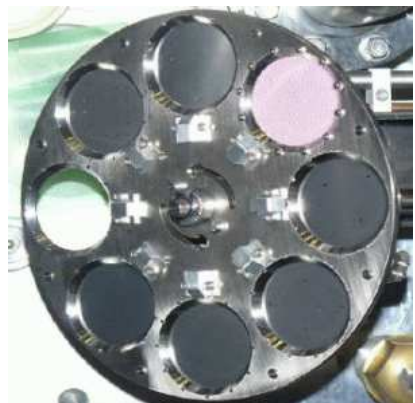}
     \vspace{2pt}
     \includegraphics[width=\linewidth]{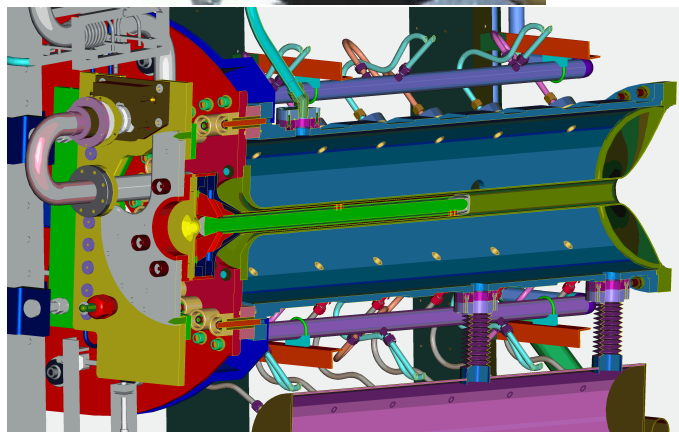}
  \end{center}
  \caption[]
  {\label{fig:photo_otr_horn1}
  Top: Photograph of the OTR carousel.
  Bottom: Cross section of the first horn and target.}
\end{figure}

The OTR has an eight-position carousel holding four titan\-ium-alloy
foils, an aluminum foil, a fluorescent ceramic foil of 100~$\mu$m
thickness, a calibration foil and an empty slot
(Fig.~\ref{fig:photo_otr_horn1}).  A stepping motor is used to rotate
the carousel from one foil to the next. The aluminum (higher
reflectivity than titanium) and ceramic (which produces fluorescent light with higher
intensity than OTR light) foils are used for low and very low
intensity beam, respectively.  The calibration foil has precisely
machined fiducial holes, of which an image can be taken using
back-lighting from lasers and filament lights.  It is used for
monitoring the alignment of the OTR system.  The empty slot allows
back-lighting of the mirror system to study its transport efficiency.

\subsubsection{Target}

The target core is a 1.9~interaction length (91.4~cm long), 2.6~cm
diameter and 1.8~g/cm$^3$ graphite rod. If a material significantly
denser than graphite were used for the target core, it would be melted
by the pulsed beam heat load.

The core and a surrounding 2~mm thick graphite tube are sealed inside
a titanium case which is 0.3~mm thick.  The target assembly is
supported as a cantilever inside the bore of the first horn inner
conductor with a positional accuracy of 0.1~mm.
% The target is cooled by 1.6~atm helium gas.  The gas flows is 250~m/s
% through the $\sim$5~mm gaps between the core and tube and between the
% tube and case.  When the 750~kW proton beam hits the target, the
% temperature at the center reaches 700~${}^\circ$C.
The target is cooled by helium gas flowing through the gaps between the core
and tube and between the tube and case. For the 750~kW beam, the flow rate
is $\sim$32 g/s helium gas with a helium outlet pressure of 0.2~MPa,
which corresponds to a flow speed of $\sim$250~m/s.  When the 750~kW
proton beam interacts with the target, the temperature at the center
is expected to reach 700$^\circ$C, using the conservative assumption
that radiation damage has reduced the thermal conductivity of the
material by a factor of four.

The radiation dose due to the activation of the target is estimated at a
few~Sv/h six months after a one year's irradiation by the 750 kW
beam~\cite{Nakadaira:2008zz}.

\subsubsection{Magnetic Horn}

The T2K beamline uses three horns. Each magnetic horn consists of two coaxial
(inner and outer) conductors which encompass a closed
volume~\cite{vanderMeer:1961sk, Palmer:1965zz}.  A toroidal magnetic
field is generated in that volume.  The field varies as $1/r$, where
$r$ is the distance from the horn axis. The first horn collects the
pions which are generated at the target installed in its inner
conductor. The second and third horns focus the pions. When the horn
is run with a operation current of 320~kA, the maximum field is
2.1~T and the neutrino flux at Super-Kamiokande is increased 
by a factor of $\sim$16 (compared to horns at 0~kA)
at the spectrum peak energy ($\sim$0.6~GeV).

The horn conductor is made of aluminum alloy (6061-T6).  The horns'
dimensions (minimum inside diameter, inner conductor thickness,
outside diameter and length respectively) are 
54~mm, 3~mm, 400~mm and 1.5~m for the first horn, 
80~mm, 3~mm, 1000~mm and 2~m for the second horn, and
140~mm, 3~mm, 1400~mm and 2.5~m for the third horn.
They are optimized to maximize the neutrino flux; the inside diameter
is as small as possible to achieve the maximum magnetic field, and the
conductor is as thin as possible to minimize pion absorption while
still being tolerant of the Lorentz force,
created from the 320~kA current and the magnetic field,
and the thermal shock from the beam.

The pulse current is applied via a pulse transformer with a turn ratio
of 10:1, which is installed beside the helium vessel in the target
station. The horns are connected to the secondary side of the pulse
transformer in series using aluminum bus-bars. The currents on the
bus-bars are monitored by four Rogowski coils per horn with a
200~kHz FADC readout.  The measurement uncertainty of the absolute
current is less than $\sim$2\%.  The horn magnetic field was measured
with a Hall probe before installation, and the uncertainty of the
magnetic field strength is approximately 2\% for the first horn and
less than 1\% for the second and third horns.

\subsubsection{Decay Volume}

The decay volume is a $\sim$96~m long steel tunnel. The cross section is
1.4~m wide and 1.7~m high at the upstream end, and 3.0~m wide and
5.0~m high at the downstream end. The decay volume is surrounded by
6~m thick reinforced concrete shielding. Along the beam axis, 40~plate
coils are welded on the steel wall, whose thickness is 16~mm, to cool
the wall and concrete to below 100$^{\circ}$C using water.

\subsubsection{Beam Dump}

The beam dump sits at the end of the decay volume. The distance
between the center of the target and the upstream surface of the beam dump
along the neutrino beam direction
for the off-axis angle of 2.5$^\circ$ is 109~m.
The beam dump's core is made of \mbox{75~tons} of graphite (1.7~g/cm$^3$),
and is 3.174~m long, 1.94~m wide and 4.69~m high.
It is contained in the helium vessel.
Fifteen iron plates are placed outside the vessel and two inside,
at the downstream end of the graphite core, 
to give a total iron thickness of 2.40~m.
%Total material is xxg/cm2 and
Only muons above $\sim$5.0~GeV/c can go through the beam dump to reach the
downstream muon pit.

The core is sandwiched on both sides by aluminum cooling modules
which contain water channels.  The temperature in the center of the
core is kept at around 150$^{\circ}$C for the 750~kW beam.

 %Nakadaira/Ichikawa/Marchionni
\subsection{Muon Monitor}
\label{mumon}
% \subsection{Muon Monitors}
% Lead Author: Nakadaira/Ichikawa/Marchionni

The neutrino beam intensity and direction can be monitored on a
bunch-by-bunch basis by measuring the distribution profile of muons,
because muons
are mainly produced along with neutrinos from the pion two-body decay.
The neutrino beam direction is determined to be the direction from the target
to the center of the muon profile. The muon
monitor~\cite{Matsuoka:2010zz,Matsuoka:2010za} is located just behind
the beam dump.
%, where the distance from the target is 118~m.
The muon
monitor is designed to measure the neutrino beam direction with a
precision better than 0.25~mrad, which corresponds to a 3~cm precision
of the muon profile center.  It is also required to monitor the
stability of the neutrino beam intensity with a precision better than
3\%.

A detector made of nuclear emulsion was
installed just downstream of the muon monitor to measure the absolute
flux and momentum distribution of muons.

\subsubsection{Characteristics of the Muon Flux}

Based on the beamline simulation package,  
described in Section~\ref{beamsoft}, the intensity of the muon flux
at the muon monitor,
for $3.3 \times 10^{14}$~protons/spill and 320~kA horn current,
is estimated to be
$1 \times 10^{7}$~charged particles/cm$^2$/bunch
with a Gaussian-like profile around the beam center and 
approximately 1~m in width. 
The flux is composed of around 87\% muons, 
with delta-rays making up the remainder. 

\subsubsection{Muon Monitor Detectors}

\begin{figure}[t]
  \begin{center}
    \includegraphics[keepaspectratio=true,width=80mm]{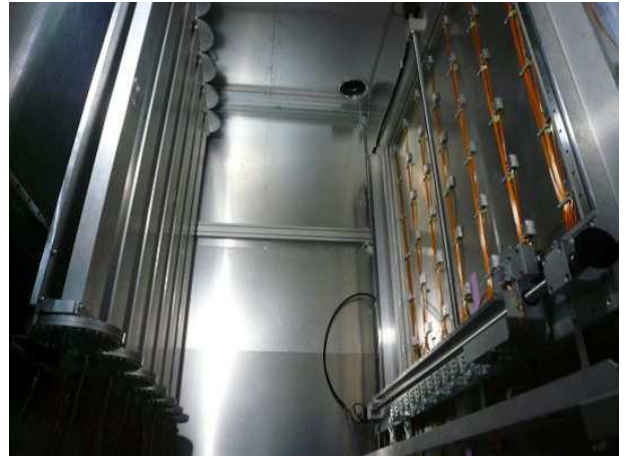}
    \caption[]
    {\label{fig:photo MUMON}
    Photograph of the muon monitor inside the support enclosure.
    The silicon PIN photodiode array is on the right side and the ionization chamber array is on the left side.
    The muon beam enters from the left side.}
  \end{center}
\end{figure}

The muon monitor consists of two types of detector arrays:
ionization chambers at 117.5~m from the target and
silicon PIN photodiodes at 118.7~m
(Fig.~\ref{fig:photo MUMON}).
Each array holds 49~sensors at 25~cm $\times$ 25~cm intervals and covers a 150
$\times$ 150~cm$^2$ area.  The collected charge on each sensor is read out
by a 65~MHz FADC.  The 2D muon profile is reconstructed in each
array from the distribution of the observed charge.

The arrays are fixed on a support enclosure for thermal insulation.
The temperature inside the enclosure is kept at around
34$^{\circ}$C
(within $\pm 0.7^{\circ}$C variation) with a sheathed heater,
as the signal gain in the ionization chamber is dependent 
on the gas temperature.

An absorbed dose at the muon monitor is estimated to be 
about 100~kGy for a \mbox{100-day} operation at 750~kW.
Therefore, every component in the muon pit is made of 
radiation-tolerant and low-activation material 
such as polyimide, ceramic, or aluminum.

\subsubsection{Ionization Chamber}
There are seven ionization chambers, 
each of which contains seven~sensors in a $150 \times 50 \times 1956$~mm$^3$ aluminum gas tube.
The $75 \times 75 \times 3$~mm$^3$ active volume of each sensor 
is made by two parallel plate electrodes on alumina-ceramic plates.
Between the electrodes, 200~V is applied.

Two kinds of gas are used for the ionization chambers according to the beam intensity: 
Ar with 2\% N$_2$ for low intensity, and He with 1\% N$_2$ for high intensity.
The gas is fed in at approximately 100~cm$^3$/min.
The gas temperature, pressure and oxygen contamination are kept 
at around 34$^{\circ}$C with a 1.5$^{\circ}$C gradient 
and $\pm$0.2$^{\circ}$C variation, at $130\pm 0.2$~kPa (absolute), 
and below 2~ppm, respectively.

\subsubsection{Silicon PIN Photodiode}
Each silicon PIN photodiode (Hamamatsu\textsuperscript{\textregistered} S3590-08) 
has an active area of $10\times 10$~mm$^2$ and a depletion layer thickness of 300~$\mu$m.
To fully deplete the silicon layer, 80~V is applied.

The intrinsic resolution of the muon monitor is less than 0.1\% for the intensity and less than 0.3~cm for the profile center.

\subsubsection{Emulsion Tracker}
The emulsion trackers are composed of two types of modules. 
The module for the flux measurement consists of eight consecutive emulsion
films~\cite{Nakamura:2006xs}. It measures the muon flux 
with a systematic uncertainty of 2\%.
The other module for the momentum measurement is made of 25 emulsion 
films interleaved by 1~mm lead plates, which can measure the momentum of each
particle by multiple Coulomb scattering with a precision of 28\% at a 
muon energy of 2 GeV/c~\cite{Lellis:2003xt, Besnier:2008zzb}. 
These films are analyzed by scanning microscopes~\cite{Armenise:2005yh, Kreslo:2008zz}.

 %Nakadaira/Ichikawa/Marchionni
\subsection{Beamline Online System}
\label{beamonline}
% \subsection{Beam Line Online System}
% Lead Author: Nakadaira/Ichikawa/Marchionni
For the stable and safe operation of the beamline, the online system
collects information on the beamline equipment and the beam measured
by the beam monitors, and feeds it back to the operators.  It also
provides Super-Kamiokande with the spill information for event synchronization by
means of GPS, which is described in detail 
in Section~\ref{GPS}.

%\subsubsection{Event Synchronization}

%At both the \mbox{J-PARC} and Super-Kamiokande sites, the beam-spill timing is
%measured by using one~pps (pulse per second) from two GPS receivers
%(for redundancy) and a 100~MHz clock from a rubidium clock with a
%precision of approximately 10~ns.  
%% Move this to the Time Synchronization Section (3.6.2)
%When the timing signal synchronized
%with the MR extraction is received, the time is recorded to a local
%time clock (LTC) module at \mbox{J-PARC}.  The LTC module counts the
%accumulated number of received signals as the spill number.  This
%time information and the spill number are sent to Super-Kamiokande through
%\mbox{L2-VPN}, and are returned from Super-Kamiokande to check consistency.  
%The LTC
%module also provides the beam trigger for the beam monitors.

\subsubsection{DAQ System}

The signals from each beam monitor are brought to one of five front-end stations
in different buildings beside the beamline.
The SSEM, BLM, and horn current signals
are digitized by a 65~MHz FADC in the COPPER
system~\cite{Higuchi:2003xs}.  The CT and ESM signals are digitized by a 160~MHz
VME FADC~\cite{UWFADC}.  The GPS for the event synchronization
and the OTR both use custom-made readout electronics.  All of these readout
systems are managed by the MIDAS framework~\cite{midas}, and the event
builder records fully concatenated events every spill, before the next
spill is issued.
MIDAS's event monitoring system locks the internal data holding buffer.
To minimize the locking time,
which can have a negative effect on the DAQ system's response time, 
an event distributor was developed.
It receives event data from the MIDAS server
and distributes the data in ROOT format to the clients.
No reduction is applied to the output from the ADCs 
and the event data size remains constant at 1.6~MB.

\subsubsection{Beamline Control System}
Information on the beamline 
(the beam monitor outputs, spill number and 
status of the beamline equipment) 
is recorded by EPICS~\cite{Clausen:2008zza}.
EPICS also controls the beamline equipment using programmable logic controllers (PLCs).

Based on the data from EPICS, 
the beam orbit and optics are simulated by SAD~\cite{SAD}, 
and the magnet currents to be adjusted are also calculated.

\subsubsection{Interlock}
The function of the interlock system is
to protect people (PPS: person protection system) 
and the machines (MPS: machine protection system).
The PPS can be fired by an emergency stop button, 
or safety sensors such as door interlocks and radiation monitors.
The MPS can be fired by a quenching of the SCFMs, 
an error from the normal conducting magnet or horn system, 
an excess in the loss monitor signal, or other machine-related causes.
 %Nakadaira/Ichikawa/Marchionni
\subsection{Beamline Simulation for Neutrino Flux Estimation}
\label{beamsoft}
%\subsection{Beamline Simulation for Neutrino Flux Estimation}
% Lead Author: Nakadaira/Ichikawa/Marchionni

%\label{jnubeam}

The neutrino flux is predicted by a Monte Carlo simulation based on 
experimental data. Specifically, hadron production by 30 GeV 
protons on a graphite target was measured by a dedicated experiment, 
NA61/SHINE~\cite{Abgrall:1292313, Abgrall:2011ae},
which fully covers the kinematic region of interest for T2K.

In the beam MC, the detailed geometry of the secondary beamline is 
described in the code. Protons with a kinetic energy of 30 GeV are 
injected into the graphite target and then secondary particles are produced 
and focused in the horn magnets. The secondaries and any un-interacted 
protons are tracked until they decay into neutrinos or are stopped at the beam 
dump. The tracks of neutrinos are extrapolated to the near and far 
detectors, providing the predicted fluxes and energy spectra at both 
detector sites. 

The primary interaction of the 30 GeV proton with carbon is 
simulated based on NA61/SHINE data. Other hadronic interactions inside 
the target are simulated by FLUKA~\cite{FLUKA:2008.3c}.
The interactions outside the 
target are simulated using GEANT3/GCALOR~\cite{GEANT3} with the interaction 
cross sections tuned to experimental data. 

 %Nakadaira/Ichikawa/Marchionni
\subsection{Global Alignment and Time Synchronization}
\label{global}
\subsubsection{Global Position Survey and Alignment}
\label{globalsurvey}
% \subsubsection{Global Position Survey and Alignment}
% Lead Author: Ishii

In a long baseline neutrino experiment,
controlling the direction of the neutrino beam is one of the most
important aspects.
For the T2K neutrino experiment, it is necessary to consider the 
three-dimensional geometry of the earth, since it covers a distance
 of $\sim$300~km from J-PARC to Super-Kamiokande.
Determining the correct direction is not simple.
Therefore, surveys were performed, including a long baseline GPS survey
between Tokai and Kamioka.
%The neutrino beamline was constructed so that both Super-Kamiokande
%and the hypothetical Hyper-Kamiokande \cite{Nakamura:2003hk} locations
%(one being the Tochibora mine in Kamioka) are at the same off-axis angle.

Based on the surveys, the primary beamline components, target, and horns were
%% ckj
%accurately 
%% ckj
aligned in order to send the neutrino beam in the right direction.
The muon monitor and the neutrino near detectors were also 
%% ckj
%accurately 
%% ckj
aligned
in order to monitor the neutrino beam direction.
A 
%% ckj
%accurate 
good
alignment of the components is also necessary in order to reduce
irradiation in a high-intensity proton beamline.

A complete neutrino beamline survey is carried out on a yearly basis.
There are five penetration holes in the neutrino beamline to connect
a ground survey network with an underground survey;
one at the preparation section, two at the final focusing section and
two at the muon pit.
In the target station, the underground survey points
were transferred to the ground level so that they can be monitored even
after the underground helium vessel is closed.
In the primary beamline, a survey and alignment is carried out
using a laser tracker
with a spatial resolution of 50~$\mu$m at a distance shorter than 20~m.
The superconducting magnets were aligned to better than 100~$\mu$m and 
the normal conducting quadrupole magnets were aligned to better than 1~mm.
In the other places, a survey is carried out using a total station
which gives a spatial resolution of
about 2~mm at a distance of 100~m.

We observed a ground sink of a few tens of millimeters during the construction
stage.
It was taken into account at the installation of the beamline components.
After the installation at the primary beamline, we still observed a sink of 
several millimeters at the final focusing section and the target station. 
Therefore the beam was tuned to follow the beamline sink.

The required directional accuracy from the physics point of view is 
$1 \times 10^{-3}$~rad.
The directional accuracy of a long baseline GPS survey is several times
$10^{-6}$~rad. That of a short distance survey is a few times $10^{-5}$~rad.
It was confirmed by surveys after construction that 
a directional accuracy
of significantly better than $1 \times 10^{-4}$~rad was attained.

The measured distance between the target and the center position of 
Super-Kamiokande is $295,335.2 \pm 0.7$~m. The measured off-axis angle 
is $2.504 \pm 0.004^{\circ}$.
 %Ishii
\subsubsection{Time Synchronization}
\label{GPS}
% \subsubsection{Time Synchronization}
% Lead Author: Wilkes

The T2K GPS time synchronization system builds on experience from K2K, 
taking advantage of subsequent advances in
commercially available clock systems and related technology. The 
system provides $O(50$ ns$)$ scale synchronization between
neutrino event trigger timestamps at Super-Kamiokande, 
and beam spill timestamps
logged at J-PARC.

The heart of the system is a custom electronics board called the local
time clock (LTC). This board uses a time base derived from a
commercial rubidium clock, and references it to GPS time using input 
from two independent commercial GPS receivers. The operational firmware 
was coded to interface efficiently with the Super-Kamiokande or J-PARC data 
acquisition systems.

 The LTC receives 1~pps (pulse per second) signals from two independent
GPS receivers. These signals have their leading edges aligned with 
the second transitions in UTC to higher precision than required for T2K. 
The primary receiver is a TrueTime (Symmetricom) rack-mounted receiver, 
and the secondary receiver is a Synergy Systems SynPaQIII receiver, 
mounted as a daughtercard on the LTC itself. The receivers are connected 
to antenna modules located with a clear view of the sky,
near the mine entrance at Super-Kamiokande, and at J-PARC. 
The Rb clock which provides a stabilized time base for the system 
in case of temporary loss of GPS signals is a Stanford Research 
Systems model FS-725. The LTC
is interfaced to the data acquisition system through a Linux PC with fast
network connections. At J-PARC, an independent optical fiber link sends 
data directly to the ND280 data acquisition system. 

% begin insert text from Section 3.4.1
When the timing signal,
synchronized with the MR extraction,
is received its time is recorded to an LTC module at \mbox{J-PARC}.
The LTC module counts the
accumulated number of received signals as the spill number.  This
time information and the spill number are sent to Super-Kamiokande through
a private network, and are returned from Super-Kamiokande to check consistency.
The LTC
module also provides the beam trigger for the beam monitors.
% end insert

At each site, two independent GPS systems run in parallel at all times 
to eliminate downtime during T2K running.
 %Wilkes
\section{Near Detector Complex (ND280)}
\label{ND280}
% \section{Near Detector Complex (ND280)}
% Lead Author: Jung

\begin{figure}[tbh]
  \begin{center}
    \includegraphics[width=\linewidth]{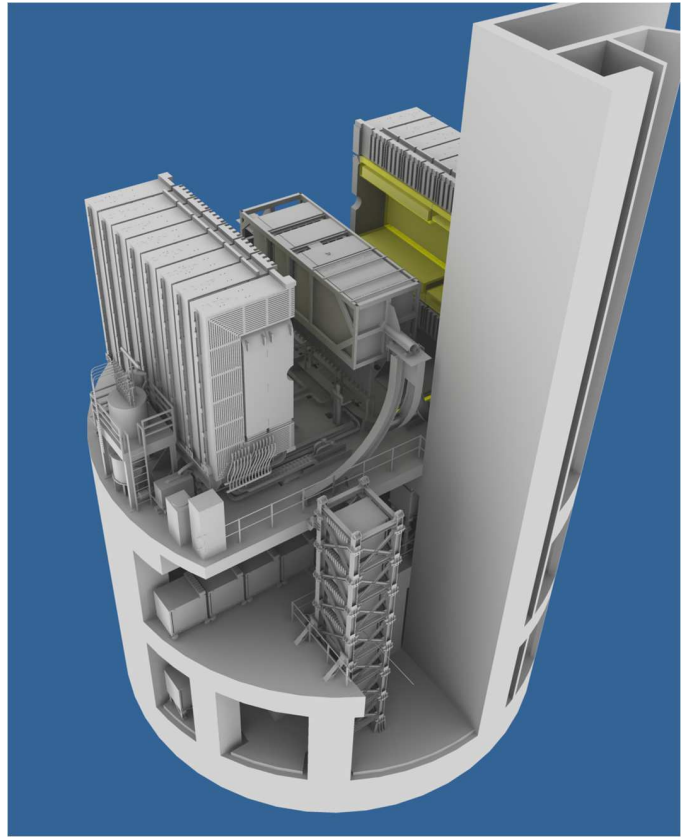}
  \end{center}
  \caption{\label{fg:ND280Pit} ND280 detector complex. 
  The off-axis detector and the magnet are located on the upper level; 
  horizontal INGRID modules are located on the level below; and the 
  vertical INGRID modules span the bottom two levels.}
\end{figure}

As stated earlier, the T2K experiment studies oscillations 
of an off-axis muon neutrino beam between the J-PARC 
accelerator complex and the Super-Kamiokande detector, 
with special emphasis on measuring the 
unknown mixing angle $\theta_{13}$ by observing the subdominant 
\numu{}\goesto{}\nue{} oscillation. The neutrino energy spectrum, 
flavor content, and interaction rates of the unoscillated beam 
are measured by a set of detectors located 280~m from the 
production target, and are used to predict the neutrino 
interactions at Super-Kamiokande. 

The primary detector at the 280~m site is a magnetized 
off-axis tracking detector. 
The off-axis detector elements are contained inside the 
magnet recycled from the UA1 experiment at CERN. 
Inside the upstream end of this magnet sits a pi-zero detector (\pod{}) 
consisting of tracking planes of scintillating bars alternating with 
either water target/brass foil or lead foil. 
Downstream of the \pod{}, the tracker, comprising  
three time projection chambers (TPCs) and 
two fine grained detectors (FGDs) 
consisting of layers of finely segmented scintillating bars, 
is designed to measure charged current interactions in the FGDs. 
The \pod{}, TPCs, and FGDs are all surrounded by an 
electromagnetic calorimeter (ECal) for detecting $\gamma$-rays 
that do not convert in the inner detectors,
while the return yoke of the magnet is
instrumented with scintillator to measure the ranges of muons 
that exit the sides of the off-axis detector. In 
addition to the off-axis detector, a separate array of 
iron/scintillator detectors called INGRID measures 
the on-axis neutrino beam profile at the 280~m site, 
while a set of muon monitor detectors located downstream 
of the beam dump monitors the beam direction and profile by detecting high 
energy muons from pion decay, as described earlier 
in Section~\ref{mumon}.
All detectors use the same coordinate convention:
$z$ is along the nominal neutrino beam axis,
and $x$ and $y$ are horizontal and vertical respectively.

These detectors are housed in a pit inside the ND280 hall
(see Fig.~\ref{fg:ND280Pit}).
The pit has a diameter of 
%19 m 
17.5~m
and a depth of 37~m, and has three floors.
The B1 floor, about 24~m below the surface, houses the off-axis detector,
which is located on the line between the target point and
the Super-Kamiokande position. 
The Service Stage, about 33~m deep, houses the horizontal modules of 
the INGRID detector. It also holds the electronics and 
many of the services for the off-axis detectors.
The B2 floor, about 37~m deep, houses the bottom modules of the vertical 
INGRID detector. 
The current off-axis angle is 2.5\degree{}, which has 
the extrapolated on-axis beam passing at about 
%4~m 
1~m above the Service Stage.
This facility design can accommodate off-axis angles in the range of  
between 2.0 and 2.5\degree{}, constrained by the requirement that
the beam axis pass through the central area of
the on-axis detector.  Outside of this area, 
the measurement of the beam axis direction would deteriorate.
A building with an internal area of 21~m $\times$ 28~m covers the pit, and has a 
10~ton crane. 
%The hut is a little
%bit shifted to the north with respect to the pit center 
%in order to allow the north area of the hut to be used for
%unloading of detector components and for the detector preparation. 
%The effective height of the
%crane is 5 m and its dead space is about 3 m from the north and south walls, 
%and 2 m from the east and west walls. The hut has an entrance shutter 
%5 m wide and 3.9 m high. There are a 6-people elevator and stairs. 
%Some area in the hut at the ground floor is used for the
%electricity and cooling water preparation purpose.
 %Jung
\subsection{Multi-Pixel Photon Counter (MPPC)}                       
\label{MPPC}
% \subsection{Multi-Pixel Photon Counter (MPPC)}
% Lead Author: Kudenko/Retiere/Yokoyama

The ND280 detectors make extensive use of scintillator  detectors
and wavelength-shifting (WLS) fiber  readout,
with light from the fibers being detected by
photosensors that must operate in a magnetic field environment 
and fit into a limited  space inside the magnet.
Multi-anode PMTs,
successfully used in other scintillator and WLS based neutrino experiments,
are not suitable for ND280 because most of  the detectors  in 
the ND280  complex have to  work in  a magnetic
field of 0.2~T.
To satisfy the ND280 experimental requirements,
a multi-pixel avalanche  photodiode was selected for the photosensor.
The device  consists of  many independent  sensitive pixels,
each  of which operates as an  independent Geiger micro-counter
with a gain of the same  order as a vacuum photomultiplier.
These novel photosensors are compact,
well  matched to  spectral emission  of WLS fibers,
and insensitive to  magnetic fields.
Detailed information and
the basic principles of operation  of multi-pixel photodiodes can be found
in a recent review paper~\cite{Renker:2009zz} and the references therein.

After R\&D and  tests provided by several groups  for three years,
the Hamamatsu  Multi-Pixel  Photon  Counter  (MPPC)  was  chosen
as the photosensor for ND280.
The MPPC  gain  is determined  by the  charge
accumulated   in  a  pixel   capacitance 
$C_{pixel}$:   $Q_{pixel}  = C_{pixel}\cdot\Delta  V$,
where  the  overvoltage  $\Delta  V$  is the
difference between  the applied voltage  and the breakdown  voltage of
the photodiode.
For MPPCs the operational voltage is about 70 V, which
is $0.8 - 1.5$ V above the breakdown voltage.
The pixel capacitance is 90 fF, which gives a gain in the range
$0.5 - 1.5$~$\times$ $10^6$.
When a photoelectron  is  produced  it   creates  a  Geiger  avalanche.
The amplitude of  a single pixel signal  does not depend on  the number of
carriers created in  this pixel.
Thus, the photodiode  signal is a sum of  fired pixels.
Each  pixel operates  as a  binary device,  but the multi-pixel photodiode 
as a whole unit is an analogue detector with a
dynamic range limited by the finite number of pixels.
 
\begin{figure}[htb]
  \begin{center}
    \begin{overpic}[width=\linewidth]%
       {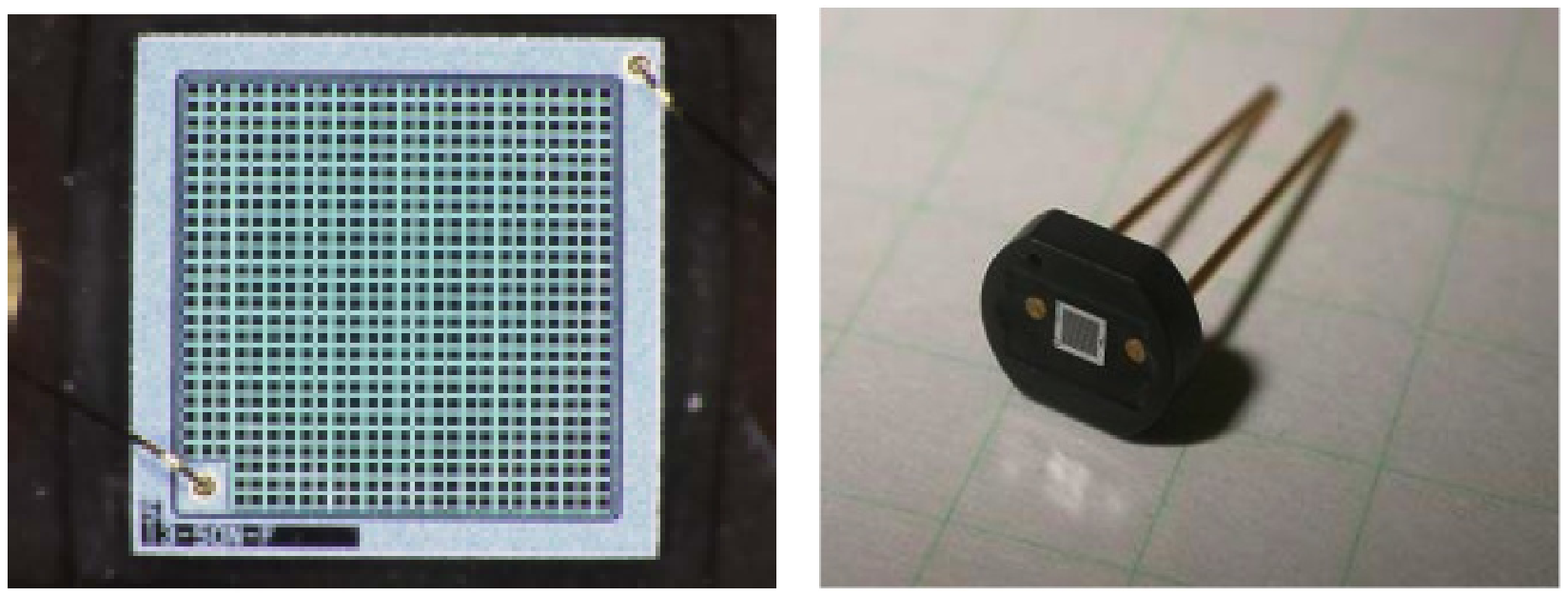}
       \put(200,10){\includegraphics[width=0.17\linewidth]%
       {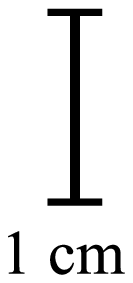}}
    \end{overpic}
  \end{center}
  \caption{ Photographs of an MPPC with a sensitive area of 
            $1.3\times 1.3$ mm$^2$: magnified face view (left)
            with 667 pixels in a $26\times 26$ array (a 9-pixel
            square in the corner is occupied by an electrode);
            the ceramic package of this MPPC (right).\label{fig:mppc}}
\end{figure}

A  customized 667-pixel MPPC,
with a sensitive  area of  $1.3\times 1.3$ mm$^2$,
was  developed  for  T2K~\cite{Yokoyama:2008hn, Yokoyama:2010qa}.  
It  is  based  on  a Hamamatsu commercial device, the sensitive area 
of which was increased to provide  better acceptance for  light 
detection from  1~mm diameter Y11 Kuraray fibers.
In total, about 64,000 \mbox{MPPCs} were produced for T2K.  
The T2K photosensor is shown in Fig.~\ref{fig:mppc}.

The main parameters of MPPCs are summarized in Table~\ref{table:mppc}.
\begin{table}
 \caption{Main parameters of the T2K MPPCs }
 \label{table:mppc}
 \begin{center} 
 \begin{tabular}{p{6cm} c}
  \hline
  Number of pixels &   667 \\
  Active area & $1.3\times 1.3$ mm$^2$ \\
  Pixel size & $50\times 50$ $\mu$m$^2$    \\
  Operational voltage & $68 - 71$ V \\ 
  Gain & $\sim10^6$ \\
  Photon detection efficiency at 525 nm  &  $26 - 30$\%   \\
  Dark rate, threshold = 0.5 p.e., T = 25 $^{\circ}$C  & $\leq 1.35$ MHz  \\ 
  \hline
 \end{tabular}
 \end{center}
\end{table}
The characterization of the MPPCs' response to scintillation light
is presented in Ref.~\cite{Vacheret:2011zz}.  
 
 %Kudenko/Retiere/Yokoyama
\subsection{INGRID On-axis Detector}
\label{INGRID}
% \subsection{INGRID On-axis Detector}
% 

\begin{figure}[tbh]
  \begin{center}
    \begin{overpic}[width=\linewidth]%
      {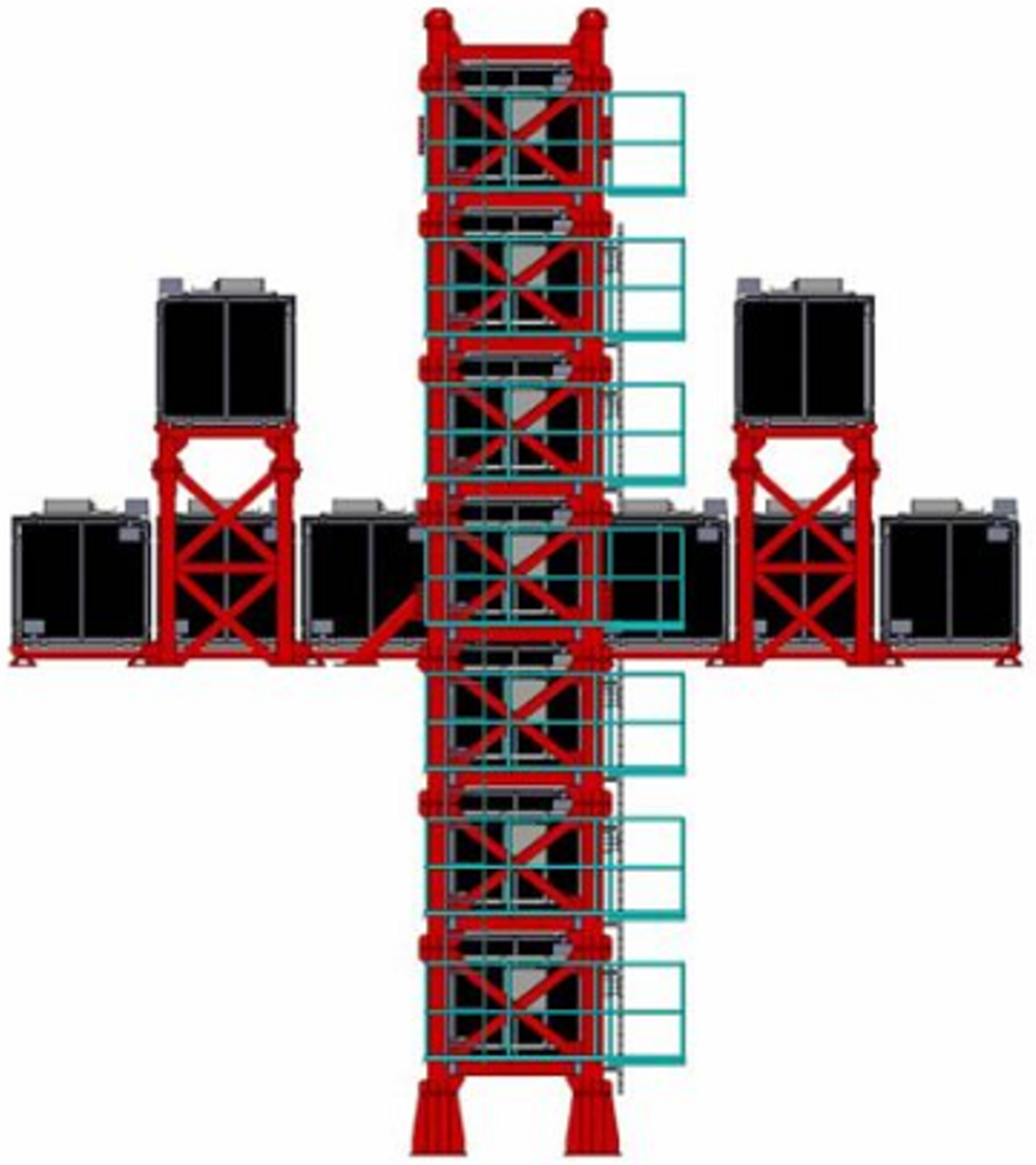}
      \put(20,0){\includegraphics[width=0.4\linewidth]%
       {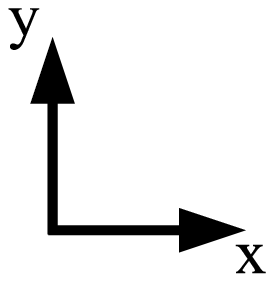}}
    \end{overpic}
  \end{center}
  \caption{INGRID on-axis detector}
  \label{fig:14mod.eps}
\end{figure}

INGRID (Interactive Neutrino GRID) is a neutrino detector
centered on the neutrino beam axis.
This on-axis detector was designed to monitor directly
the neutrino beam direction and intensity
by means of neutrino interactions in iron,
with sufficient statistics to provide daily measurements
at nominal beam intensity.
Using the number of observed neutrino events in each module,
the beam center is measured to a precision better than 
%28~cm.
10~cm.
This corresponds to 
%1~mrad 
0.4~mrad
precision at the near detector pit,
280~meters downstream from the beam origin.
The INGRID detector consists of 14 identical modules
arranged as a cross of two identical groups
along the horizontal and vertical axis, and two additional separate
modules located at off-axis directions outside the main cross, 
as shown in Fig.~\ref{fig:14mod.eps}.
The detector samples the neutrino beam in a transverse section
of 10~m $\times$ 10~m.
The center of the INGRID cross, with two overlapping modules,
corresponds to the neutrino beam center,
defined as 0\degree{} 
with respect to the direction of the primary proton beamline.
The purpose of the two off-axis modules is to 
check the axial symmetry of the neutrino beam.
The entire 16 modules are installed in the near detector pit
with a positioning accuracy of 2~mm in directions
perpendicular to the neutrino beam.

\begin{figure}[tbh]
  \begin{center}
    \includegraphics[width=\linewidth]{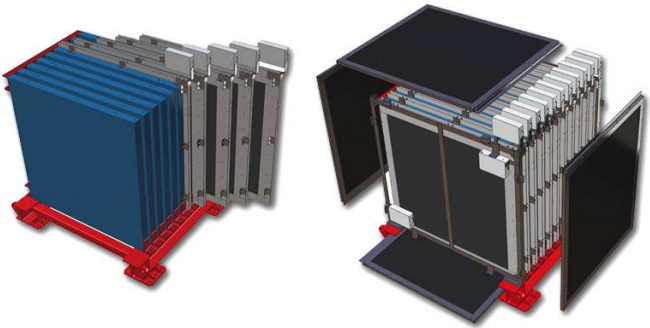}
  \end{center}
  \caption{An INGRID module. The left image shows the tracking planes (blue) and iron plates.  The right image shows veto planes (black).}
  \label{fig:HModule2.eps}
\end{figure}

The INGRID modules consist of a sandwich structure
of nine iron plates and 11 tracking scintillator planes
as shown in Fig.~\ref{fig:HModule2.eps}.
They are surrounded by veto scintillator planes,
to reject interactions outside the module.
The dimensions of the iron plates are 124~cm $\times$ 124~cm
in the $x$ and $y$ directions and 6.5~cm along the beam direction.
The total iron mass serving as a neutrino target is
7.1~tons per module.
Each of the 11 tracking planes consists of 24 scintillator bars
in the horizontal direction glued to 24 perpendicular bars
in the vertical direction with Cemedine PM200, for a total number of 8,448.
No iron plate was placed between the 10th and 11th tracking planes
due to weight restrictions, but this does not affect the tracking performance.
The dimensions of the scintillator bars used for the tracking planes
are 1.0~cm $\times$ 5.0~cm $\times$ 120.3~cm.
Due to the fact that adjacent modules can share one veto plane
in the boundary region,
the modules have either three or four veto planes.
Each veto plane consists of 22 scintillator bars 
segmented in the beam direction.
The dimensions of those scintillator bars are
1.0~cm $\times$ 5.0~cm $\times$ 111.9~cm (bottom sides)
and 1.0~cm $\times$ 5.0~cm $\times$ 129.9~cm (top, right and left sides).
The total number of channels for the veto planes is 1,144,
which gives a total of 9,592 channels for INGRID as a whole.

The extruded scintillator bars used for the tracking and veto planes
are made of polystyrene doped with
1\% PPO and 0.03\% POPOP by weight.
The wavelength of the scintillation light at the emission peak is 420 nm (blue).
They were developed and produced at Fermilab~\cite{PlaDalmau:2001en}.
A thin white reflective coating, 
composed of TiO$_{2}$ infused in polystyrene,
surrounds the whole of each scintillator bar.
The coating improves light collection efficiency by
acting as an optical isolator.
A hole with a diameter of about 3~mm
in the center of the scintillator bar
allows the insertion of a WLS fiber for light collection.

The WLS fibers used for INGRID are 1~mm diameter Kuraray double-clad Y-11.
The absorption spectrum of the fiber is centered 
at a wavelength of 430~nm (blue).
The emission spectrum is centered at 476~nm (green),
and the overlap between the two is small,
reducing self-absorption effects in the fiber.
One end of the fiber is glued to a connector
by epoxy resin (ELJEN Technology EJ-500).
The surface of the connector was polished with diamond blades.
An MPPC is attached to each fiber using the connector.
A detailed description of the MPPCs can be found in Section~\ref{MPPC}.
Some characterization of the MPPCs used for INGRID can be found 
in \cite{Yokoyama:2010qa, Moreau:2010zz}.

Finally, the set of scintillators, fibers and photosensors is contained
in a light-tight dark box made of aluminum frames
and plastic plates.
The readout front-end electronics boards, the Trip-T front-end boards (TFBs),
are mounted outside the dark box and each connected to 48 MPPCs 
via coaxial cables.
This forms one complete tracking scintillator plane.

INGRID was calibrated using cosmic ray data taken on the surface
and, during beam, in the ND280 pit.
The mean light yield of each channel 
is measured to be larger than ten photoelectrons per 1~cm of MIP tracks
which satisfies our requirement.
Furthermore
the timing resolution of each channel is measured to be 3.2~ns.

\begin{figure}[tbh]
  \begin{center}
    \includegraphics[width=50mm]{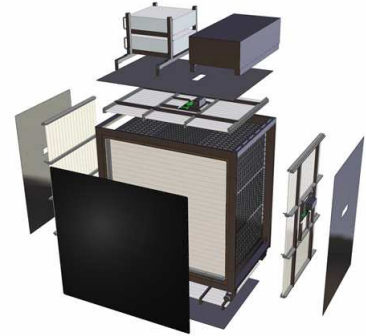}
  \end{center}
  \caption{The Proton Module.  Similar to the INGRID modules, but with finer grain scintillator and without the iron plates.}
  \label{fig:Proton_Module.eps}
\end{figure}

An extra module, 
called the Proton Module, 
different from the 16 standard modules, has been added
in order to detect with good efficiency the muons together with the protons
produced by the neutrino beam in INGRID.
The goal of this Proton Module is to identify the quasi-elastic channel
for comparison with Monte Carlo simulations of
beamline and neutrino interactions.
It consists of scintillator planes without any iron plate
and surrounded by veto planes.
A different size scintillator bar was used 
to improve tracking capabilities.
A schematic view of the Proton Module can be seen in Fig.~\ref{fig:Proton_Module.eps}.
It is placed in the pit in the center of the INGRID cross
between the standard vertical and horizontal central modules.

\begin{figure}[tbh]
  \begin{center}
    \includegraphics[width=1.0\linewidth]{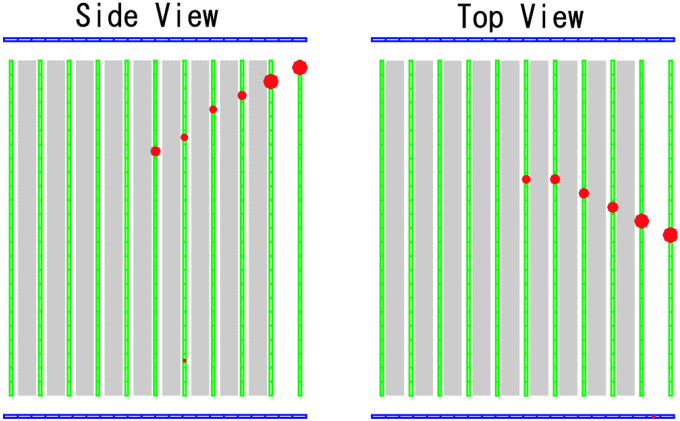}
  \end{center}
  \caption{A typical neutrino event in an INGRID module.
  A neutrino enters from the left and interacts
  within the module, producing charged particles.
  One of them makes a track which is
  shown as the red circles. Each of the green cells in this figure
  is a scintillator, and the size of the red circles indicates 
  the size of the observed signal in that cell.
  Blue cells and gray boxes indicate veto scintillators and 
  iron target plates, respectively.}
  \label{fig:INGRID_event_display.eps}
\end{figure}

\begin{figure}[tbh]
  \begin{center}
    \includegraphics[width=65mm]{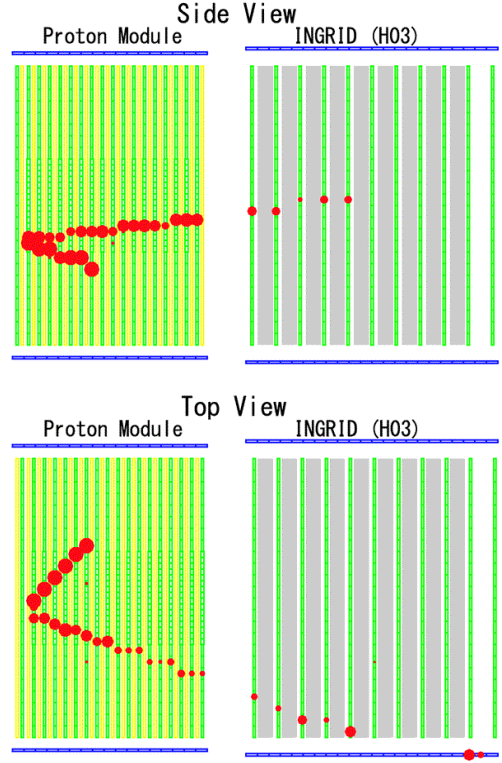}
  \end{center}
  \caption{A typical neutrino event in the Proton Module.
  A neutrino enters from the left and interacts within the module,
  producing charged particles whose tracks are shown as the red circles.
  One of them exits the Proton Module and enters
  the central INGRID horizontal module.
  Each of the green cells in this figure is a scintillator,
  and the size of the red circles indicates the size of the observed signal
  in that cell.
  Blue cells indicate veto scintillators.}
  \label{fig:Proton_Module_event_display.eps}
\end{figure}

Typical neutrino events in the INGRID module and the Proton Module
are shown in Figs.~\ref{fig:INGRID_event_display.eps}
and \ref{fig:Proton_Module_event_display.eps}.

%  \bibitem {scinti_paper}
%    A. Pla-Dalman, FERMILAB-Conf-00/343 (2001).

% \bibitem {PlaDalmau:2001en}
% A. Pla-Dalman, FERMILAB-Conf-00/343 (2001).

% \bibitem {Yokoyama:2010qa}
%  M. Yokoyama et al.,
%  "Performance of multi-pixel photon counters
%  for the T2K near detectors",
%  Nucl. Instr. and Meth. A \textbf{622} (2010) 567.

% \bibitem {Moreau:2010zz}
%  F. Moreau et al.,
%  "Mass characterization of Multi-Pixel-Photon-Counters for the T2K 280m 
% near detectors", 
%  Nucl.Instr. and Meth. A, \textbf{613} (2010) 46.
  
 %Minamino/Gonin 
\subsection{Off-axis Detector}
\label{offaxis}
% \subsection{Off-axis Detector}
% Lead Author: Jung

\begin{figure}[tbp]
  \begin{center}
    \begin{overpic}[width=\linewidth]%
      {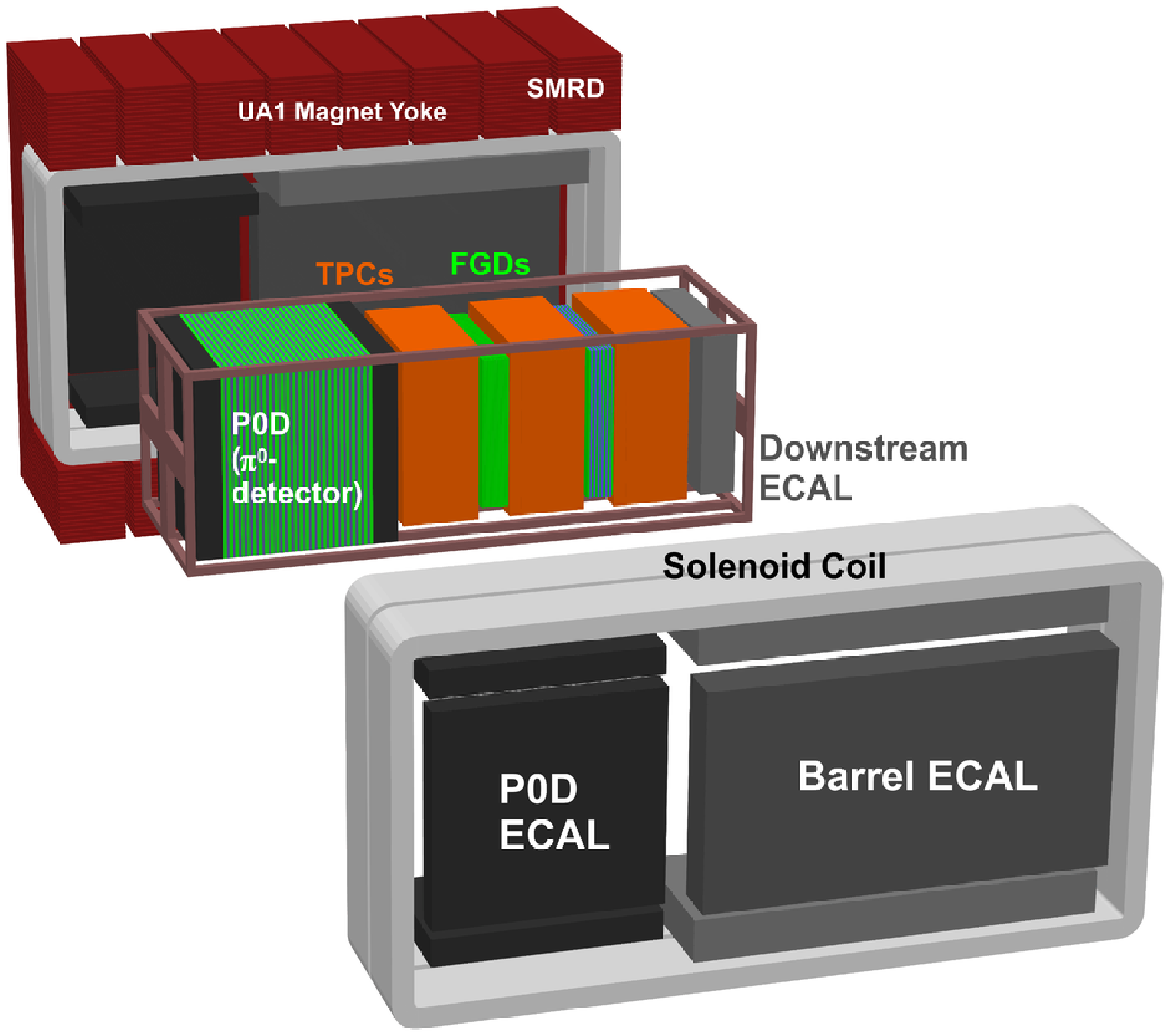}
      \put(150,160){\includegraphics[width=0.3\linewidth]%
        {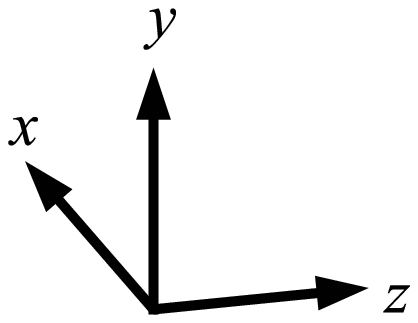}}
      \put(10,60){\includegraphics[width=0.2\linewidth]%
        {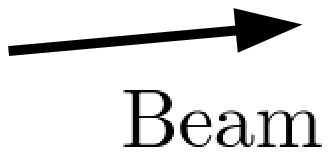}}
    \end{overpic}
  \end{center}
  \caption{\label{fg:nd280-exploded} An exploded view of the ND280
    off-axis detector.}
\end{figure}

A large fine grained off-axis detector (see
Fig.~\ref{fg:nd280-exploded}) serves to
measure the flux, energy spectrum and electron neutrino contamination in
the direction of the far detector, along with measuring rates for exclusive
neutrino reactions.  This characterizes
signals and backgrounds in the Super-Kamiokande detector.

The ND280 off-axis detector must satisfy several requirements. Firstly, it
must provide information to determine the \numu{} flux at the
Super-Kamiokande detector.  Secondly, the \nue{} content of the beam must be
measured as a function of neutrino energy.  The beam \nue{} background is
expected to be approximately 1\% of the \numu{} flux and creates a
significant non-removable background.  Thirdly, it must measure \numu{}
interactions such that the backgrounds to the \nue{} appearance search at
Super-Kamiokande can be predicted.  These backgrounds are dominated by
neutral current single \pizero{} production.  To meet these goals the ND280
off-axis detector must have the capability to reconstruct exclusive event
types such as \numu{} and \nue{} charged current quasi-elastic, charged
current inelastic, and neutral current events, particularly neutral current
single \pizero{} events.  In addition, the ND280 off-axis detector should
measure inclusive event rates.  All of these requirements were considered 
in designing the off-axis detector.
  
The constructed off-axis detector consists of: the
\pod{} and the TPC/FGD sandwich (tracker), both of which are 
placed inside of a metal frame container, called the ``basket''; an 
electromagnetic calorimeter (ECal) that surrounds the basket; and 
the recycled UA1 magnet instrumented with scintillator to 
perform as a muon range detector (SMRD). 
(See Fig.~\ref{fg:nd280-exploded}).  

The basket has dimensions of 6.5~m $\times$ 2.6~m $\times$ 2.5~m
(length $\times$ width $\times$ height). 
It is completely open at the top, to allow the insertion of the 
various detectors. Two short beams are fixed at
the center of the two faces of the basket perpendicular to the beam axis.
The short beam at each end of the basket connects to
an axle which runs through the hole in the magnet coil
originally intended to allow passage of the beam-pipe in the UA1 experiment.
The axle is in turn supported by an external support frame
which is bolted to the floor.
When opening the magnet, the
half yokes and the coils move apart, while the basket 
and the inner detector remain fixed in the position chosen
for data taking.
In the following sections, more detailed descriptions of 
these elements are provided.
\subsubsection{UA1 Magnet}
\label{UA1}
% \subsubsection{UA1 Magnet}
% Lead Author: Rubbia

The ND280 off-axis detector is built around the old CERN UA1/NOMAD magnet providing a dipole magnetic field of 0.2~T,
to measure momenta with good resolution and determine the sign of charged particles produced by neutrino interactions.

The magnet consists of water-cooled aluminum coils, which create the horizontally oriented dipole field, and a flux return yoke.
The dimensions of the inner volume of the magnet
are 7.0~m $\times$ 3.5~m $\times$ 3.6~m.
The external dimensions are 7.6~m $\times$ 5.6~m $\times$ 6.1~m and 
the total weight of the yoke is 850 tons. 
The coils are made of aluminum bars with 5.45~cm $\times$ 5.45~cm square cross sections, with a central 23~mm diameter bore
for water to flow. The coils are composed of individual ``pancakes'' which are connected hydraulically in parallel and electrically in series.

The magnet consists of two mirror-symmetric halves. 
The coils are split into four elements, two for each half, and are mechanically supported by, but electrically
insulated from, the return yoke. The two half yoke pieces each consist of
eight C-shaped elements, made of low-carbon steel plates, which
stand on movable carriages. The carriages are fitted on rails and operated
by hydraulic movers, so that
each half magnet is independent of the other and can be separately moved
to an open or closed position. 
When the magnet is in an open position, the inner volume is 
accessible, allowing access to the detectors.

The magnet yoke and coils were reused from UA1/NOMAD, while the movers
were obtained from the completed HERA-B experiment at DESY. 
In order to comply with
seismic regulations, detailed FEM static and dynamic analyses were
performed and cross-checked with measurements of 
deformation and modal frequency
of the yoke elements. As a result of this, the carriages were mechanically
reinforced by additional steel bars to increase their lateral strength.
Additional components
had to be specially designed and built for the ND280 magnet operation. These were: the 
power supply (PS), the cooling system (CS),  the magnet safety system (MSS), and 
the magnet control system (MCS). Finally, the magnetic field map was determined {\it in situ}
with a dedicated measurement campaign.

The PS, specially made for ND280, was designed and manufactured by Bruker to provide
the DC current to energize the magnet. The nominal current is 2900~A with a voltage drop
of 155~V. The requirements for the DC current resolution and stability were 300~ppm and
$\pm$ 1000~ppm over 24~hours respectively. The PS is also able to cope with AC phase imbalance
($\pm$ 2\%) and short voltage drops.
A thyristor switch mode was employed, with digital current regulation
via a DCCT captor (ULTRASTAB series from Danfysik). The power supply can be controlled
locally or remotely via the MCS.

The CS, assembled by MAN Ferrostaal AG (D),
provides up to 750~kW of cooling power 
via two independent demineralized water circuits 
to compensate for the heat loss from the coils and in the power supply. 
The cold source consists of a primary glycol circuit 
maintained at 8$^{\circ}$C by a chiller (built by Friotherm, D).
The secondary pumping circuit units and their heat exchangers, 
the water purification units and the main panel controller 
are mounted in an ISO container, suitable for easy road and sea transport. 
They were assembled and tested in Europe before shipment to J-PARC. 
The secondary circuit demineralized water for the magnet coils 
has a flow of 30~L/s and a pressure of 10~bar 
to compensate for the 7~bar pressure drop across the coil bore holes.
  
The MSS, based on a hardwired fail-safe interface, 
was built to ensure the operational safety of the magnet. 
It continuously monitors a set of input signals from 
the thermo-switches mounted on the magnet coils, 
fault signals from the power converter, cooling and magnet control systems, 
and magnet emergency stop signals from manual buttons located in the ND280 building. 
A Boolean OR of all fault signals is generated and logically combined with the on/off magnet status. 
When the magnet is off, 
the system issues a power convert permit signal only if none of the input signals is in a fault state. 
When the magnet is operating, a fast abort signal is generated 
and sent to the power converter in less than 1~ms when any of the input signals switches to a fault state. 
All input and output signals of the MSS are monitored by a VME computer, 
and any change in the status of the signals is recorded with 1~ms timing resolution, 
meaning that the detailed sequence of events leading up to a fast abort can be understood.

The aim of the MCS is to monitor the behavior
of the magnet and cooling system, to control the current set point of
the magnet power supply and to interface all the information and
control parameters with the global slow control (GSC).  The system is based on an industrial
programmable logic controller (PLC) that reads: the coil temperature at
52 points; the water flow, input and output temperature and pressure
on each half of the magnet; the voltage drop through each half of the
magnet; the power converter voltage and current; and the status flags of
the power converter, CS and MSS.
The PLC is linked via PROFIBUS DP (Process Field Bus for Decentralized Peripherals) 
with the power converter, in order to switch
on and off, and to read and write, the current and other settings.
All this information is processed and analyzed several times per
second. If any subsystem should exceed the operational
parameters, the MCS will switch off the magnet and trigger the
corresponding alarms for later diagnostics. 
All the information in the PLC can be accessed via an 
open connectivity standard for industrial automation (OPC server).
The OPC server is interfaced with the GSC for monitoring and control of the magnet. 
The measured current is used offline to define the magnetic field for 
data analysis.

%The refurbishing, packing and transportation of the magnet 
%yokes and aluminum coils from CERN to J-PARC started in 2007. 
%The magnet system was shipped from CERN in January 2008 and 
%arrived in Japan in March 2008. Then, reassembly of items and 
%installation in the ND280 pit was performed. 

The refurbishing of the magnet yokes and aluminum coils was performed
at CERN. Then, they were packed and shipped to Japan, and reassembled and 
installed in the ND280 pit. During the installation
particular attention was paid to take into account the constraints of 
alignment coming from the later insertion of the SMRD modules within 
the gaps of the magnet yokes, 
which required that the 16 individual yoke elements, each weighing 53 tons, 
be aligned with a precision of better than 1~mm. 
After successful magnet yoke and coil assembly, 
the installation of the services was performed. 
In August 2009 the B-field mapping device was installed in the basket 
and the magnet was then ready to be closed. 
The commissioning and B-field mapping procedure during August/September 
2009 allowed for comprehensive testing of many aspects of the magnet.

During the dedicated mapping procedure
the magnetic field of the ND280 magnet was precisely measured 
with a computer-controlled movable device equipped with 89  electronic cards, 
each holding three orthogonal Hall probes and the corresponding readout electronics. 
The overall mechanical structure of the device had dimensions of
2~m $\times$ 2~m $\times$ 1~m 
and could be moved throughout the whole instrumented region of the basket. 
Special care was taken in the region of the TPCs. 
The intrinsic uncertainty of each Hall probe in our region of interest is 0.2~G 
and the systematic error of the measurements is as low as 0.5~G. 
This was ensured by a calibration method 
employing a very well-known homogeneous field, from a magnet designed for the purpose. 
Higher-order magnetic moments and second-order Hall effects were taken into account for the measurements.

An important element of the field mapping was 
surveying the position and skewing of the measurement device before, during, and after the mapping. 
Two surveys were carried out.
One related the reference frame of the mapping machine to the ND280 main reference frame;
the accuracy of this survey is better than 1~mm in position offset and better than 1~mrad in rotational uncertainty. 
The second survey defined the internal reference frame of the mapping device, 
which is even more accurate, with the possibility of measuring changes in position up to a precision of 10~$\mu$m. 
The alignment of the Hall probes with respect to each other was ensured at a level of 1~mrad.
The above-mentioned systematic error of 0.5~G was achieved by exploiting the ND280 magnet's Cartesian symmetry.

The measurements were performed at a magnetic field value of 0.07~T---lower
than the nominal field during neutrino data taking due to limited electrical power
available at the time from the ND280 facility. 
It was therefore necessary to rescale to the nominal field, 
also taking hysteresis and saturation effects into account. 
By using a quadratic function as a first-order correction, 
a field uncertainty of $10^{-3}$ was obtained after scaling. 
The errors of the measurements scale in the same way, 
leading to a final magnetic field uncertainty of 2~G for each field component at the nominal field of 0.2~T. 
This very precise knowledge of the magnetic field map, especially for the
transverse field components, helps to reduce the systematic uncertainty of 
the momentum determination, which is intended to be below 2\% for 
charged particles below 1 GeV/c.

\begin{figure}[tbh]
        \centering\includegraphics[clip, width=\linewidth]{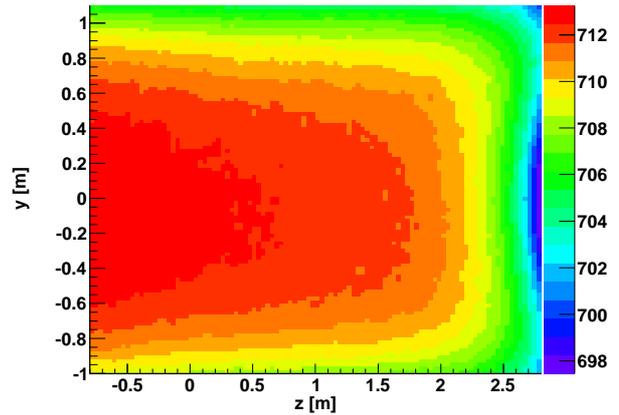}
    	\caption{The color plot shows a slice ($x=0$, the basket central plane) of the mapped B-field (in Gauss) in the TPC region. The neutrino beam is entering the picture from the left.}
	\label{fig:x0_slice} 
\end{figure}

\begin{figure}[tbh]
        \centering\includegraphics[clip, width=\linewidth]{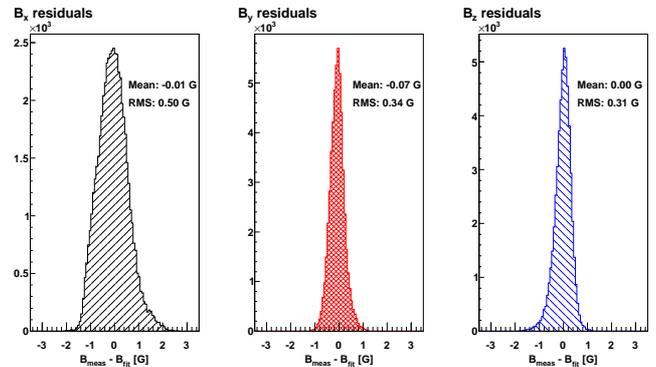}
    	\caption{For each B-field component ($x$, $y$ and $z$ respectively) the residuals between a fit of the data and the actual measurements is shown. The RMS of the distributions is taken as a measure of the systematic uncertainty of the mapping. The fit is performed in the center region.}
	\label{fig:residuals} 
\end{figure}

Figs.~\ref{fig:x0_slice} and \ref{fig:residuals} 
show two performance plots of the magnetic field mapping. 
Fig.~\ref{fig:x0_slice} shows a slice of the TPC region. 
The field is quite homogeneous in the center of the magnet but increasingly varies
the closer one comes to the edges of the TPC region. 
In Fig.~\ref{fig:residuals}, the difference between our fit to the data and 
the actual measured values is plotted for each B-field component. 
The widths of the resulting distributions are a measure of the systematic error of the B-field measurement. 
They indicate residuals less than 1~G for each field component.

%residuals for each B-field component between measured and fitted value showing the systematic uncertainty of the map
 %Rubbia
\subsubsection{Pi-zero Detector (\pod{})}
\label{P0D}
% \subsubsection{Pi-zero Detector (\pod{})}
% Lead Author: McGrew/Toki

The primary objective of the \pod{} is to measure the neutral
current process $\nu _{\mu }+N\rightarrow \nu _{\mu }+N+\pi ^{0}+X$ on a
water (H$_{2}$O) target with the same neutrino beam flux as reaches
Super-Kamiokande.
These aims were realized by a design using $x$ and $y$ planes of
scintillator bars,
with each bar read out with a single WLS fiber.
The planes of scintillator bars are interleaved with fillable water target
bags and lead and brass sheets.  This arrangement forms a
neutrino target where the \pod{} operates with the water target bags filled
or emptied, enabling a
%simple 
subtraction method to determine the water target
cross sections.  The scintillator bars provide sufficiently fine segmentation
to reconstruct charged particle tracks (muons and pions) and
electromagnetic showers (electrons and photons from $\pi ^{0}$'s).

\begin{figure}[tbh]
  \includegraphics[width=\linewidth]{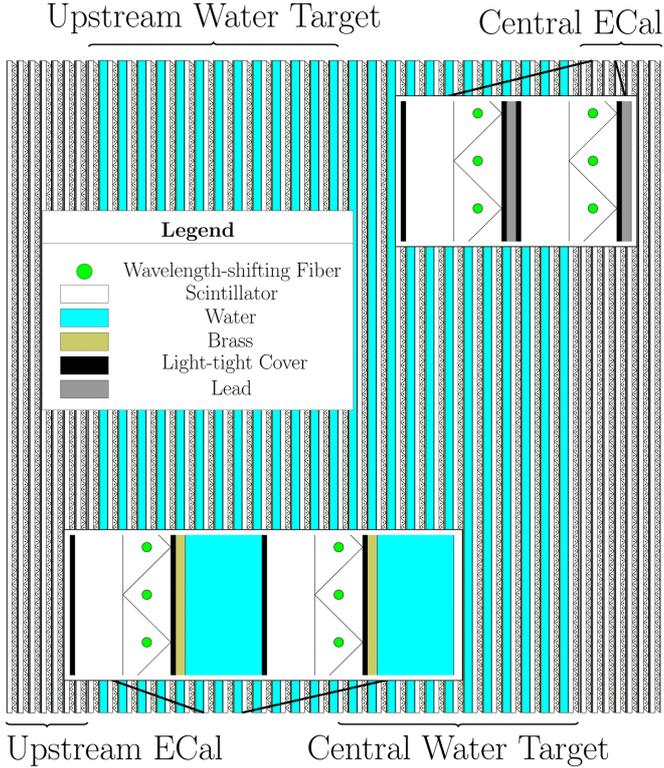}
  \caption{\label{fg:p0dSchematic} A schematic of the pi-zero detector.
    The beam is coming from the left and going right.   Insets show details
    of the Water Target super-\pod{}ule layers and Central ECal layers.}
\end{figure}

The main features of the \pod{} design are shown in
Fig.~\ref{fg:p0dSchematic}.  The central section, composed of the
``upstream water target'' and ``central water target'', uses alternating
scintillator planes, HDPE water bags, and brass sheets.  The front and rear
sections, the ``upstream ECal'' and ``central ECal'', use alternating
scintillator planes and lead sheets.  This layout improves the containment
of electromagnetic showers and provides a veto region before and after
the water target region to provide effective rejection of particles 
entering from interactions outside the \pod{}.

There are a total of 40 scintillator modules in the \pod{}.  Each \pod{}
module, or \pod{}ule, has two perpendicular arrays of triangular
scintillator bars.  There are 134 vertical bars (2200~mm long) and 126
horizontal bars (2340~mm long) in each \pod{}ule.  Each bar has a single
hole filled with a WLS fiber (Kuraray double-clad Y11 of 1~mm diameter).
Each fiber is mirrored on one end and the
other end is optically read out using a Hamamatsu MPPC (see
Section~\ref{MPPC}).  Each photodetector is read out with TFB
electronics (see Section~\ref{ND280elec}).  There are 40 \pod{}ules, each
with 260 scintillator bars and fibers being read out,
totaling 10,400 channels for the
entire \pod{} detector.  The \pod{}ules were formed into four ``super-groups''
called super-\pod{}ules.  The two ECal super-\pod{}ules are a sandwich
of seven \pod{}ules alternating with seven stainless steel clad lead sheets
(4~mm thick).  The upstream (central) water target super-\pod{}ule is a
sandwich of 13 \pod{}ules alternating with 13 (12) water bag layers (each 28~mm
thick), and 13 (12) brass sheets (each 1.5~mm thick).
The water target layers each have two bags,
for a total of 50 in the \pod{} detector,
each with dimensions of 1006~mm $\times$ 2062~mm $\times$ 28~mm.
The dimensions of the
active target of the entire \pod{} are 2103~mm $\times $ 2239~mm $\times $
2400~mm (width $\times $ height $\times $ length) and the mass of the detector
with and without water is 16.1~tons and 13.3~tons respectively.

The \pod{} polystyrene scintillator bars were identical to bars originally
developed for the MINERvA experiment~\cite{PlaDalmau:2005}.  The bulk
polystyrene is Dow Styron 663 (W), a commercial grade, general-purpose
polystyrene without additives.  Wavelength-shifting dopants, 1\% PPO and
0.03\% POPOP, were added into the bulk polystyrene.  The cross section of
the extrusion is an isosceles triangle with a 33~mm base and 17~mm height.
There is a hole centered in both dimensions, with a diameter of
approximately 1.5~mm, through which a WLS fiber may be inserted.  A
thin (0.03~mm on average) co-extruded layer of polystyrene with 20\%
TiO${}_2$ was added to the outside of the strip in order to reflect
escaping light back into the bulk and increase the probability of capture
by the center fiber.

The WLS fibers were mounted in the scintillating bars by
gluing a custom ferrule over one end of each fiber so that a
small portion of the fiber and epoxy extended past the ferrule.  The fiber
and epoxy were then diamond-polished.  The MPPCs were mounted in custom
sleeves designed to snap-fit to a ferrule, allowing them to be installed
and removed as necessary.

The \pod{} construction was done in three stages.  First, the scintillator bars
were glued into arrays of $15-17$ bars on a template mounted on an optical
table.
The arrays were cured at room temperature, 
under a vacuum film, for a minimum of four hours.
These pre-glued bar arrays were called ``planks''.  Each \pod{}ule uses 16
planks and a total of 640 are required for the entire detector.

In the next stage, the \pod{}ules were constructed on a gluing table.  The
\pod{}ules were assembled as a sandwich of an outer lower PVC skin, eight
$x$-scintillator planks, eight $y$-scintillator planks, and an outer upper PVC
skin.  All four edges of the assembly were enclosed with PVC frames, which
had been drilled with precision holes to allow the fibers to be inserted
and connected to the MPPCs after the \pod{}ules were assembled.  The
assembly was coated with epoxy and cured under a vacuum film overnight.
After the \pod{}ules were assembled, the fibers were inserted into each
bar, and the MPPCs were attached to the fibers and connected via
mini-coaxial cables to the TFB electronics boards.  Then the \pod{}ule was
scanned with a movable $^{60}$Co source to characterize the signal from
every channel.

In the last stage, the instrumented \pod{}ules were assembled into
super-\pod{}ules by laying a \pod{}ule with lead plates (for the ECals) or
water bags plus brass sheets (for the water targets) on a horizontal
strongback table.  This strongback table was lifted to a vertical
position to assemble an upright super-\pod{}ule.
%Moving this earlier, during other dimensions
%The water target layers each have two bags,
%for a total of 50 in the \pod{} detector,
%each with dimensions of 1006~mm $\times$ 2062~mm $\times$ 28~mm.
Finally, the TFB electronics
boards were mounted onto aluminum plates attached to two aluminum cooling
extrusions in which a closed loop of negative pressure cooling water flows
at $\sim$5~L/min.  
The electronics plus cooling assembly was mounted on the top
and one side of the super-\pod{}ule.  A light injection system was added
that strobes the opposite end of the fiber with an LED flasher.
Final testing of the super-\pod{}ules, using a cosmic ray trigger,
the water bag filling system and the light injection system,
was done at J-PARC, prior to installation into the ND280 off-axis detector.

%% CKJ - remove chronological reference
%In September 2009, all four super-\pod{}ules were lifted with the NM building
%crane and lowered into position into the support basket frame of the ND280
%detector.  
After installation of the super-\pod{}ules in the pit, 
%% CKJ
airtight aluminum cover panels were placed over the
electronics and dry air was circulated to moderate temperature fluctuations
while preventing condensation on the electronics cooling system.

Determining the amount of water in the fiducial volume is critical to the
\pod{} physics goals.  The required precision is achieved by first
measuring the mass vs. depth in an external 
buffer tank, filling the water targets to predetermined levels, and then
observing the water volume removed from the tank.
The water target volume is
instrumented using a combination of binary (wet or dry) 
level sensors and pressure sensors, 
allowing the depth of the water to be determined to $\pm$5~mm.  The
water target fiducial region is designed to contain 1944 $\pm$ 53~kg of
water, and the measured mass is $1902\pm{}16$~kg.

During initial operations, all but seven of the 10,400 \pod{} detector channels
were operational.  The detector was calibrated with minimum ionizing
tracks from cosmic ray muons.
An average of 19 photoelectrons was obtained for the scintillator
bars and 38 photoelectrons per $x/y$ layer.  The average attenuation of the
pulse height in the scintillator bars from opposite ends is approximately
30\%.  The internal alignment of scintillator bars was checked using 
through-going muons with the magnet field off, and was determined to be
approximately 3 mm.
 %McGrew/Toki
\subsubsection{Time Projection Chamber (TPC)}
\label{TPC}
% \subsubsection{TPC}
% Lead Author: Karlen/Zito

The TPCs perform three key functions in the near detector.
Firstly, with their excellent imaging capabilities in three dimensions, the
number and orientations of charged particles traversing the detectors
are easily determined and form the basis for selecting high purity
samples of different types of neutrino interactions.
Secondly, since they operate in a magnetic field, they are used
to measure the momenta of charged particles produced by neutrino
interactions elsewhere in the detector, and therefore determine the
event rate as a function of neutrino energy for the neutrino beam,
prior to oscillation.
Finally, the amount of ionization left by each particle, when combined
with the measured momentum, is a powerful tool for distinguishing different
types of charged particles, and in particular allows the relative abundance of
electron neutrinos in the beam to be determined.

\begin{figure}[htp]
\centering
\includegraphics[width=\linewidth]{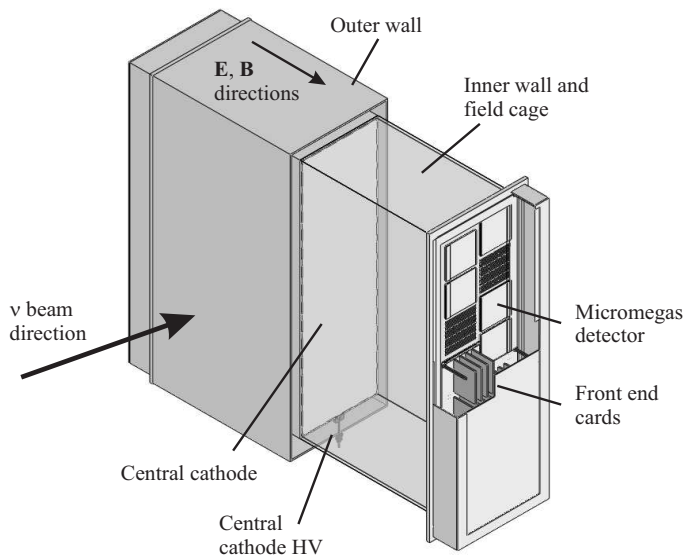}
\caption{Simplified cut-away drawing showing the main aspects of
the TPC design. The outer dimensions of the TPC are approximately 
2.3~m $\times$ 2.4~m $\times$ 1.0~m.
}\label{fig:tpc-design}
\end{figure}

Each TPC consists of an inner box that holds an argon-based
drift gas, contained
within an outer box that holds CO$_2$ as an insulating gas.
The inner (outer) walls are made from composite panels with copper-clad G10
(aluminum) skins.
The inner box panels were precisely machined to form an 11.5~mm pitch copper
strip pattern which, in conjunction with a
central cathode panel, produces a uniform electric drift field in the
active drift volume of the TPC, roughly aligned with the
field provided by the near detector magnet.
A simplified drawing of the TPC design is shown in 
Fig.~\ref{fig:tpc-design}.

Charged particles passing through the TPCs produce ionization electrons 
in the gas that
drift away from the central cathode and toward one of the readout planes.
There, the electrons are multiplied and sampled with 
bulk micromegas~\cite{Giomataris:2004aa}
detectors with 7.0~mm $\times$ 9.8~mm (vertical $\times$ horizontal)
anode pad segmentation.
The pattern of signals in the pad plane and 
the arrival time of the signals combine to
give complete 3D images of the paths of the traversing charged particles.
Twelve 342~mm $\times$ 359~mm micromegas modules tile each
readout plane, for a total of 72 modules and nearly 
9~m$^2$ of active surface for the three TPCs, the first to use
micropattern gas detectors in a physics experiment.
%%%CKJ: replace the following sentence. 
%The modules are arranged in two vertical columns 
%with small horizontal (vertical) offsets between vertically (horizontally)
%adjacent modules so that the small inactive regions between
%modules are not aligned.
The modules are arranged in two vertical columns that are offset so that the
small inactive regions between modules are not aligned.
%%%CKJ

Blind vias are used to route connections between the readout pads and
connectors on the back side of the micromegas printed circuit boards.
Six front-end electronics cards, each using four custom ASICs 
called ``AFTER'', plug into the connectors and sample and 
digitize signals from the 1,728 pads.
Each AFTER ASIC shapes the signals and buffers 72 pad signals into 511
time-bin switched capacitor arrays.
The six front-end cards connect to a single front-end mezzanine card
that aggregates the data, performs zero suppression, and sends
the remaining data off detector over a 2~Gb/s optical link.

The gas system was designed to maintain a stable mixture
in the inner volume, a constant positive pressure with respect to the
outer volume, and a constant pressure between the outer volume and the
atmosphere.
The inner gas mixture, Ar:CF$_4$:iC$_4$H$_{10}$ (95:3:2) 
was chosen for its high speed, low diffusion, and good performance
with micromegas chambers.
Each of the three TPC volumes contains 3000~liters, and each of the three 
gap volumes contains 3300~liters.
The TPC gas system was designed for an operating flow of 10~L/min/TPC 
(30~L/min total flow), corresponding to five TPC-volume flushes per day.
To reduce gas operating costs, the system was designed to purify and 
recycle roughly 90\% of the TPC exhaust gas.

A calibration system produces a control pattern of electrons
on the central cathode
in order to measure and monitor important aspects of the electron transport
in the TPCs.
Photoelectrons are produced from
thin aluminum discs glued to the copper surface of the cathode by
flashing the cathode with a diffuse pulse of 266 nm light.
Data from this system are used to precisely determine the electron 
drift velocity and to measure distortions in the electron drift due
to inhomogeneous and misaligned electric and magnetic fields.

Since late 2009, the three TPCs have been in place within the 
off-axis near detector, and
the TPC systems operated stably during the first physics run.
After correcting
for atmospheric pressure variation, the residual gain variation, 
due to other factors such as gas composition,
is below 1\% and therefore does not degrade particle identification
performance.

\begin{figure}[htp]
\centering
\includegraphics[width=\linewidth]{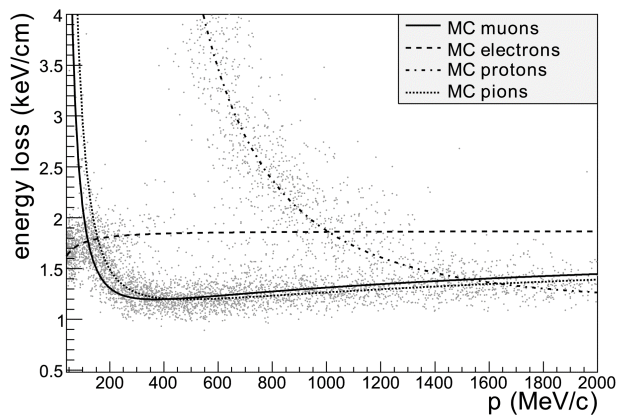}
\caption{Each point shows measurements by a single TPC of
the energy loss and momentum of positively charged 
particles produced in neutrino interactions. The expected
relationships for muons, positrons, protons, and pions are shown by the curves.
}\label{fig:tpc-dedx}
\end{figure}

Particle identification is done with a
truncated mean of measurements of energy loss of charged particles in the gas.
The linear charge density of the track is estimated for each cluster 
by taking into account the length of the track segment corresponding
to a pad column. 
The lowest 70\% of the values are used to compute the truncated mean,
an optimized approach found through Monte Carlo simulation and 
test beam studies.
The resolution of deposited energy obtained using this method is 
about 7.8\% for minimum ionizing particles, 
better than the design requirement of 10\%. 
Fig.~\ref{fig:tpc-dedx} demonstrates the TPC particle
identification capability by comparing energy loss and momentum
for positively charged particles recorded during the first T2K physics
run.

The point spatial resolution is estimated by comparing the 
transverse coordinate resulting from the global track fit 
to the one obtained with information from a single column of pads.
The resolution is found to be 
typically 0.7~mm per column, in line with expectations, and
degrades with increasing track angle with respect to the horizontal
due to the ionization fluctuations along the track.
The observed spatial resolution is sufficient to achieve the momentum
resolution goals for the detectors.

More information about the design, construction, and performance
of the TPC systems can be found in a recent publication~\cite{T2KtpcNIM2010}.
 %Karlen/Zito
\subsubsection{Fine Grained Detector (FGD)}
\label{FGD}
% \subsubsection{Fine Grained Detector (FGD)}
% Lead Author: Oser

Two fine grained detectors (FGDs) provide target mass for
neutrino interactions as well as tracking of charged particles coming
from the interaction vertex. The FGDs are constructed from 9.61~mm
$\times$ 9.61~mm $\times$ 1864.3~mm bars of extruded polystyrene
scintillator, which are oriented perpendicular to the beam
in either the $x$ or $y$ direction.  Each scintillator bar has a
reflective coating containing TiO$_2$ and a WLS fiber
going down a hole in its center. One end of each fiber is mirrored by vacuum
deposition of aluminum, while the other end is attached to an MPPC and associated electronics,
which digitize the light signal produced by scintillation inside the
bar.  

\begin{figure}
\begin{center}
        \includegraphics[width=0.7\linewidth]{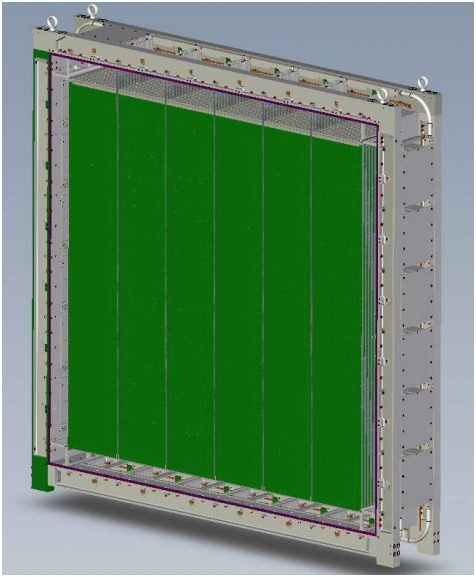}
\end{center}
\caption{
%Photograph of an FGD with the front cover removed, supported
%by a portable brown cart.  The surface of the outermost XY
%scintillator module is clearly visible, along with the five steel straps
%that support its weight.
View of an FGD with the front cover removed.  XY scintillator modules
(green) hang perpendicular to the direction of the neutrino beam.  Along the
top, six mini-crates with electronics can be seen without their cooling
lines, while on the right side the cooling lids covering the mini-crates are
shown.
\label{fig:fgd_mech}
}
\end{figure}

Each FGD (see Fig.~\ref{fig:fgd_mech})
has outer dimensions of 2300~mm $\times$ 2400~mm
$\times$ 365~mm (width $\times$ height $\times$ depth in beam direction), and contains
1.1~tons of target material.  The first FGD consists of 5,760
scintillator bars, arranged into 30 layers of 192 bars each, with each
layer oriented alternatingly in the $x$ and $y$ directions
perpendicular to the neutrino beam.  The scintillator provides the
target mass for neutrino interactions, and having alternating $x$ and
$y$ layers of fine grained bars allows for tracking of charged
particles produced in those interactions.    An ``XY module''
consists of one layer of 192 scintillator
bars in the horizontal direction glued to 192 perpendicular bars in
the vertical direction, with thin G10 sheets glued to the outer
surfaces to add structural stability. 
The photosensors are mounted along all four sides of the XY module on
photosensor bus-boards that are screwed directly into the edges of the
XY module.  Each fiber is read out from one end, and within an $x$ or
$y$ layer alternating fibers are read out from alternating ends.  An
LED-based light injection system that flashes the exposed far ends of
the WLS fibers permits {\em in situ} calibration of
photosensor response, saturation, and non-linearity.

The second FGD is a water-rich detector consisting of seven XY modules
of plastic scintillator alternating with six 2.5~cm
thick layers of water (for a total of 2,688 active scintillator bars
and 15~cm total thickness of water). These layers are made from sheets
of thin-walled hollow corrugated polycarbonate, 2.5~cm thick, whose
ends have been sealed with HE 1908 polyurethane sealant.
The modules are then filled with
water to provide a layer of water target.  The water is maintained
under sub-atmospheric pressure by a vacuum pump system so that if a leak
develops the system will suck air into the modules rather than
spilling water inside the FGD. Comparing the interaction rates in the
two FGDs permits separate determination of cross sections on carbon
and on water.

Both FGDs were built with the same geometry, mounting, and
readout for interoperability.  Each FGD is contained in a light-tight
dark box that contains the scintillator, fibers, and photosensors,
while the FGD electronics are mounted in mini-crates around the
outside of the dark box. The modules are
supported by several stainless steel straps that loop around the bottom
of each module and attach to anchor points in the top side of the dark box.
The dark box itself is a sturdy aluminum frame that supports the
weight of the FGD modules and transfers that weight to the detector
basket.  The walls of the dark box are made of thin opaque panels to
keep its interior light-tight.

The FGD's front-end electronics resides
in 24 mini-crates that attach to the outside of the four sides of the
dark box.  Signals from the photosensors inside the dark box are
carried from the photosensor bus-boards to the electronics by ribbon
cables that attach to the crates' backplanes, which are mounted over
cutouts on the four sides of the box.  The mini-crates are cooled by a
negative-pressure water cooling system running along the sides of the
frame of the dark box, and power is carried to the mini-crates by a
power bus mounted on the frame.  The electronics is arranged so that
all heat-producing elements are located outside of the dark box in the
mini-crates where they can be readily cooled by the cooling system,
while only elements with negligible power outputs (the photosensors
themselves) are present inside the dark box.

Each mini-crate contains four front-end boards and one crate master board (CMB),
and can read out 240 photosensors.  The front-end boards
use the ``AFTER'' ASIC (Section~\ref{TPC}), to
shape and digitize high and low attenuation copies of the photosensor
signals at 50~MHz, storing the waveform in a 511-deep switched
capacitor array.  In addition the front-end boards provide the
photosensor bias voltages and analog trigger primitives.  Data from
each crate is read out over optical fiber links to data collector
cards (DCCs) located outside of the magnet.  Slow control systems use a
separate data and power bus for redundancy.

 %Oser
\subsubsection{Electromagnetic Calorimeter (ECal)}
\label{ECAL}
% \subsubsection{Electro-magnetic Calorimeter (ECal)}
% Lead Author: Touramanis

The ND280 ECal is a sampling electromagnetic calorimeter surrounding 
the inner detectors (\pod{}, TPCs, FGDs). It uses layers of 
plastic scintillator bars as active material with lead absorber sheets 
between layers, and it provides near-hermetic coverage for all 
particles exiting the inner detector volume. 
Its role is to complement the inner detectors in full event 
reconstruction through the detection of photons and measurement of 
their energy and direction, as well as the detection of charged 
particles and the extraction of information relevant for their 
identification (electron-muon-pion separation). A key function of 
the ECal is the reconstruction of $\pi^0$'s produced in neutrino 
interactions inside the tracker detectors. In the case of $\pi^0$ 
production inside the \pod{}, the \pod{}-ECal complements the 
\pod{} reconstruction with information on escaping energy.

\begin{figure}[tb]
  \begin{center}
    \includegraphics[width=\linewidth]{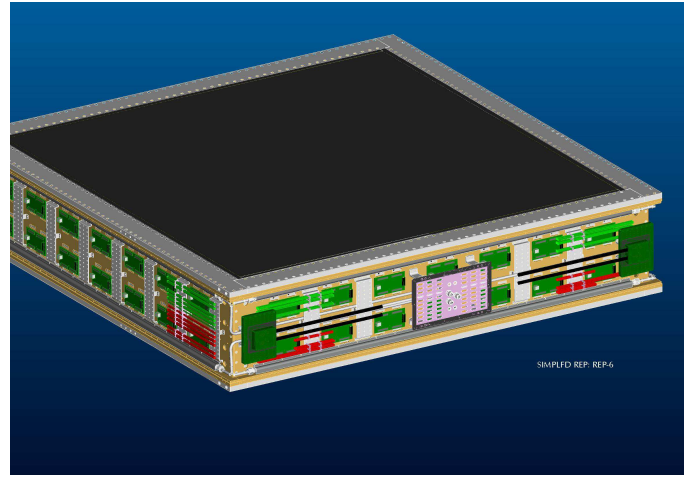}
  \end{center}
  \caption{\label{fg:ECAL-module} External view of one ECal module. 
The scintillator bars run
horizontally inside the module as shown. The readout electronics, signal and
power cables, and cooling pipes can be seen mounted on the aluminum plates
on the sides of the module. The gray surface at the top is the carbon fiber
sandwich front plate, which in the final module position is facing towards
the inner subdetectors (\pod{}, FGDs and TPCs).}
\end{figure}

The ECal is made of 13 independent modules of three different 
types arranged as in Fig.~\ref{fg:nd280-exploded}: 
six Barrel-ECal modules surround 
the tracker volume on its four sides parallel to the $z$ (beam) axis; 
one downstream module (Ds-ECal) covers the downstream exit of the 
tracker volume; and six \pod{}-ECal modules surround the \pod{} detector 
volume on its four sides parallel to the $z$ axis. 
Each module is made of consecutive layers of scintillator bars glued 
to a sheet of lead converter. The Ds-ECal is located inside the basket 
carrying the inner subdetectors of the off-axis detector.
The other 12 ECal modules are mounted inside of the UA1 magnet.
A drawing of a completed module is shown in Fig.~\ref{fg:ECAL-module}.

All ECal scintillator bars have a 4.0~cm $\times$ 1.0~cm cross section 
with a 2.0~mm $\times$ 3.0~mm elliptical hole running along their 
full length in the middle. The bars were extruded at a dedicated 
Fermilab facility and the material used was polystyrene doped with 1\% PPO 
and 0.03\% POPOP. A 0.25~mm thick layer of TiO$_2$ was co-extruded at 
the surface of the bars providing light reflection and isolation. 
A Kuraray 1~mm diameter double-clad Y11 WLS fiber 
runs along the hole in the center of each bar as described in 
Section~\ref{INGRID} for the INGRID. The fibers were cut to length 
and diamond-polished at a dedicated Fermilab facility. 
The light is read out at one or both ends of each fiber with MPPCs 
(see Section~\ref{MPPC}). 
Those read out at one end only are mirrored at the other end 
with vacuum deposition of aluminum, performed at the same facility. 
Fiber-MPPC optical coupling is 
achieved using custom-made plastic connector assemblies.
A ferrule is glued to the fiber end using Saint-Gobain BC600 silicon-based 
optical epoxy resin. The ferrule secures mechanically inside an MPPC holder
that includes a layer of elastic foam to ensure good 
contact between the fiber end and the MPPC entrance window. 
The MPPC signal is read out with TFB cards (see Section~\ref{ND280elec}). 
The lead sheets are made of lead with 2\% antimony to provide 
some stiffness and were primed on both sides with a quick-drying metal primer.

Each ECal module was assembled from pre-made scint\-illator-lead layers. 
Each layer was prepared by laying the appropriate scintillator 
bars on a special aluminum tooling plate, aligning them  within 
a border made of precision-machined aluminum bars, 
and gluing them to the lead sheets using Araldite 2011 cured for a few hours 
under vacuum bagging. 
For each module, the front of the mechanical frame is made of a 
carbon fiber sandwich front plate, which provides support while putting minimal 
material in the way of incoming photons, glued to an aluminum frame.
The frame is screwed to side plates made of thick aluminum with holes to allow 
the WLS fibers to exit.
Attached to the side plates are back frames made of aluminum plates 
which provide support and a system for securing the modules 
in their final position. A separate side wall made of thinner 
aluminum plates carries the TFB readout cards, special extruded 
pipes carrying cooling water, and a dry-air circulation system, 
and supports power distribution bars and signal and communication cables. 
Each complete module is closed off with thin aluminum covers for protection 
and light tightening.

For every module, the mechanical structure was 
assembled and aligned and then the scintillator-lead layers were 
installed one at a time. For each layer the fibers were threaded 
and glued to their ferrules. After the glue had cured, all fibers 
of the layer were connected to a test system of MPPCs and readout 
electronics of well-known performance and scanned with a $^{137}$Cs 
source moved along each bar in predefined spatial intervals. 
An automated scanner, data acquisition system and analysis software 
suite allowed this process to be completed within a few hours 
for each layer. In this manner, a very small percentage of fibers were 
found to be damaged; these were
replaced before the next layer was installed in the module. 
Once all layers of each module were in place the module was 
fitted with the readout electronics, cables, and other services, 
tested, and shipped to J-PARC for installation. 

The Ds-ECal module consists of 34 layers with lead sheets of 
1.75~mm thickness for a total of 10.6~$X_0$. There are 50 bars,
each 2.04~m long, 
in every layer and each one is read out with an MPPC at each end. 
Consecutive layers have their bars at 90$^{\circ}$ to allow three-dimensional 
reconstruction of electromagnetic clusters and charged particle tracks. 

Limited by available space inside the UA1 magnet and structural 
considerations, the Barrel-ECal modules have 31 layers each with the 
same lead sheets of 1.75~mm thickness for a total of 9.7~$X_0$. 
Layers are again assembled with bar orientation alternating at 90$^{\circ}$. 
The bars running in the $z$ direction are 3.84~m long and are read by 
MPPCs at each end. Bars running in the $x$ ($y$) directions in the 
top/bottom (side) modules are 1.52~m (2.36~m) long and are each read by a 
single MPPC at one end.

The \pod{}-ECal modules are not intended for $\pi^0$ reconstruction 
as this takes place inside the dedicated \pod{} detector which 
they surround. However, their presence is required to detect 
photons that either do not convert in the active \pod{} volume or that produce 
showers only partially contained within the \pod{}. They can also confirm 
the passage of charged tracks, identify MIPs, and act as a veto for 
incoming backgrounds. This allows a simpler construction. 
Each module is made of six active scintillator layers separated by 
five layers of 4~mm thick lead converter giving 3.6~$X_0$. 
All bars are 2.34~m long and run along the $z$ direction for all layers. 
This simplifies construction while meeting the requirements as verified 
by Monte Carlo simulations. Each bar is read out by a single MPPC at one end. 

The Ds-ECal module was constructed in 2008 and was used for beam 
tests at the CERN T9 PS mixed electron-muon-hadron beam in April--June 2009. 
It was installed in ND280 in October 2009. The Barrel-ECal 
and \pod{}-ECal modules were constructed in 2009--10 and were 
installed in ND280 in July--October 2010. The complete ECal has been 
integrated with the rest of ND280 and is taking data.
 %Touramanis
\subsubsection{Side Muon Range Detector (SMRD)}
\label{SMRD}
% \subsubsection{Side Muon Range Detector (SMRD)}                   
                        
% Lead Author: Kutter        

The SMRD performs multiple functions. Firstly, 
it records muons escaping with high angles with respect to the 
beam direction and measures their momenta. Secondly, it triggers on cosmic 
ray muons that enter or penetrate the ND280 detector. Thirdly,
it helps identify beam-related event interactions in the 
surrounding cavity walls and the iron of the magnet.

\begin{figure}[htp]
  \begin{center}
    \includegraphics[width=0.5\linewidth]{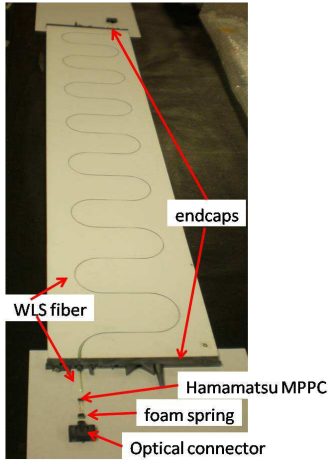}
  \end{center}
  \caption{View of SMRD scintillator counter components prior to assembly.}
  \label{fig:SMRD}
\end{figure}

The SMRD consists of a total of 440 scintillator modules which are 
inserted in the 1.7~cm air gaps between 4.8~cm thick steel plates 
which make up the UA1 magnet flux return yokes. 
The UA1 magnet consists of 16 C-shaped flux return yokes which are 
grouped in pairs to form a ring surrounding the inner detectors on four sides. 
Each yoke consists of 16 steel plates and hence has 15 air gaps 
in the radial direction.
Pairs of yokes are labeled 1 through 8 from upstream to downstream. 
The SMRD consists of three layers of scintillator modules on the 
top and bottom for all yokes. Both sides are instrumented with 
three layers for yokes 1 through 5, four layers for yoke 6 and six layers 
for yokes 7 and 8. 
All of the SMRD modules 
populate the innermost gaps so as to be able to detect particles 
escaping the inner detectors.
Due to the differently sized spaces for horizontal and vertical gaps, 
horizontal modules are composed of four scintillation counters with 
dimensions 875~mm $\times$ 167~mm $\times$ 7~mm 
(length $\times$ width $\times$ height) 
and vertical modules consist of five scintillation counters with 
dimensions 875~mm $\times$ 175~mm $\times$ 7~mm. The counter sizes 
have been optimized to maximize the active area in each magnet gap. 
The inter-module spacing in the beam direction is determined by the 
geometry of the inter-yoke spacing.
%In the directions transverse to the beam the inter-module space is 
%determined by the size of spacers which separate the 4.8~cm steel 
%plates of the UA1 magnet and measure between 1~cm and 3~cm.
The scintillation counters consist of extruded polystyrene and 
dimethylacetamide with admixtures of POPOP and para-terphenyl. 
The surface of each scintillator counter features a white diffuse 
layer which acts as a reflector. 
An S-shaped groove, with a bending radius of 
2.9~cm and a depth of 2.5~mm which deepens at the counter ends to 
4~mm, has been machined into each scintillator in multiple passes to 
ensure good surface quality. A 1 mm diameter Kuraray Y11 double-clad
%\cite{Y11-WLS} 
WLS fiber is glued with Bicron BC600  
into each groove \cite{Izmaylov:2009jq}.
As shown in Fig.~\ref{fig:SMRD},
the WLS fiber exits both sides of the scintillator through a ferrule 
which is part of an endcap. The endcaps, which are custom injection 
molded and made out of black Vectra, are glued and screwed to 
the end face of each scintillator counter,
%\cite{vectra} 
and were 
tested to be light-tight at the few photon level.

The outward-facing side of each endcap features a snap-on mechanism 
which allows for a connector holding a foam spring backed MPPC to couple reliably 
to the polished WLS fiber ends. The scintillator counters are 
wrapped in a layer of Tyvek
%\textsuperscript{\textregistered}
to increase the light yield by about 5\%. 
A stainless steel foil wrapping provides a light-tight and mechanically 
durable protection layer. Individual SMRD counters
are connected into modules
by means of aluminum C-channel profiles which are tapered towards 
the ends to facilitate
installation. Special cut-outs in the C-channels are used as 
attachment points for bronze springs which serve to center and 
fix the modules inside the magnet gaps.

All 4,016 MPPCs are connected to miniature printed circuit boards (PCBs) 
which are free to slide along rails in the backside of the optical 
connector to allow for optimal WLS fiber and MPPC interconnections. 
The miniature PCBs couple the MPPC signals into 
%\cite{hirose_cables} 
mini-coaxial cables, manufactured by Hirose,  
which lead the signals 
to the Trip-T front-end boards (TFBs) mounted on the vertical 
sections of the magnet yokes. 
The mini-coaxial cables are routed between the magnet yokes 
and measure 3.5~m and 3.0~m for 
horizontal modules and are 2.2~m long for all vertical modules. 

The SMRD front-end electronics consists of 128 TFBs which are described 
in Section~\ref{ND280elec}. Each of the SMRD TFBs features 
a mezzanine board to connect to and read out up to 12 temperature 
sensors which are embedded in the scintillator counter
endcaps close to the MPPCs. 
With temperature sensors on both sides of each module, the SMRD 
supports a network of 880 temperature sensors in total. 
The TFBs are interconnected via Cat~6 cables to four 
back-end electronics boards called readout merger modules (RMMs) and 
two clock trigger modules (CTMs). All of the electronics boards are 
powered by means of a power distribution system which is mounted 
along the lower parts of the magnet sides and connected to two Wiener 
power supply units. 
%\cite{wiener}. 

%The SMRD installation was completed in July 2009 and 
%it was the first ND280 off-axis subdetector to be installed.
%The SMRD has been triggering on and recording cosmic muon data 
%since fall 2009 and beam data from January 2010. 
SMRD calibration procedures are shared with other Trip-T 
based ND280 off-axis subdetectors. 
The average light yield of individual counters in response to a minimum 
ionizing particle amounts to about 50 p.e. for the summed 
signal from both ends of a counter. The beam-related SMRD event rate of  
coincidence hits has been observed to be stable to 
within 3\% after temperature corrections and 
a normalization of spills to protons on target (POT)
has been applied. The relative timing information of both sensors results in a 
position resolution of about 7~cm in the direction of the beam.
More than one year after construction, the percentage of 
dead MPPCs in the SMRD is 0.07\%.

A more detailed description of the SMRD can be found in \cite{SMRD_NIM}.

 %Kutter
\subsection{ND280 Electronics and DAQ}
\label{ND280elec}
% \subsection{ND280 Electronics and DAQ}
% Lead Author: Weber/Pearce/Poutissou

% contributions from Weber
% \subsection{ND280 Electronics and DAQ}
% Lead Author: Weber/Pearce/Poutissou
% Trip-T Electronics
\subsubsection{Readout Electronics}

The \pod{}, ECal and SMRD subdetectors and 
the INGRID detector use identical
electronics to read out the \mbox{MPPCs}. This electronics is based on the
Trip-T ASIC~\cite{tript}. Signals from 64 MPPCs are routed to
custom-designed front-end boards (Trip-T front-end boards or TFBs) that house
four Trip-T ASICs using miniature coaxial cables.  The signals from the
photosensors are capacitively split (1:10) and routed to two separate
channels of the ASIC, to increase the dynamic range of the electronics.
Depending on the MPPC gain,
a one photoelectron signal corresponds to around 10 ADC counts in the 
high-gain channel, while the full-scale signal in the low-gain channel 
corresponds to 500 p.e.

The Trip-T chip integrates the charge in programmable integration
windows, which are synchronized with the neutrino beam structure. There
is a programmable reset time after each integration cycle, which is at
least 50~ns long. The chip can store the result of 23 integration
cycles in a capacitor array. Once the 23 integration cycles have been
recorded, the data is multiplexed onto two dual-channel 10-bit ADCs,
which digitize the data. Signals from the high-gain
channel are routed to a discriminator, which is part of the Trip-T
ASIC. The front-end board is controlled by an FPGA, which also
timestamps the output of the discriminator 
with an accuracy of 2.5~ns. The discriminator threshold is
programmable from 0 to 5 p.e.
The ADC and timestamp data is assembled by the FPGA and
sent to a back-end board for data concentration and buffering. The
output from the discriminators is also used to calculate trigger
primitives, which are used to initiate the readout of the detector for
cosmic ray muons. Monitoring information (mainly temperature and
voltages) is also recorded by the TFB and asynchronously transmitted
to the back-end board.
More details regarding the front-end part of the electronics can
be found in~\cite{Vacheret:2007tfb}.

The back-end of the electronics system consists of readout merger
modules (RMMs), cosmic trigger modules (CTMs), several slave clock
modules (SCMs) and a master clock module (MCM). All the boards were
developed at the Rutherford Appleton Laboratory using a common
hardware platform, which has been built around a high-end Vertex II
Pro FPGA from Xilinx, which is clocked at 100 MHz. 
The board can drive 14 high-speed optical links
via its RocketIO and up to 192 LVDS links.

The signals from up to 48 TFBs, which are mounted on the detector and typically less
than 1~m away from the photosensors, are routed to one RMM via Cat~5e
cables. The RMM controls the TFBs, distributes the clock and trigger
signals and receives the data after a trigger signal is received by
the TFBs. It sends this data asynchronously via a Gigabit Ethernet
link to a commercial PC that collects and processes the data. The RMM is
equipped with 500~MB DDR2 memory and can buffer up to 128 triggers.
Each RMM receives trigger and timing signals from the SCMs (see below).

\begin{figure}[tb]
  \begin{center}
    \includegraphics[width=\linewidth]{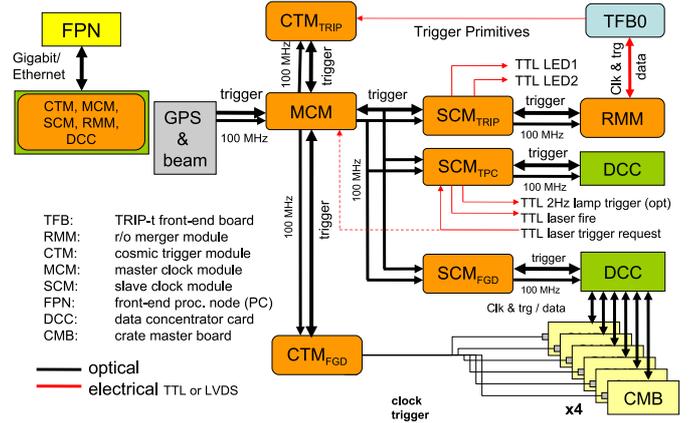}
  \end{center}
  \caption{\label{fig:ND280ElecSchematics} The general layout of the ND280 
  electronics.}
\end{figure}

The master clock module receives signals from the accelerator that
determine when the neutrino spill happens and also from a GPS-based
clock. The latter signals are used to synchronize the electronics to
UTC. The MCM is also connected to two cosmic trigger modules, which
receive signals from up to 192 TFBs or from 48 crate master
boards (in the case of the FGD). Based on these signals the CTM will
decide whether there was a cosmic event in the detector and trigger the
readout. The MCM can also generate pedestal and calibration triggers
at a programmable rate.  All timing and trigger signals are
transmitted via the RocketIO-driven optical link to the slave clock
modules. There is one SCM for each of the subdetectors (SMRD, ECal, \pod{},
FGD, TPC), which allows the electronics to be configured for independent
operation of each subsystem.  The INGRID is operated independently
from ND280 and only uses one MCM and a single CTM.
The general layout of the electronics has been visualized in Fig.~\ref{fig:ND280ElecSchematics}.

%Bibliography

%[tript]
%"Bench Test of First Trip-T Prototypes"
%Leo Bellantoni, Paul Rubinov 
%D0 note 4845

%or 

%J. Estrada, C. Garcia, B. Hoenison, and P. Rubinov, ÒMCM II and the
%Trip chip
%,Ó D0 note 4009, Fermilab-TM-2226, Dec 2003.

%[tfb]
%"The front-end readout system for the T2K-ND280 detectors"
%Vacheret, A.;   Greenwood, S.;   Noy, M.;   Raymond, M.;   Weber, A.; 
%Nuclear Science Symposium Conference Record, 2007. NSS '07. IEEE
%Issue Date: Oct. 26 2007-Nov. 3 2007
%On page(s): 1984 - 1991 

% contribution from Poutissou
% \subsection{ND280 Electronics and DAQ}
% Lead Author: Weber/Pearce/Poutissou
% DAQ and GSC intro section
\subsubsection{DAQ and Global Slow Control}

The ND280 data acquisition system has been divided into two components: DAQ 
and global slow control (GSC).
The DAQ component takes care of the main data stream, collecting the 
data banks from each subdetector front-end
system, storing the data in files and providing online histogramming. 
The GSC component runs in parallel to the DAQ using 
the same software 
framework: MIDAS~\cite{midas}.

% waiting for contributions from Pearce
% \subsection{ND280 Electronics and DAQ}
% Lead Author: Weber/Pearce/Poutisou
% DAQ

The ND280 off-axis and INGRID detectors
are equipped with independent DAQ systems
which are described in detail in~\cite{Thorpe:2010}.
These have a common architecture and are based on the 
MIDAS DAQ framework~\cite{midas},
operating on commercially available computing hardware running the 
Scientific Linux operating system.
MIDAS provides the system with a number of standard components necessary 
for operation and is interfaced
to the experimental hardware through custom C/C++ front-end client 
applications. 

\begin{figure}[tb]
  \begin{center}
    \includegraphics[width=\linewidth, clip]%
    {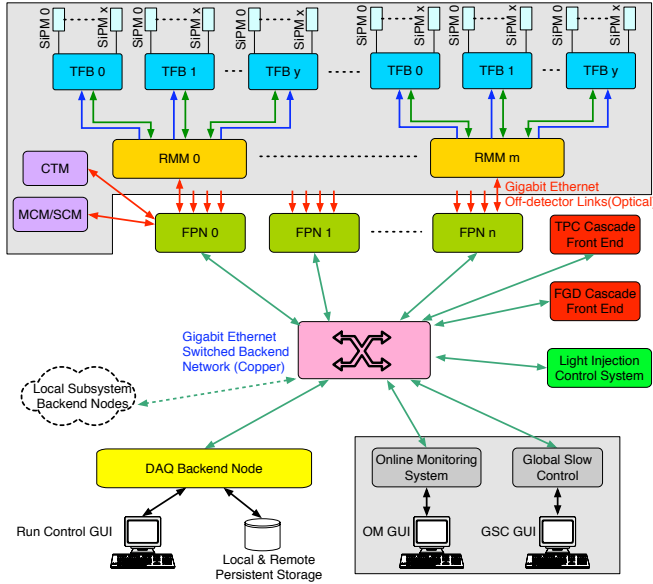}
  \end{center}
  \caption{\label{fig:nd280daq} The architecture of the T2K ND280 off-axis and INGRID detectors' DAQ systems.}
\end{figure}

The system architecture is shown in Fig.~\ref{fig:nd280daq}.
In both the INGRID and off-axis systems the back-end electronics modules 
on the Trip-T based subdetectors
are interfaced to the DAQ by point-to-point optical Gigabit Ethernet links.
The MIDAS processes are distributed across a number of nodes to provide 
the necessary performance
and to allow flexible partitioning of the system.
An additional MIDAS instance is implemented for each of the FGDs and 
TPCs in the off-axis detector system. These assemble
data from the respective readout electronics and transmit it to the global DAQ.
A pair of commercial Gigabit Ethernet switches
interconnect the nodes in the DAQ system, providing local infrastructure 
and data transfer networks. The DAQ networks are isolated from the main 
T2K experimental network by a further node acting as a gateway.

The Trip-T based subdetectors are controlled and read out by the front-end
processor nodes (FPNs),
each of which serves up to two back-end boards.
The FPN is implemented as three tasks running as separate processes, 
interconnected by shared memory data buffers and communicating via 
standard inter-process mechanisms.
Readout and configuration of the electronics and all connected hardware 
is provided by the readout task (RXT).
The readout is parallelized across electronics boards in a 
multi-threaded manner
and data is buffered for access by the data processing task (DPT).
The RXT additionally receives periodic monitoring data from the TFBs 
which it passes to the global slow control.
The DPT performs data reduction and basic data processing. 
It decodes the TFB raw data blocks,
associates amplitude and timing information for individual hits,
performs pedestal subtraction on a channel-by-channel basis,
applies zero suppression to the unsparsified data 
and formats the data for output.
To preserve monitoring information,
the DPT also performs per-channel histogramming of signal amplitudes 
for specific trigger types prior to zero suppression
and periodically inserts the histograms into the output data stream.
The processed event fragments are buffered and dispatched to the 
DAQ back-end by the third process,
which implements the MIDAS front-end functionality. 
% Need to insert paragraph about TPC/FGD front-ends here.
% Waiting for Renee's input

The MIDAS framework provides the core components necessary for the DAQ system
and is used to gather the event fragments from the Trip-T FPNs, 
and the TPC and the FGD front-ends.
An event-building process collects the fragments,
performs some basic consistency checks 
and writes the fully assembled events to a system buffer for output 
to a local RAID array.
A custom archiver process transfers completed files to
the KEK HPSS~\cite{nd280data-hpss} mass storage facility over the network
and additionally makes a preview copy available to a local semi-offline system for 
fast-turnaround analysis.

Additional clients are provided to interface the DAQ to the light 
injection (LI) hardware for the ECal and \pod{} subdetectors. 
These configure the LI hardware according to predefined calibration 
sequences and monitor the status of the LI system.
This information is provided in real-time to the DPT, where it is 
used to manage the formation of
calibration histograms.
A custom online-monitoring server based around the ROOT framework
retrieves built events from the system buffer and generates a range of plots
for data and detector quality monitoring.

%[t2kdaq]
%IEEE Paper .. will fill in appropriate reference

%[midas]
%MIDAS (Maximum Integration Data Acquisition System) 
%Online: http://midas.psi.ch

% contribution from Poutissou
% \subsection{ND280 Electronics and DAQ}
% Lead Author: Weber/Pearce/Poutissou
% GSC section

In terms of hardware, the GSC consists of two PCs
and two network switches connecting to subdetector-specific equipment. 
Front-end tasks running on the GSC PCs or on a subdetector computer 
connect to the various 
equipment of the magnet and the subdetectors. 
A central shared memory database named ODB contains the settings and 
the readback information 
from all front-end tasks. This data is in turn stored in a MySQL 
database at regular intervals. 
Users interact with the GSC through web pages and can display history 
plots of any of the
stored variables. Several customized web pages have been developed 
for controlling the magnet, 
power supplies, electronics parameter settings, and the TPC gas system, or for 
displaying detector feedback, such as temperature
readings or the \pod{} water target levels. 
Alarms are set up interactively to catch readback variables going 
out of range, alerting
the shift person as well as sending messages to remote experts. 
The GSC system is the main tool used by the shift person responsible 
for safety matters. 

The MySQL database containing the GSC information expands at 
the rate of 0.5 GB per week while
taking data. It is an essential element of the calibration process 
and offline analysis. 
This database is currently replicated from J-PARC to a server 
located at TRIUMF. Thanks to
this server, collaborators can query the GSC history plots in 
real-time from a remote location
and can connect to the database from computers and grid systems 
all over the world for data processing. 

An electronic logbook (Elog) service is used to complement the GSC system.
It runs on a computer in the counting room, with a replica hosted at TRIUMF
and is read-write accessible worldwide, with email capabilities. 
This provides a record of each shift and any maintenance work, and is
extremely useful for communication of problems
between J-PARC shift people and remote detector experts.

 %Weber/Pearce/Poutissou
\subsection{ND280 Software and MC Simulation}
\label{ND280soft}
% \subsection{ND280 Software and MC Simulation}
% Lead Author: Lindner/Uchida

\subsubsection{Introduction}
The ND280 offline software suite consists of about 60 software packages
totaling over 600,000 lines of original code.

Work on the ND280 offline software commenced in 2004
in order to allow it to be used to aid the early discussions
on the design of the ND280 detectors.
The design principles that have been followed include
an emphasis on the use of standard particle physics software libraries and conventions where possible,
structural modularity, good documentation and openness.
Hence ROOT~\cite{Brun:1997pa} is used for the underlying framework and data storage model,
with Geant4~\cite{Agostinelli:2002hh} as the basic simulation library.
The CMT~\cite{Arnault:2000vu} software configuration management system provides the packaging mechanism,
with CVS~\cite{software-cvs} used to store all versions of the source code files in a repository.

It was anticipated that a large number of physicists,
in groups spread across the nations that form the ND280 Collaboration,
would write and maintain the software,
so a modular structure with well-defined interfaces was chosen
such that individual components could be developed independently of each other.
Approximately 100 collaborators have actively contributed to the code base as of this writing.

\subsubsection{Structure}
\begin{figure}[htb]
\centering\includegraphics[width=\linewidth]{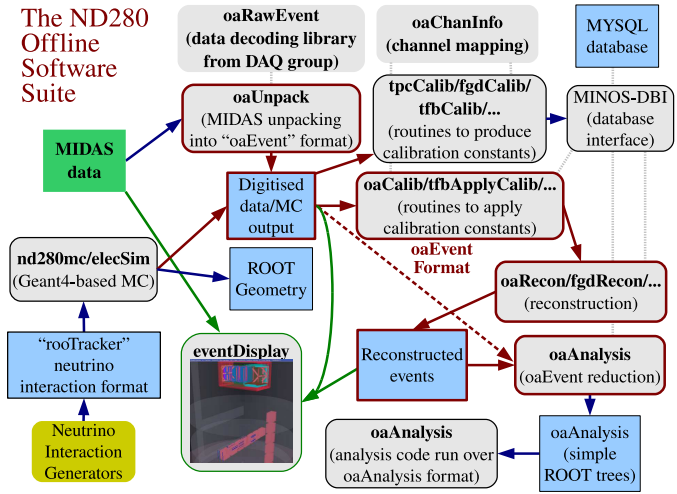}
\caption{Schematic of the package structure of the ND280 Software Suite. Only the most representative packages are included.}
\label{fig:nd280software}
\end{figure}

The general structure of the software suite is indicated in Fig.~\ref{fig:nd280software}.

The I/O library ``oaEvent'' defines the file format for the offline software,
which is used from the point that the raw MIDAS~\cite{midas} data files are converted for offline use,
up to the production of summarized format files at the end of the chain of processing.
It is actively managed to be small and stable. 
``oaRawEvent'' interfaces with the readout data format that is provided by the DAQ group,
and allows the raw MIDAS files to be read directly by the offline software.
Calibration constants for the detectors are stored on a centralized MySQL database,
and are applied by ``oaCalib'' and its sub-packages at processing time.
The access routines for the database are based on those developed for the MINOS experiment.

A representation of the geometry of the detectors is constructed in Geant4 code,
and is converted to ROOT TGeoManager~\cite{Brun:2003xr} format and stored in version-controlled files.
These are retrieved from a central repository to be used in the interpretation of raw data.

For Monte Carlo simulations,
interfaces have been built between 
the neutrino beam simulation, 
the neutrino interaction generation packages,
GENIE~\cite{Andreopoulos:2009rq} and NEUT~\cite{Hayato:2009zz} (see Section~\ref{SKsoft}),
and the ND280 software.
The neutrino fluxes estimated from beam MC are passed through the detector geometries,
and neutrino cross sections specific to the nuclei present in the geometries
are used to generate interactions
that are appropriate for the distribution of materials in the detector.

Geant4 is used to simulate the energy deposits from the final state particles
that pass through the detector,
and the response of the active detectors 
(scintillator bars, fibers, MPPCs and electronics, 
and TPC electron drift and electronics)
is simulated through custom-written code in the elecSim package.

Individual subdetectors have dedicated packages designed to reconstruct event information internal to them.
The RecPack toolkit~\cite{CerveraVillanueva:2004kt} is used as the framework for event reconstruction across the off-axis detector.
It is an independent software package, but has been developed in close conjunction with the ND280 software to meet its needs.

The full event information contained in the oaEvent format files
is distilled by the oaAnalysis package into files based on ``trees'' which are built up from pure ROOT objects.
An accompanying library of analysis tools helps end-users to process the summarized output using standardized C++ routines and Python macros.

An overall software control package allows for the fully automated running of the software,
based on simple configuration files which list the inputs and processing steps.

\subsubsection{Automated Support Tools}
A number of tools have been used by the ND280 software group to assist
in  simultaneous code development  across a  large number  of packages.
The Buildbot software~\cite{software-buildbot} performs automated 
builds of the full software suite on multiple computing platforms
to test for compilation problems and allow tests to be run.
The TUT framework~\cite{software-tut} provides a structure 
for regression tests of
code in the low-level packages,
to test performance and compliance to specifications.
Also, higher-level validation tests were written in multiple packages,
to flag problems that are introduced during development.

\subsubsection{Management}
One individual acts as the release manager,
overseeing the packages as they are combined to form ``releases'' several times a year.
In addition to this active management,
several tools are used to assist users in contributing to the overall evolution of the software:
Bugzilla~\cite{software-bugzilla}, a widely used management utility for tracking the development of software,
allows developers and end-users to file bugs and feature requests;
ViewVC~\cite{software-viewvc}, a browser-based tool to access all past versions of each file that forms the software suite,
tagged with the comments submitted as each change was committed to the repository;
and LXR~\cite{software-lxr}, a cross-referenced source code browser.

\subsubsection{Documentation}
The Doxygen~\cite{software-doxygen} system is used to generate documentation from comments that are embedded in the code.
An online workbook is also maintained to provide higher-level documentation on overall procedures and information for new users of the software and developers.

\subsubsection{Performance}
For the dataset from the first data-taking period in 2010,
the neutrino beam events in a single ``subrun'' file,
corresponding to approximately ten minutes of data,
take approximately one hour to process fully on a typical CPU.

% Put INGRID event display here
%Fig.~\ref{fig:INGRIDevent} shows an event display of an neutrino 
%event where a muon track originating from the INGRID proton module 
%and  continuing  into a regular INGRID module.

%\begin{figure}[htb]
%\centering\includegraphics[width=\linewidth]{./INGRID_event.eps}
%\caption{A schematic of a neutrino's journey from the neutrino beamline
%at J-PARC, through the near detectors (yellow dot) which are used to
%determine the properties of the neutrino beam, and then 295 km underneath
%the main island of Japan to Super-Kamiokande.}
%\label{fig:INGRIDevent}
%\end{figure}
%

\begin{figure*}[htb]
\centering\includegraphics[width=\linewidth]{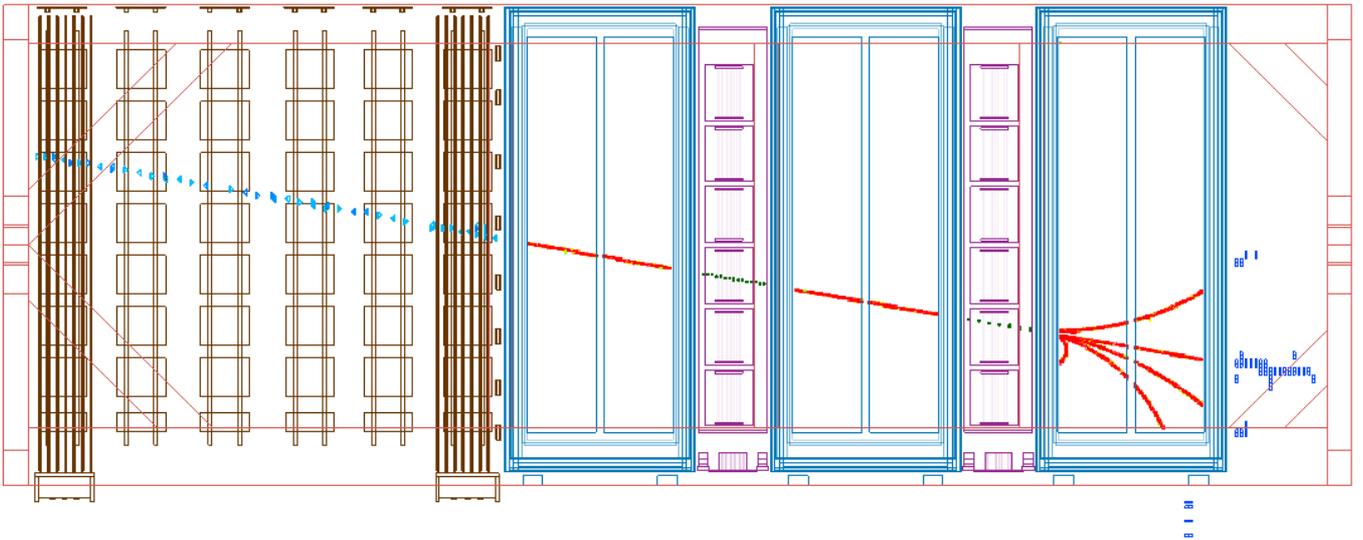}       
\caption{This event display shows an event with a muon track 
entering via the front face of the \pod{} detector, continuing
to the tracker (TPC and FGD) region and 
producing secondary particles on the way.
The secondary particles are then stopped 
in the ECal detectors.
}
\label{fig:ND280event}                                                        
\end{figure*}

Fig.~\ref{fig:ND280event} shows an event display of an
event with a muon track entering into the \pod{} and continuing into 
the tracker (TPC and FGD) region.
Multiple secondary particles are produced in the FGD,
all of which are finally stopped in the ECal detectors.

This event display illustrates an overall successful performance of the 
ND280 off-axis detector system (in terms of both hardware and software). 

%\input ND280_data.tex %Ushida
 %Linder/Uchida
\subsection{ND280 Data Processing and Distribution}
\label{ND280data}
% ND280 Data Processing and Distribution
% Originally data.tex 
% Lead Author: Uchida
% Rearranged and modified by Jung

\subsubsection{ND280 Data}
The ND280 detector produces raw data during normal data taking on the order of several MB a second.
Single raw data files are approximately 1 GB in size, and are recorded to disk approximately every ten minutes.
The DAQ group writes these to the HPSS storage system at the KEK Computing Center (KEKCC) as the primary archive for ND280 data.

\subsubsection{Data Distribution}
\begin{figure}[htb]
\centering\includegraphics[width=\linewidth]{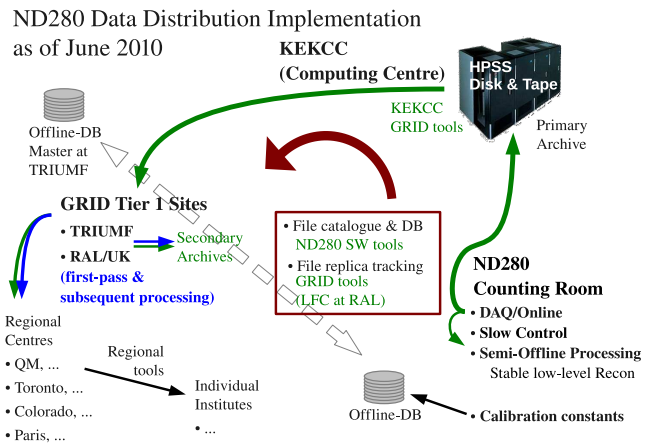}
\caption{Schematic of the flow of data from the ND280 counting room to the primary and secondary archive sites and the individual collaboration institutions.}
\label{fig:nd280datadistribution}
\end{figure}

Fig.~\ref{fig:nd280datadistribution} is a schematic of the flow of data from the ND280 counting room to the end-users, via the primary archive.
From the primary archive onwards, tools that were created for the GRID~\cite{Bird:2005js} are used to manage the flow and storage of data files.
LHC Computing GRID (LCG) utilities are used to transfer files to GRID storage elements at the RAL or TRIUMF laboratories in the U.K. and Canada respectively, where the data is also copied to long-term storage, for secondary archiving. 
Once the files are made available on the GRID, they are further distributed to different sites for processing using the LCG tools.

\subsubsection{Data and Monte Carlo Processing}
Subsequent processing of these files using the ND280 software suite may be either GRID-based, or based on independent computing clusters with their own batch-processing systems, e.g. SciNet~\cite{1742-6596-256-1-012026}.
Monte Carlo files are generated at computing sites across the collaboration, with different tasks assigned to suit the hardware capabilities of each site.
Once these files are generated, they are copied to the GRID, archived at RAL and TRIUMF, and distributed across the collaboration, in the same way as the raw data files.

\subsubsection{File Cataloging}
The procedure described above results in a large number of files, residing on a large number of data storage sites across the international collaborating institutes. 
Many of these are replicas of each other which are identical in content.
These are recorded and tracked using the LCG File Catalog (LFC) tools, which are based around a central catalog for all replica files.
This allows end-users to choose a replica that is situated closest to them for working on, or to copy files to a local storage element and register them on the LFC as replicas for subsequent use.
Processing jobs can also be sent to locations in which replica files already exist, to minimize the need to transfer data between sites.

 %Uchida                                           
\section{Super-Kamiokande Far Detector}
\label{SK}
% \section{Super-Kamiokande Far Detector}
% Lead Author: Shiozawa/Konaka/Walter

The world's largest land-based water Cherenkov detector, Super-Kamio\-kande,
serves as the far detector in the T2K experiment.
The detector is located 295 km west of the beam source
where it is used to sample the beam's flavor composition
and look for $\nu_{\mu} \to \nu_e$ appearance and $\nu_\mu$ disappearance.
Built 1 km deep within the center of Mt. Ikenoyama,
Super-Kamiokande
is a cylindrical cavern filled with 50 kton of pure water
within which the detector's roughly 13,000 photomultiplier tubes (PMTs)
image neutrino interactions.
Super-Kamiokande has been running since 1996
and has produced data for a number of well-known results that include
world-leading limits on the proton 
lifetime~\cite{Shiozawa:1998si, Hayato:1999az, Nishino:2009gd}
and the measurement of flavor oscillations
in atmospheric, solar and accelerator-produced 
neutrinos~\cite{Fukuda:1998mi, Fukuda:1998ua, Ahn:2006zza, Ashie:2005ik, Hosaka:2006zd}.
Over this time there have been four running periods:
SK-I, SK-II, SK-III, and SK-IV.
The latest period, SK-IV, is still in progress
and features upgraded PMT readout electronics.
SK-IV is also the period in which the T2K experiment takes place. 

Because of the detector's long-running operation,
the behavior of Super-Kamiokande is well understood.
The calibration of the energy scale is known to the percent level,
and the software for modeling events in the detector matches
calibration samples to the percent level as well.  

This section will review the components and operation of Super-Kamiokande,
describe the upgraded front-end readout electronics of the detector,
and provide an overview of the Monte Carlo simulation of events in Super-Kamiokande.
For a much more detailed description of the Super-Kamiokande detector,
refer to~\cite{Fukuda:2002uc}.

%\begin{figure}
%\begin{center}
%\includegraphics[width=\linewidth]{./T2K2SK.eps}
%\end{center}
%\caption{Placeholder for some similar figure? The figure comes from ~\cite{JHF%2001Itow}}
%\label{T2K2SK}
%\end{figure}

 %Shiozawa/Konaka/Walter
\subsection{Super-Kamiokande Detector Overview}
\label{SKoverview}
% \subsection{Super-Kamiokande Detector Overview}
% Lead Author: Shiozawa/Konaka/Walter

\begin{figure}
\begin{center}
\includegraphics[width=\linewidth]{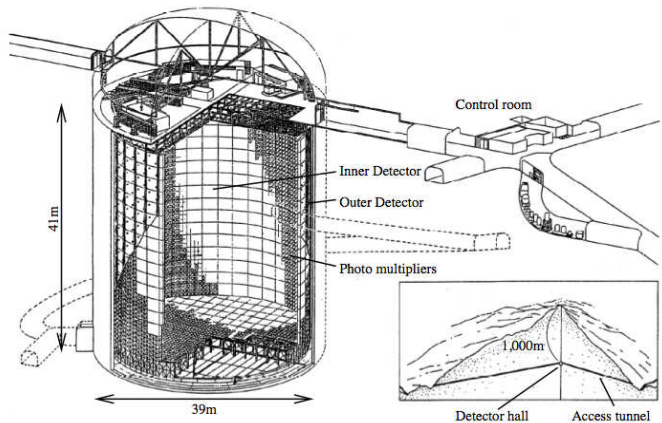}
\end{center}
\caption{Diagram of the Super-Kamiokande Detector. The detector is mainly comprised of two segments, the inner and outer detectors.  The boundary between the two segments is defined by a cylindrical scaffold used to mount photomultiplier tubes and optically separate the segments.  The figure comes from~\cite{Itow:2001ee}.}
\label{SK_JHFDiagram}
\end{figure}

The geometry of the Super-Kamiokande detector consists of two major volumes,
an inner and an outer detector
which are separated by a cylindrical stainless steel structure.
Fig.~\ref{SK_JHFDiagram} gives a schematic of the Super-Kamiokande detector geometry.
The inner detector (ID) is a cylindrical space
33.8 m in diameter and 36.2 m in height
which currently houses along its inner walls
11,129 inward-facing 50 cm diameter PMTs.
Enclosing the ID is the outer detector (OD)
which is a cylindrical space about 2 m thick radially and on the axis at both ends.
The OD contains along its inner walls 1,885 outward-facing 20 cm diameter PMTs.
The ID and OD boundaries are defined by a cylindrical structure about 50 cm wide.
This structure consists of a stainless steel scaffold covered by plastic sheets
which serve to optically separate the ID and OD.
The wall facing into the ID is lined with a black sheet of plastic
meant to absorb light and minimize the number of photons which 
either scatter off of the ID wall back into the ID volume, or pass through from the ID to the OD.
The walls facing the OD, however, are lined with the highly reflective material
Tyvek\textsuperscript{\textregistered}, in order to compensate for the OD's sparse instrumentation.
With the Tyvek\textsuperscript{\textregistered},
photons reflect off of the surface of the OD walls
and have a higher chance of finding their way to one of the OD PMTs.
Finally, within the stainless steel scaffold there is a 50 cm ``dead space'',
which combines with the ID and OD to make Super-Kamiokande a total of 39 m in diameter and 42 m in height.

The ID is well instrumented, with ~40\% PMT cathode surface coverage,
so that there is sufficient spatial resolution to infer
a number of physical quantities from the imaged neutrino interactions.
The ID is lined with Hamamatsu R3600 hemispherical PMTs,
which feature a combined quantum and collection efficiency of about 20\%.
Neutrino interactions often produce charged particles which,
if above an energy threshold,
produce a cone of Cherenkov photons as they traverse the water.
When the photons reach the PMTs on the detector walls
they produce a ring-shaped hit pattern
which is used to 
extract information about the interaction such as the event vertex position
and momenta of product particles.

The primary strategy to measure the flavor composition of the T2K neutrino beam at Super-Kamiokande,
and thereby observe the oscillation of $\nu_{\mu}$ to either $\nu_e$ or $\nu_{\tau}$,
is to count charged current quasi-elastic (CCQE) interactions for muon and electron neutrinos,
both of which produce leptons of their respective flavor.
Muons, counted to measure $\nu_\mu$ disappearance,
are resilient to changes in their momentum due to their relatively large mass.
As a result, muons that travel through the detector produce
a well-defined cone of Cherenkov radiation
which leads to a clear, sharp ring of PMT hits seen on the detector wall.
In contrast, electrons, used to search for $\nu_e$ appearance,
scatter more easily because of their smaller mass and 
almost always induce electromagnetic showers at the energies 
relevant to Super-Kamiokande.
The result of an electron-induced shower is a ``fuzzy'' ring pattern seen 
by the PMTs,  which can be thought of as the sum of many overlapping 
Cherenkov light cones.
The routines in the Super-Kamiokande event reconstruction software,
sketched out in Section~\ref{SKsoft},
use this difference between sharp and fuzzy to designate whether
the rings imaged in the detector derived from muon-like or electron-like particles.

In contrast to the ID, the OD is only sparsely instrumented
due to its original purpose as an active veto of cosmic ray muons and other backgrounds.
The PMT array in the OD, made of up 611 Hamamatsu R1408 PMTs and 1,274 R5912 PMTs,
is capable of an almost 100\% rejection efficiency of cosmic ray muon backgrounds.
However, by selecting events in coincidence with the T2K beam,
neutrino-induced events that illuminate the OD can be efficiently selected from background events.
From volume considerations alone,
the number of OD events is expected to roughly equal the number of events in the ID.
However, because of the OD's narrow geometry and sparse instrumentation,
the OD lacks spatial and temporal resolution,
which prevents detailed event reconstruction in the OD.
Nonetheless, the OD provides a sample of events
which can be categorized into three types:
events that produced light only in the OD;
events where both detector segments are illuminated
and the particles seem to originate from an interaction vertex inside the ID;
and events where both segments are illuminated
but the particles seem to originate from an interaction vertex 
outside of the ID.

The detector is calibrated through a number of sources,
both from introduced laser light and cosmic ray particles.

 %Shiozawa/Konaka/Walter
\subsection{Super-Kamiokande Electronics and DAQ Upgrade}
\label{SKelec}
% \subsection{Super-Kamiokande Electronics and DAQ Upgrade}
% Lead Author: Shiozawa/Konaka/Walter

In 2008, the Super-Kamiokande collaboration completed
an upgrade to the detector's readout electronics~\cite{SK2008NishinoElectronics, SK2009YamadaElectronics}
between the Super-Kamiokande data-taking periods SK-III and SK-IV.
This upgrade introduced a new scheme in the acquisition of hits from the detector's 13,014 PMTs.
In the old system,
once the total number of PMT hits within a 200 ns window exceeded a threshold,
a hardware trigger would fire
and direct the readout electronics to record data over a specified time window.
However, the old front-end's data throughput was too low to accommodate a number of neutrino searches at low threshold,
in particular those for solar neutrinos,
because the trigger rate at the required threshold level would overload the front-end electronics.
The new upgraded electronics, therefore,
includes a new front-end capable of a higher data processing rate.
Furthermore, it improves upon the triggering method of the old system.
In the new system, the arrival time and charge of each PMT hit
is sent to a cluster of PCs that organizes the hit data
and searches for event candidates based on programmable software triggers.
The new electronics' combination of higher throughput and flexible 
triggers, along with other improvements such as better impedance 
matching and a larger front-end dynamic range, improved Super-Kamiokande's 
ability to better accommodate a larger range of neutrino studies.  
For example, supernova relic neutrino searches require not only a 
lower threshold but also a more complicated delayed-coincidence trigger.  
The new electronics is also capable of implementing a coincidence trigger 
with a beam arrival time as in the case of the T2K experiment.

The new front-end boards are named QBEE
which stands for QTC Based Electronics with Ethernet.
The name describes the units at the start and end of the boards' signal processing chain.
The QTC (Charge to Time Converter)
is a custom ASIC that responds to input PMT pulses
by producing a square-wave pulse~\cite{Nishino:2009zu}.
The front edge of the QTC's output coincides with the arrival time of the PMT signal
and the length is proportional to the integrated charge of the PMT pulse.
The output of the QTC is then fed to a TDC (Time to Digital Converter)
that digitizes the QTC pulses' times and lengths.
Finally, the digitized data from the TDCs is sent to readout PCs using Ethernet technology
which provides the needed high rate of data transfer.
Custom-made network interface cards, which transfer the data,
consist of a TCP/IP firmware, called SiTCP~\cite{SK2006UchidaSiTCP},
and other interface logic routines that are installed on an FPGA chip.
The whole circuit on the QBEE board is able to transfer 11.8 MB/s of data
according to a test where analog pulses are sent through the QBEE.
This throughput corresponds to an input pulse rate of 80 kHz/channel
and is an order of magnitude improvement over the old system
which had a maximum hit rate of 1.4 kHz/channel.
Each QBEE has eight QTC chips,
and the whole DAQ system employs 550 QBEE boards
which together read out Super-Kamiokande's 13,014 PMTs
and send their hit information to a cluster of online PCs.

The online PCs' role in the DAQ system is to organize the PMT hit information from the QBEEs
and produce data files of candidate events which later undergo more offline analysis.
The PCs fall into three groups based on their task.
The first group consists of 20 ``Frontend'' PCs.
Each PC collects data from 30 ID QBEEs (20 OD QBEEs),
and then sorts the PMT hit information in order of time.
The second group of PCs, called ``Mergers'',
collects all hits into a time-ordered list of PMT hits.
They also apply a set of software triggers
to select event candidates from these lists.
There are ten Merger PCs, which each collect data from 30 QBEE boards.
For each candidate event,
a window is defined around the time of the event trigger
and all the information for hits falling within that window
is sent to a single ``Organizer'' PC.
The Organizer PC collects all of the candidate events, eliminating overlaps,
and writes them to disk for later offline analysis.
During a typical period of detector operation,
about 470 MB/s of data flows from the Super-Kamiokande PMTs through to the Merger PCs.
That stream of hit information results in a software trigger rate of 3 kHz
and eventually 9 MB/s worth of candidate event information being written to disk.

For the T2K experiment,
the DAQ system was extended to trigger in time with the beam spills produced by the J-PARC accelerator.
Each beam spill is given a GPS timestamp that is passed to the online Super-Kamiokande PCs.
Each timestamp is used to define an additional software trigger
that records all the hit information in a 1~ms window around the T2K beam arrival time.
These spill events are then collected and written to disk.
Later the events are fed into offline processing
which applies the usual Super-Kamiokande software triggers used to search for neutrino events,
and any candidate events found are extracted for further T2K data analysis.
%Removed June 2nd, with comment from Obayashi-san
%Fig.~\ref{T2KSKElectronics} displays a schematic of the DAQ setup.

%\begin{figure}
%\begin{center}
%\includegraphics[width=\linewidth]{./ElectronicsSchematic.eps}
%\end{center}
%\caption{A schematic of the T2K-SK DAQ setup. This is a place-holding figure for now.}
%\label{T2KSKElectronics}
%\end{figure}
 %Shiozawa/Konaka/Walter
\subsection{Super-Kamiokande Software and MC Simulation}
\label{SKsoft}
% \subsection{Super-Kamiokande Software and MC Simulation}
% Lead Author: Shiozawa/Konaka/Walter

The Super-Kamiokande software can be divided into four categories:
(1) neutrino event generators, NEUT and GENIE, used to simulate neutrino interactions in the Super-Kamiokande detector,
(2) SKDETSIM, which is responsible for modeling Super-Kamiokande's response to particles propagating through the detector,
(3) the T2K reduction software which selects neutrino candidate events from detector backgrounds and calibration events,
and (4) the Super-Kamiokande event reconstruction library.

\subsubsection{NEUT Neutrino Event Generator}
The first package, NEUT,
produces a list of neutrino interactions and their product particles from a given neutrino flux and energy spectrum.
Each interaction NEUT generates is the result of a primary neutrino-nucleon interaction
which produces particles that undergo a number of secondary interactions as they propagate out of the nucleus.
The primary interactions include
quasi-elastic scattering with both free and bound nucleons,
single meson production,
coherent pion production,
and deep inelastic scattering.
The secondary interactions are modeled using a particle cascade routine.
In the rest of this subsection, a quick overview of the interactions is given.
For a more complete description of NEUT, including the details on all of the interactions employed, see~\cite{Hayato:2002sd}.

The cross section for CCQE interactions is computed using one of two different methods
depending on whether the scattered nucleon is considered 
free or bound within a nucleus.
For free nucleons, which in the case of water are the protons in hydrogen,
the CCQE cross section is calculated using an effective hadronic current and can be found in~\cite{Smith:1971zm}.
Values for the form factors in the cross section are taken from experimental data~\cite{Grabosch:1988gw, Bernard:2001rs}.
For interactions on bound nucleons,
NEUT follows the method of~\cite{Smith:1972xh}
for computing the cross section,
in which the momentum distribution of the nucleons is dictated by the Fermi gas model and the outgoing nucleon momentum takes into account Pauli blocking effects.
The Pauli blocking model is implemented
by requiring that the momentum of the recoiling nucleon be
greater than the Fermi surface momentum.
This final momentum includes the cost of the nucleon first having to escape
the nuclear potential, whose values are from~\cite{Brieva:1977dv}.

NEUT simulates both incoherent and coherent single pion production.
Calculation of the incoherent production cross section follows from the method in~\cite{Rein:1981wg}
in which the decay of baryon resonances produces pions.
However, NEUT also uses this method to simulate the production of kaons and etas.
The excited resonances are restricted to an invariant mass below 2 GeV,
within which pions make up the majority of the outgoing mesons.
As for coherent pion production, the calculation of the cross section comes from~\cite{Rein:1983pf}.
In this interaction, the neutrino can be thought of as interacting with the entire nucleus.
The result is an enhancement of pion production in the forward direction,
because with the nucleus much heavier than the neutrino most of the incoming momentum is carried away by the outgoing pion.

The deep inelastic scattering (DIS) cross section is computed over an invariant mass energy, W,
greater than 1.3 GeV and employs the nucleon structure functions from 
GRV98~\cite{Gluck:1998xa}, with corrections suggested by Bodek and Yang.
Because a model of single pion production, described above,
is already implemented below 2 GeV,
the cross section calculation for 1.3 GeV $<$ W $<$ 2 GeV
only includes the probability function for pion multiplicities greater than 1 in order to avoid double-counting of this process.
The outgoing particles resulting from the DIS interaction
are modeled using PYTHIA/JetSet for interactions with W above 2 GeV
and custom-written routines for interactions below 2 GeV.
The custom routines were needed because PYTHIA/JetSet was developed for higher energy interactions.
For the interactions below 2 GeV,
experimental results are used to model the DIS event
with KNO scaling employed to interpolate between data points at different values of W~\cite{Saarikko:1979jr}. 

In addition to the primary interactions above,
NEUT models the secondary interactions between the nucleus and any products of the primary interaction,
which include mesons and recoiling nucleons.
For all of the particles,
an initial position of the neutrino interaction in the nucleus is generated in proportion to the Woods-Saxon type nucleon density distribution.
From there, a cascade model tracks the initial particle and any of its interaction products out of the nucleus.
The interactions during the cascade also include the effects of Fermi motion and Pauli blocking.
Of all of the particles, the secondary interactions for pions are the most important
due to both the large pion production cross section and the secondary interaction cross sections between the pion and the nucleus.
Intranuclear, final state interactions (FSI) of the hadrons produced by the primary neutrino
interaction are simulated by a microscopic cascade model. For $p_{\pi} < 500$~MeV/c, the mean
free paths for quasi-elastic scattering, charge exchange and absorption interaction channels
are calculated by~\cite{Salcedo:1987md}, while the scattering kinematics is determined
by~\cite{Rowe:1978fb} with in-medium corrections. For $p_{\pi} > 500$~MeV/c, hadron production
is also considered, and the mean free paths for all interactions, as well as the scattering
kinematics, are determined from fits to pion-proton cross section data~\cite{Arndt:2003if}.
The model reproduces pion scattering and photoproduction
data for nuclei with $A=12$~to~209~\citep[e.g.][]{Lee:2002eq, Ashery:1981tq, Ingram:1982bn, Allardyce:1973ce, Arends:1981ed}.
Secondary interactions for kaons, etas and recoil nucleons are also modeled by a cascade. 

\subsubsection{GENIE Neutrino Event Generator}
GENIE \cite{Andreopoulos:2009rq} simulates neutrino interactions,
for all neutrino flavors and all nuclear targets,
over the energy range from a few MeV to several hundred GeV.
Over this broad kinematic regime
a variety of scattering mechanisms are important.

Charged current quasi-elastic scattering is modeled using
an implementation of the Llewellyn-Smith model \cite{Smith:1971zm}.
The vector form factors are related,
via the CVC hypothesis,
to electromagnetic form factors measured in electron elastic scattering,
with the BBA2005 form factors \cite{Bradford:2006yz} used by default.
The pseudo-scalar form factor in the Llewellyn-Smith model 
has the form suggested by the PCAC hypothesis \cite{Smith:1971zm},
while the axial form factor is assumed to have a dipole form.
The default quasi-elastic axial vector mass is set to 0.99 GeV/c$^{2}$ \cite{Kuzmin:2006dh}.

Neutral current elastic scattering is simulated
according to the model described by Ahrens et al. \cite{Ahrens:1986xe}.
The axial form factor is again assumed to have dipole form
with the default axial vector mass set to 0.99 GeV/c$^{2}$.
The factor $\eta$
accounting for possible isoscalar contributions to the axial current
is set to a value of 0.12.

Charged and neutral-current baryon resonance production is described
using an implementation of the Rein-Sehgal model \cite{Rein:1981wg}.
From the 18 baryon resonances of the original paper,
\mbox{GENIE} includes the 16
that are listed as unambiguous in the PDG baryon tables \cite{Yao:2006px} 
and all resonance parameters have been updated.
Resonance interference and lepton mass terms
are ignored in the cross section calculation, 
although lepton mass terms are taken into account
in determining the phase space boundaries.
The default value of the resonance axial vector mass is 1.12 GeV/c$^{2}$,
as determined from global fits \cite{Kuzmin:2006dh}.

Charged and neutral-current deep inelastic scattering is described
using the effective leading order model of Bodek and Yang \cite{Bodek:2002ps}.
The model provides corrections
to be applied on the GRV98 LO parton distributions \cite{Gluck:1998xa}
in order to yield better agreement with low Q$^{2}$
electron scattering data.
Higher twist and target mass corrections are accounted for
through the use of a new scaling variable.
The longitudinal structure function is taken into account
using the Whitlow parametrization \cite{Whitlow:1990gk}.
An overall scale factor of 1.032
is applied to the predictions of the Bodek-Yang model
to achieve agreement with the measured value of
the neutrino cross section at high energy (100 GeV).
The same model is extended in the transition region
(hadronic invariant mass below 1.8 GeV/c$^{2}$)
to simulate non-resonance backgrounds.
To avoid double-counting in the transition region,
the extrapolated non-resonance background is scaled
so that the sum of the resonance and non-resonance contribution 
fits exclusive one-pion and two-pion cross section data.

Charged and neutral-current coherent pion production is modeled
using an implementation of the Rein-Sehgal model \cite{Rein:1983pf}.
The model uses the PCAC form at Q$^{2}$
and assumes a dipole dependence for non-zero Q$^{2}$,
with an axial mass of 1.99 GeV/c$^{2}$.
GENIE uses a recent revision of the Rein-Sehgal model \cite{Rein:2006di}
where the PCAC formula was updated to take into account lepton mass terms. 

Quasi-elastic and deep-inelastic charm production is also simulated.  
Quasi-elastic charm production is simulated using
an implementation of Kovalenko's local duality inspired model \cite{Kovalenko:1990zi}
tuned to recent NOMAD data \cite{Bischofberger:2005ur}.
Deep-inelastic charm production is modeled according to
the Aivazis, Olness and Tung model \cite{Aivazis:1993kh}.
Charm production fractions for neutrino interactions
are taken from \cite{DeLellis:2002pr} and
the Peterson fragmentation functions are used \cite{Peterson:1982ak}.
The default value of the charm mass is set to 1.43 GeV/c$^{2}$.

To calculate neutrino cross sections for scattering off nuclear targets,
GENIE employs the impulse approximation.
The nuclear environment is described using a simple
Fermi gas model with a modification by Bodek and Ritchie
to include nucleon-nucleon correlations \cite{Bodek:1981wr}. 

Neutrino-induced hadronization is simulated using the AGKY model \cite{Yang:2009zx}.
At low invariant masses, AGKY uses an empirical model anchored to bubble chamber
measurements of the average hadronic multiplicity and
multiplicity dispersion for
neutrino/antineutrino scattering off hydrogen and deuterium.
For hadronic invariant masses above 3 GeV/c$^{2}$,
AGKY integrates the PYTHIA-6 fragmentation model \cite{Sjostrand:2000wi}. 
There is a smooth transition between the models
to ensure continuity of all simulated observables.

Hadrons produced in the nuclear environment
do not immediately re-interact with their full cross section.
During the time it takes for quarks to materialize as hadrons, 
they propagate through the nucleus with a dramatically reduced interaction probability.
This is implemented in GENIE using a step during which no re-interaction can occur.
The step size is determined from a formation time of 0.523~fm/c,
according to the SKAT model \cite{Baranov:1984rv}.

Once hadrons are fully formed,
the effect of nuclear matter in hadron transport is taken into account
using a GENIE sub-package called INTRANUKE \cite{dytman-ladek}. 
INTRANUKE is an intranuclear cascade MC which tracks pions and nucleons.
The hadron re-interaction probability is calculated based on the local density of 
nucleons \cite{DeJager:1987qc}
and a partial wave analysis of
the large body of hadron-nucleon cross sections \cite{Arndt:2003if}. 
Once it has been decided that a hadron re-interacts in the nucleus,
the type of interaction is determined based on experimentally measured cross sections. 
Where data is sparse,
cross section estimates are taken from calculations of the CEM03 group \cite{Mashnik:2005ay}.
A number of interaction types are simulated: 
elastic and inelastic scattering,
single charge exchange,
and pion production and absorption followed by multi-nucleon knockout.

\subsubsection{SKDETSIM}
The propagation of the particles generated by NEUT across the Super-Kamiokande detector is handled by SKDETSIM,
a program library based on the GEANT3 particle propagation package.
For hadronic interactions in water,
the CALOR physics package is used because of its success at reproducing pion interactions around 1 GeV~\cite{Zeitnitz:1994bs}.
For pions with momentum below 500 MeV, however, custom routines have been written~\cite{Nakahata:1986zp}.
For the propagation of light in water,
SKDETSIM considers
absorption, Rayleigh scattering, and Mie scattering
as possible interactions.
The parameters employed in the model of these processes have been tuned using a number of laser calibration sources.

\begin{figure}[htb]
  \centering
  \includegraphics[width=0.9\linewidth]{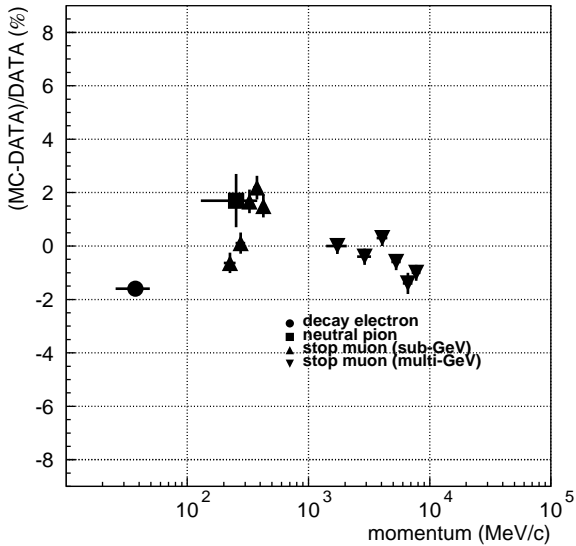} 
   \caption{Comparison between data and the SKDETSIM prediction of the reconstructed
     momentum of different event samples derived from cosmic ray
     events observed in Super-Kamiokande. For some samples, the comparison is made over different
     momentum values.  `Decay Electron' refers to the
     reconstructed momentum spectrum of electrons from the
     decay of cosmic ray muons.  `Neutral Pion' is
     the reconstructed momentum of neutral pions created from 
     atmospheric neutrinos.  The `Stop Muon' points indicate the
     reconstructed momentum of cosmic ray muons which have been
     determined to stop inside the inner detector.}
  \label{fig:skdetsim_data_comparison}
\end{figure}

The overall agreement for SKDETSIM is checked
using a number of different cosmic ray samples.
For example, the reconstructed momentum spectrum of
electrons produced from the decay of cosmic ray muons
is compared to the predicted reconstructed spectrum from SKDETSIM.  
For all samples, the agreement between data and simulation is
within a few percent (Fig.~\ref{fig:skdetsim_data_comparison}).

\subsubsection{T2K Data Reduction}

The T2K reduction software first selects
events tagged with the T2K beam trigger from the raw data
and then categorizes the beam data into three mutually exclusive samples. 
Events that are above a certain energy threshold 
and contain particle tracks that originate and terminate within the ID 
are labeled as ``fully contained'', or FC, events.  
The number of hits seen in the OD is used to determine whether 
any particles leave or enter the ID, 
and so the FC events are required to have no more than 15 hits in the largest OD hit cluster. 
Those events which exceed this limit are instead categorized as
``outer detector'', or OD, events.  
Events with energy below the FC energy threshold 
that still display PMT hit patterns suggestive of neutrino interactions, 
are classified as `low energy'', or LE, events.  
After categorization, the reduction software then applies a set of cuts 
unique to each sample to remove backgrounds.  
The events which pass the cuts are used for physics analyses.

\subsubsection{Event Reconstruction}
The reconstruction of FC and OD events in the detector occurs in four stages.  
First,  an initial event vertex is established from the timing of the PMT hits, 
and an initial track direction is calculated by searching for a well-defined edge in the PMT charge pattern.  
Second, an iterative algorithm utilizes a Hough transform of the PMT charge distribution to search for Cherenkov ring candidates.   
In the third step, a particle identification (PID) algorithm then 
classifies all the candidate rings observed as either muon-like or electron-like 
by comparing the observed pattern of charge to 
an analytically calculated expected pattern in the case of muons 
and an MC-generated expected pattern in the case of electrons.  
Finally, the reconstructed momentum for each particle is determined 
from the distribution of observed charge assigned to the particle's Cherenkov ring. 
The relationship between the observed charge and the particle's momentum 
is established using Monte Carlo simulation and 
detector calibrations from a number of different sources. 
More details on the event reconstruction algorithm can be found in~\cite{Ashie:2005ik}.

\begin{figure*}[htb]
  \centering
  \subfloat[muon-like event]{ \label{fig:mu-like} \includegraphics[width=0.45\linewidth]{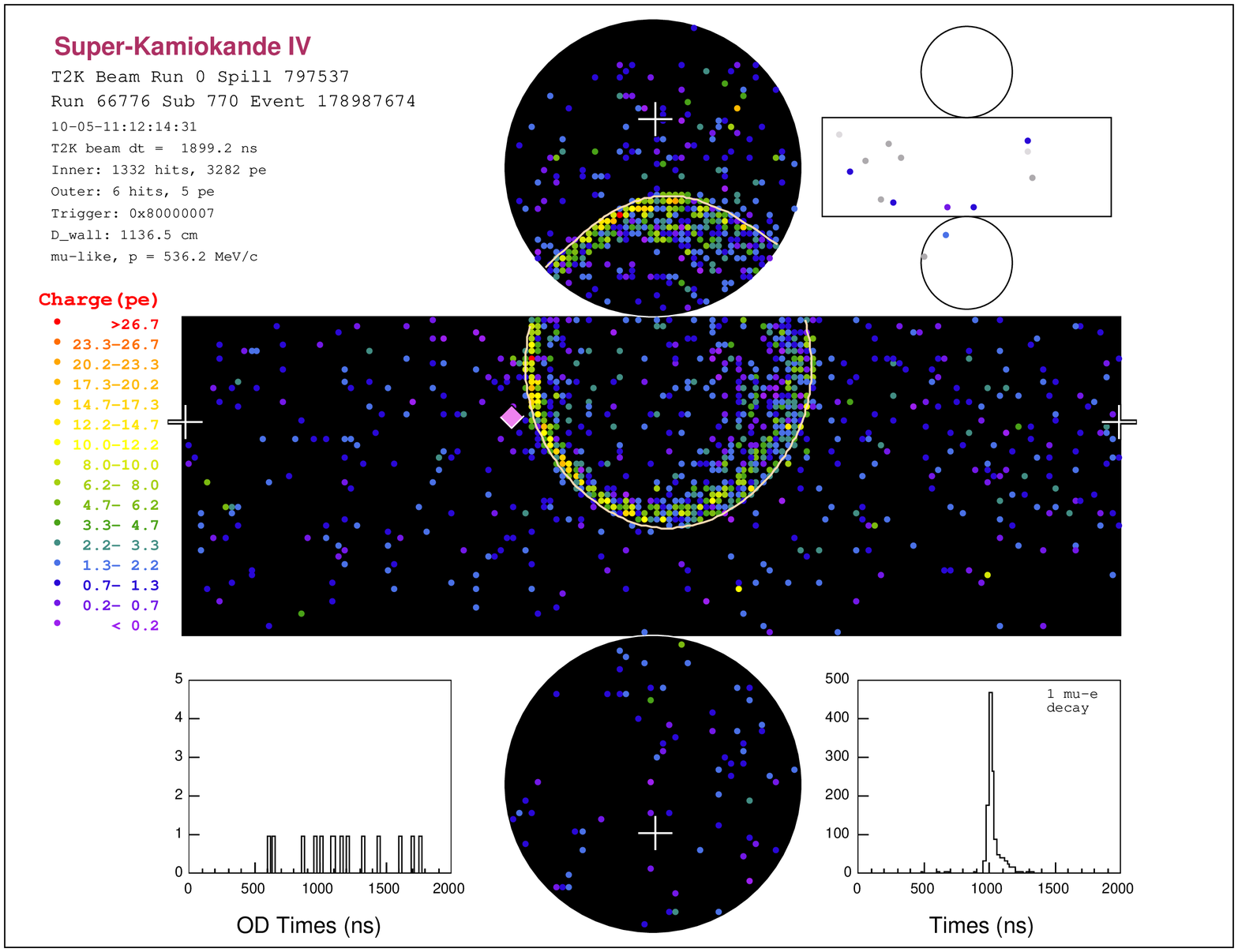} }
  \subfloat[electron-like event]{ \label{fig:e-like} \includegraphics[width=0.45\linewidth]{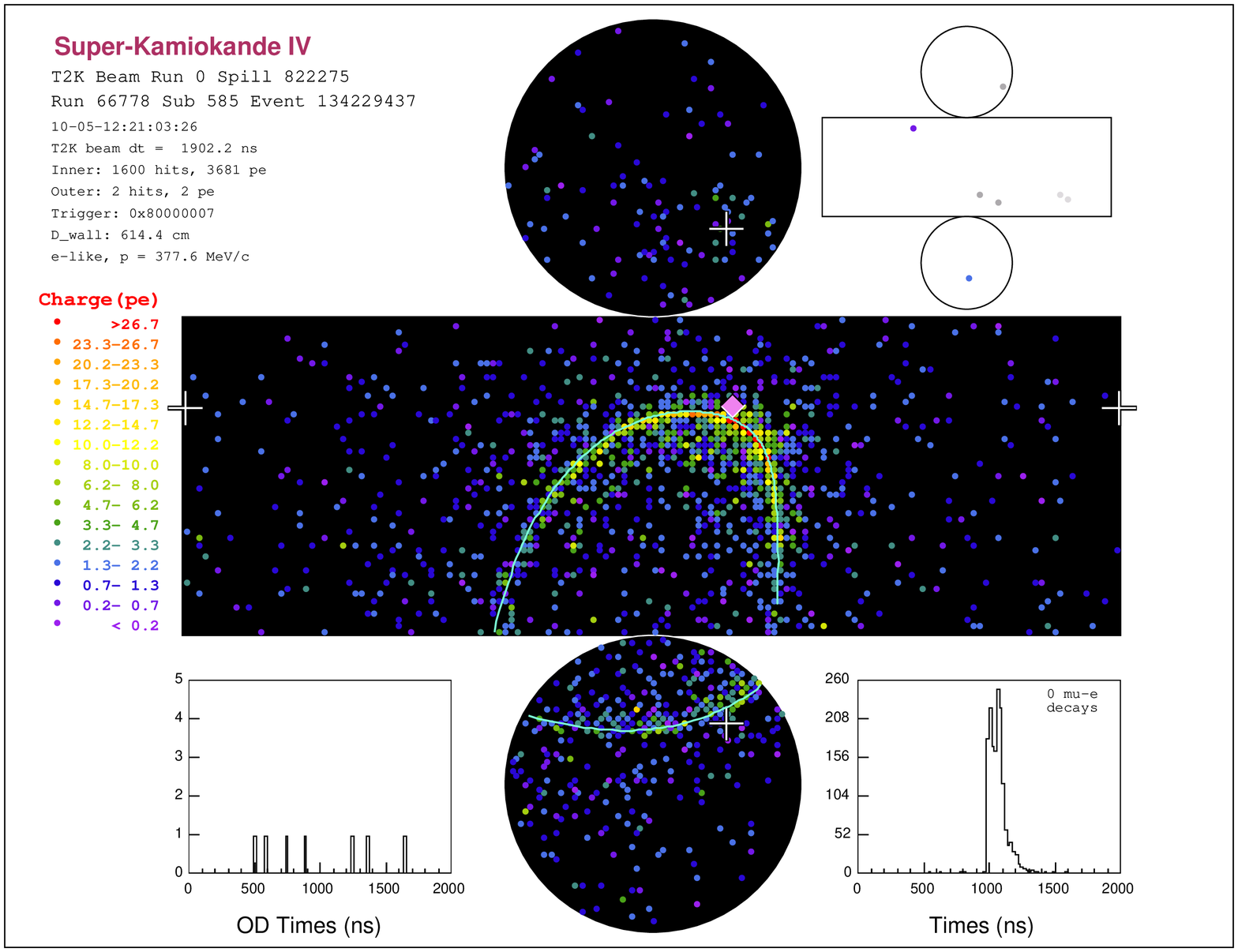} }                
  \caption{Example of reconstructed T2K events in Super-Kamiokande for
    (a) a muon-like ring and
    (b) an electron-like ring.
    Both figures show the cylindrical detector, unrolled onto a plane.
    Each colored point represents a PMT, with the color corresponding to the amount of charge, and the reconstructed cone is shown as a white line.
    The second figure in the upper right corner shows the same hit map for the OD.
    The white crosses indicate the location of the reconstructed vertex.
    The diamond marks the location  where a ray
    starting from the event vertex and
    heading in the direction of the beam would intersect the detector wall.}
\label{fig:SK_event_display}
\end{figure*}

Fig.~\ref{fig:SK_event_display} shows an event display of two T2K neutrino 
beam interaction events. The left display shows a ``muon-like'' Cherenkov 
ring with a sharp outer ring edge, which can be compared to the 
right display's ``electron-like'' Cherenkov ring with its characteristic 
fuzzy edges.

These event displays illustrate an overall successful performance
of the Super-Kamiokande detector system (in terms of both hardware 
and software).

%\begin{figure}[htb]
%  \centering
%  \includegraphics[width=1.0\linewidth]{./run066776_770_0178987674.eps%}
%  \caption{A reconstructed muon-like T2K event in Super-Kamiokande.  The white crosses %indicate the location of the reconstructed vertex.  The diamond marks the loca%tion  where a ray starting from the event vertex and heading in the direction %of the beam would intersect the detector wall. The hit map in the upper right %hand corner is for the Outer Detector.}
%\end{figure}
%
%\begin{figure}[htb]
%  \centering
%  \includegraphics[width=1.0\linewidth]{./run066778_585_0134229437.eps%}
%  \caption{A reconstructed electron-like T2K event in Super-Kamiokande.  The white cros%ses indicate the location of the reconstructed vertex.  The diamond marks the %location  where a ray starting from the event vertex and heading in the direct%ion of the beam would intersect the detector wall.  The hit map in the upper r%ight hand corner is for the Outer Detector.}
%\end{figure}

 %Shiozawa/Konaka/Walter
% Move these to ND280
%\section{T2K Data Processing and Distribution}
%\label{data}
%\input data.tex %Uchida --> ND280_data.tex
\section{Conclusion}
\label{conclusion}
% \section{Conclusion}
% Lead Author: Jung

%The T2K experiment will deepen our understanding of the mixing 
%in the lepton sector and provide us with critical information that will 
%guide us in our quest to discover charge-parity non-conservation, 
%known as CP violation, in the lepton sector.   

The T2K experiment, which comprises a new neutrino beamline and new 
near detector complex at 
%the 
J-PARC 
%site in Tokai, Japan, 
and the upgraded Super-Kamiokande detector, 
%at the University of Tokyo Kamioka Observatory in Kamioka, Japan, 
has been constructed successfully.
%and for the most part on schedule and on budget. 
The construction of the neutrino beamline started in
April 2004. The complete chain of accelerator and
neutrino beamline was successfully commissioned
during 2009. T2K began accumulating the neutrino beam data
for physics analysis in January 2010.
%The total project cost was approximately \$200M (in U.S. dollars).
Essentially all components of the experiment perform as expected and there 
have been no major problems. For example, 
during the 2010 data taking run, the proportion of good spills for neutrino
beams to the total number of spills delivered on target was 99.0\%. 
The overall live times of the INGRID 
%(ND280 on-axis) 
detector, the ND280 off-axis detector system and 
the Super-Kamiokande detector with regard to the delivered neutrino beam were
99.9\%, 96.7\% and 99.9\% respectively, demonstrating reliable operation of 
all of the subcomponents of the T2K experiment.

In this paper, we have described the basic structure and parameters
of the detector hardware, electronics, online DAQ system, and offline data
reduction and distribution scheme. More detailed descriptions of subdetectors
can be found in separate papers, 
some of which have already been published while
others are being prepared.

%The T2K collaboration is excited by the prospect of anticipated 
%fruitful scientific findings resulting from this magnificent experiment.

\bigskip
\noindent
{\bf Acknowledgments}

The authors thank the Japanese Ministry of Education, Culture, Sports, 
Science and Technology (MEXT) for their support for T2K.
%the J-PARC, KEK and JAEA 
%Directorates for their strong support and encouragement.
%T2K is made possible by the inventiveness, diligence and dedication of the
%J-PARC accelerator group. 
%The authors gratefully acknowledge the cooperation of
%the Kamioka Mining and Smelting Company. 
The T2K neutrino beamline, the ND280 detector and the Super-Kamiokande 
detector have been built and operated using funds provided by:
the MEXT, Japan; the Natural Sciences and Engineering Research
Council of Canada, TRIUMF, the Canada Foundation for Innovation, and the
National Research Council, Canada;
Commissariat \`a l'Energie Atomique, and 
Centre National de la Recherche Scientifique -- 
Institut National de Physique Nucl\'eaire et de Physique des Particules, 
France;
Deutsche Forschungsgemeinschaft, Germany;
Istituto Nazionale di Fisica Nucleare, Italy;
the Polish Ministry of Science and Higher Education, Poland;
the Russian Academy of Sciences, the Russian Foundation for Basic Research,
and the Ministry of Education and Science of the Russian Federation, Russia;
the National Research Foundation, and the Ministry of Education, 
Science and Technology of Korea, Korea;
Centro Nacional De F\'isica De Part\'iculas, Astropart\'iculas y Nuclear, 
and Ministerio de Ciencia e Innovacion, Spain;
the Swiss National Science Foundation
and the Swiss State Secretariat for Education and Research, Switzerland;
the Science and Technology Facilities Council, U.K.; and 
the Department of Energy, U.S.A.

We are also grateful to CERN for their generous donation of the 
UA1/NOMAD magnet for our use and their assistance with the preparation and 
operation of the magnet; and to DESY for their generous donation of the 
HERA-B magnet movers.
% ; and to Fermilab for their assistance in the production of 
% Trip-T chips and scintillator bars, and treatments of wavelength-shifting 
% fibers.

In addition, the participation of individual researchers
and institutions in the construction of the T2K experiment has been
further supported by funds from: the European Research Council;
the Japan Society for the Promotion of
Science Fellowship for Foreign Researchers program;
the U.S. Department of Energy Outstanding Junior Investigator (OJI) Program 
and Early Career program; and the A. P. Sloan Foundation.

The authors also wish to acknowledge the tremendous support provided by 
the collaborating institutions, including but not limited to:
%(*** remove this in the final version: Add institutions that have 
%provided at least about \$100k of support from the institutional 
%resources not including the supports from funding agencies.***)
Ecole Polytechnique -- Palaiseau;
and the State University of New York at Stony Brook, 
Office of the Vice President for Research.
 
 %Jung



%% References


\bibliographystyle{model1a-num-names}
\bibliography{NIMpaper}

\begin{thebibliography}{116}
\bibitem[{Thorpe et~al.(2010)}]{Thorpe:2010}
\bibinfo{author}{M.~Thorpe}, et~al., in: \bibinfo{booktitle}{Proceedings of the
  17th IEEE Real Time Conference}. \bibinfo{note}{To be published}.
\end{thebibliography}





% \end{thebibliography}

%\end{multicols}
\end{document}